\documentclass[aps,prl,twocolumn,superscriptaddress]{revtex4-1}
\usepackage{color}

\usepackage{graphicx}
\usepackage{mathtools}
\usepackage{verbatim}
\usepackage{amsmath}
\usepackage{caption}
\usepackage{subcaption}
\usepackage{appendix}
\usepackage{esint}
\usepackage[font={small,it}]{caption}

\begin{document}

\title{Broadband Electrical Action Sensing Techniques with conducting wires for low-mass dark matter axion detection}

\author{Michael E. Tobar}
\email[]{michael.tobar@uwa.edu.au}
\affiliation{ARC Centre of Excellence For Engineered Quantum Systems, Department of Physics, School of Physics and Mathematics, University of Western Australia, 35 Stirling Highway, Crawley WA 6009, Australia.}
\author{Ben T. McAllister}
\affiliation{ARC Centre of Excellence For Engineered Quantum Systems, Department of Physics, School of Physics and Mathematics, University of Western Australia, 35 Stirling Highway, Crawley WA 6009, Australia.}
\author{Maxim Goryachev}
\affiliation{ARC Centre of Excellence For Engineered Quantum Systems, Department of Physics, School of Physics and Mathematics, University of Western Australia, 35 Stirling Highway, Crawley WA 6009, Australia.}

\date{\today}

\begin{abstract}

Due to the inverse Primakoff effect, it has been shown that when axions mix with a DC $\vec{B}$-field, the resulting electrical action will produce an AC electromotive force, which oscillates at the Compton frequency of the axion. As in standard electrodynamics, this electromotive force may be modelled as an oscillating effective impressed magnetic current boundary source. We use this result to calculate the sensitivity of new experiments to low-mass axions using the quasi-static technique, defined as when the Compton wavelength of the axion is greater than the dimensions of the experiment. First, we calculate the current induced in a straight conducting wire (electric dipole antenna) in the limit where the DC $\vec{B}$-field can be considered as spatially constant and show that it has a sensitivity proportional to the axion mass. Following this we extend the topology by making use of the full extent of the spatially varying DC $\vec{B}$-field of the electromagnet. This is achieved by transforming the 1D conducting wire to a 2D winding with inductance, to fully link the effective magnetic current boundary source and hence couple to the full axion induced electrical action (or electromotive force). We investigate two different topologies: The first uses a single winding, and couples to the effective short circuit current generated in the winding, which is optimally read out using a sensitive low impedance SQUID amplifier: The second technique uses multiple windings, with every turn effectively increasing the induced voltage, which is proportional to the winding number. The read out of this configuration is optimised by implementing a cryogenic low-noise high input impedance voltage amplifier. The end result is the realisation of new Broadband Electrical Action Sensing Techniques with orders of magnitude improved sensitivity over current low-mass axion experiments, with a sensitivity linearly proportional to the axion photon coupling and capable of detecting QCD dark matter axions in the mass range of $10^{-12}-10^{-8}eV$ and below.

\end{abstract}

\pacs{}

\maketitle

\section{Introduction}

Axions are neutral spin-zero bosons, which should exist to solve the strong charge-parity problem in QCD and have been postulated to be cold dark matter \cite{PQ1977,Wilczek1978,Weinberg1978,wisps,K79,Kim2010,Zhitnitsky:1980tq,DFS81,SVZ80,Dine1983,Preskill1983,Sikivie1983,Sikivie1983b}. Axions modify electrodynamics through the axion two photon coupling \cite{Wilczek:1987aa} so intrinsically the system consists of three degrees of freedom, two photon and one axion. If we were to consider the whole three degrees of freedom of the axion coupled to two photons, the overall system would be conservative. In fact, a recent analysis of the conserved quantities in the entire three degrees of freedom of the interaction was undertaken \cite{ORT20}, which concluded that the conserved charges are the sum of the ``Noether'' electric and magnetic charge induced by the dynamical axion field.

Modified axion electrodynamics represents the equations of motion from only the point of view of the two photonic degrees of freedom. Recently it was shown that the equations of motion can be described by non-conservative electrodynamics, where the axion dynamics enters as a forcing function to standard electrodynamics through an extension of the constitutive relations generated by an external impressed boundary source, such as an impressed effective magnetic current \cite{TobarModAx19}. Here, the forcing function terms are mixing terms between the axion and photons, which present as a product of the axion scalar field, $a(t)$, with either the electric field, $\vec{E}$, or magnetic field, $\vec{B}$ of the photons. However, for axion experiments to reach predicted QCD model limits, it is common that one of the photonic degrees of freedom becomes a large solenoidal or toroidal DC $\vec{B}$-field, $\vec{B}_{DC}(\vec{r})$ \cite{Wuensch,hagmann1990,Bradley2003,ADMXaxions2010,ADMX2011,Sikivie2014a,7390193,McAllisterFormFactor,McAllister:2016fux,Klash,JEONG2018412,Baker2012,ABRACADABRA,DMRadio17,HOANG2017,PhysRevD.96.061102,FirstAbra}. For this configuration $\vec{B}_{DC}(\vec{r})$ acts as a mediator to convert axions to photons through the second photonic degree of freedom (inverse Primakoff effect), creating photons oscillating at the Compton frequency of the axion, $\omega_a$. To create $\vec{B}_{DC}(\vec{r})$, an impressed DC electrical current, $\vec{J}_{DC}^i$, is necessary to drive an electromagnet solenoidal or toroidal coil of many turns, and in this case the axion-induced forcing function driving the system is an electrical action proportional to, $a(\omega_at)\vec{B}_{DC}(\vec{r})$.

It is well-known in standard electrodynamics, that an ideal voltage source, which converts an external non-electromagnetic energy source into electromagnetic energy via an electromotive force (emf), can be modelled by an effective impressed magnetic current localised at the boundary of the voltage source (without invoking magnetic monopoles) \cite{RHbook2012,Balanis2012,TobarElectret}. Such a voltage source may be described by a non-conservative electric field (or impressed electric field) with a vector direction governed by the left hand rule of the effective magnetic current, and can also be characterized with an electric vector potential. Axion modified electrodynamics exhibit the same characteristic, and can be described by a non-conservative process with respect to the photonic degrees of freedom, and has been shown to have a similar localised effective impressed magnetic current boundary source with an electric vector potential \cite{TobarModAx19,Kinsler20,VISINELLI}. Thus, the dynamic action of the axion creates oscillating electromagnetic fields in a similar way to an AC voltage source in electrodynamics, except with no $\pm$ voltage terminals. 

The purpose of this work is to lay the foundations of some new Broadband Electric Action Sensing Technique (BEAST) for low-mass axion dark matter detection when applying DC magnetic fields. We achieve this by solving the axion modified equations in the quasi-static limit to explore experiments, which are directly sensitive to the axion generated emf. These new techniques implement a conducting wire antenna or coil windings, which are predicted to have very small currents and voltages induced by the QCD axion. However, we calculate that these can be measured by using low noise detection techniques based on Superconducting Quantum Interference Devices (SQUIDs) or High Impedance Amplifiers (HIAs).

Other low-mass axion detection experiments, which utilise axion-photon coupling are underway by utilising magnetic field or phase sensing \cite{Sikivie2014a,ABRACADABRA,DMRadio17,freqmetrology}, with first results of some of these experiments recently published \cite{FirstAbra,Nicol19,Cat19}. We show that our sensitivity calculations are consistent with others, which calculate that the electric field produced by the axion current in the low-mass limit is suppressed \cite{Ouellet2018,Beutter18,Younggeun18}. \textit{Our technique does not detect this suppressed electric field, but directly detects the electric action (or emf) from the non-conservative process}. This is a much more sensitive physically measurable technique because it is not suppressed by the mass of the axion and puts axion-photon coupling low-mass dark matter experiments on a similar footing to axion-gluon coupling experiments \cite{Graham11,altmethods2013} such as the CASPEr \cite{BudkerPRX,Budker19}, Sussex-RAL-ILL nEDM experiments \cite{RALnEDM17}, which are not suppressed by the axion mass and have also been projected to be sensitive enough to detect QCD axions.

\section{Quasi-Static Solution under DC Magnetic field}

In this section we summarise the quasi-static solution of axion electrodynamics under DC magnetic field, which has been presented more generally in \cite{TobarModAx19}. More details on how we arrived at equations (\ref{ExStatic})-(\ref{EaB(t)1}) are also given in Appendix A. 

\begin{figure}[t]
\includegraphics[width=1.0\columnwidth]{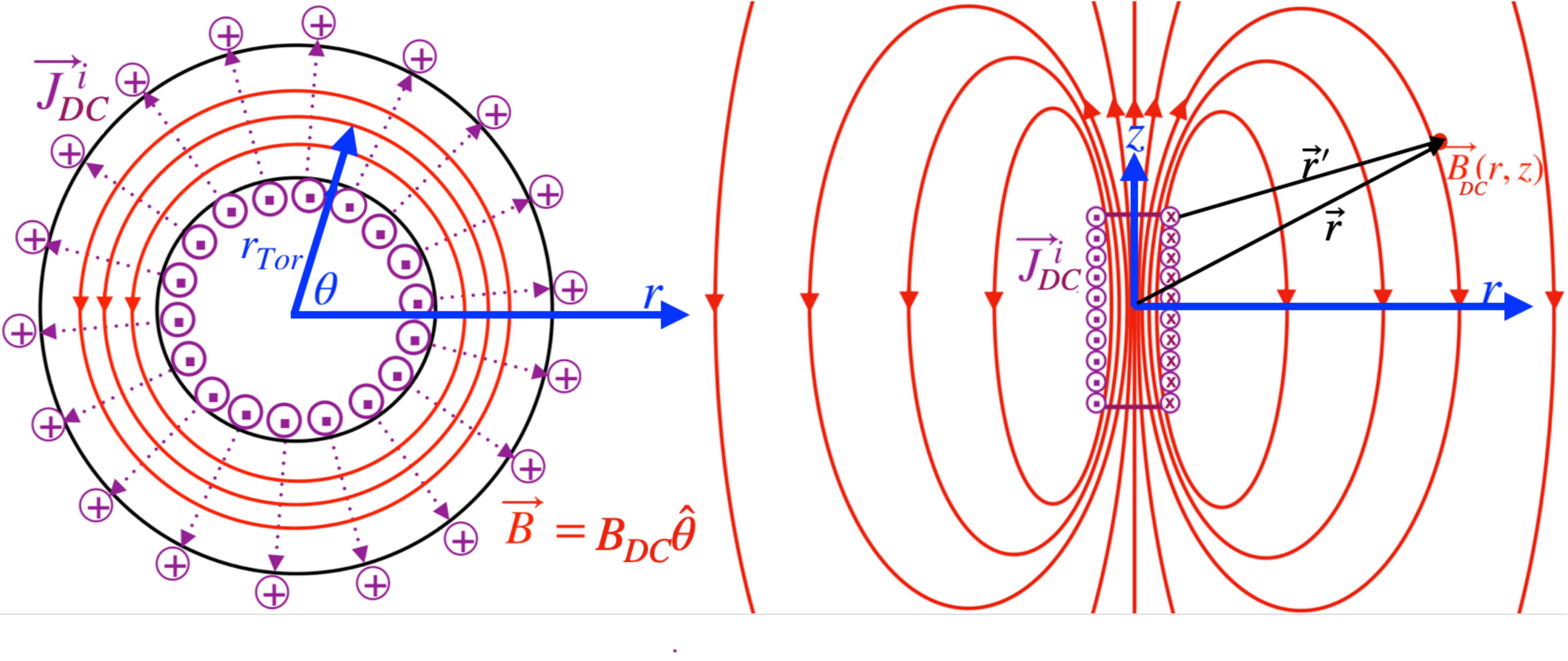}
\caption{Left: The $r-\theta$ plane of a toroidal magnet, with the impressed DC electrical current density, $J_{{DC}}^i$, creating a DC magnetic field of $\vec{B}=\vec{B}_{DC}\hat{\theta}$ in this plane. Right: The $z-r$ plane of a solenoidal magnet, with the impressed DC electrical current density, $J_{{DC}}^i$, creating a DC magnetic field of $\vec{B}_{DC}(r,z)$.}
\label{TorSol}
\end{figure}

To become sensitive to axions it is common to have one of the photonic degrees of freedom represented by a large DC magnetic field, and to create such a field a large impressed DC electrical current, $\vec{J}^i_{DC}$, is necessary. Assuming there is no impressed DC electric field or DC static charge in the system, then the external field is given by,
\begin{equation}
\begin{aligned}
&\vec{\nabla} \cdot \vec{B}_{DC}(\vec{r})=0 \\
&\vec{\nabla} \times \vec{B}_{DC}(\vec{r})=\mu_{0} \vec{J}^i_{DC},
\end{aligned}
\label{ExStatic}
\end{equation}
Here, equation (\ref{ExStatic}) describes the excitation of the coil winding of the electromagnet with an impressed free current, $\vec{J}^i_{{DC}}$, which can be considered as a boundary source that creates the DC magnetic field through the static Ampere's law, so that,
\begin{equation}
\vec{B}_{DC}(\vec{r})=\frac {\mu_0}{ 4\pi} \int_{\Omega } \frac {\vec{\nabla} \times \vec{J}^i_{DC} \left( \vec{ r } ^ { \prime } \right) } { \left| \vec{r}-\vec{r} ^ { \prime } \right| } \mathrm{d}^{3}\vec{r }^{\prime},
\label{Bo}
\end{equation}
Here $\vec{B}_{DC}(\vec{r})$ at point $\vec{r}$ is calculated from the impressed electrical current at distant position $\vec{r}^{\prime } $. The location $\vec{r}^{ \prime } $ is a source point within volume $\Omega$ that contains the free current distribution. The integration variable, ${d} ^{3}\vec{r}^ { \prime }$, is a volume element around position $r^{\prime}$. Examples of the impressed current exciting a coil and producing a magnetic field are shown in Fig. \ref{TorSol}.

\begin{figure}[t]
\includegraphics[width=1.0\columnwidth]{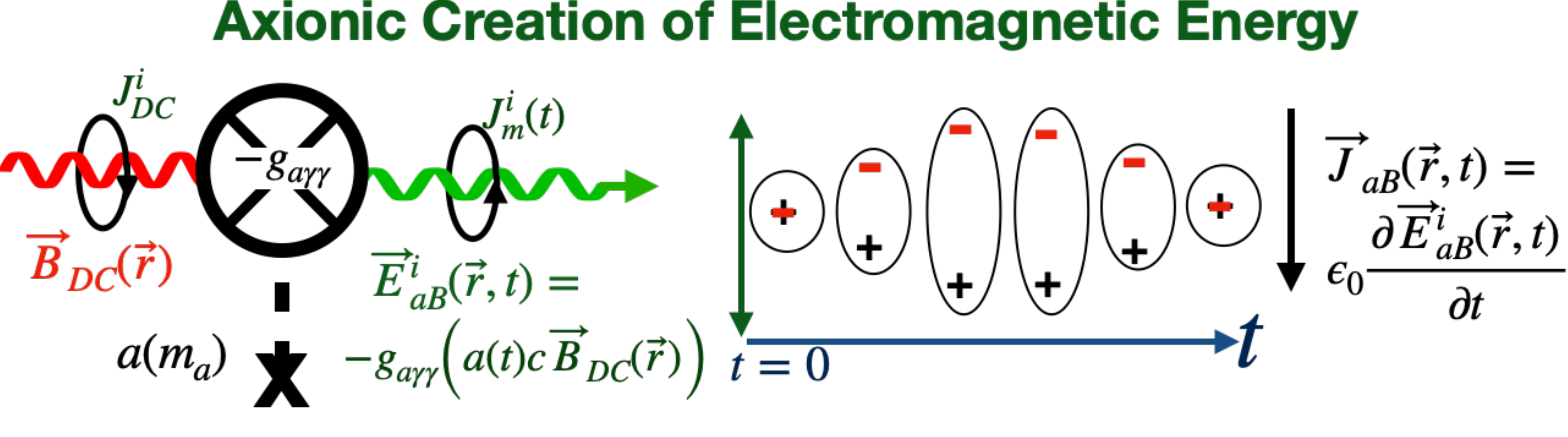}
\caption{Schematic of the creation of electromagnetic energy through the mixing process between axion particles, $a(m_a)$ and a DC $\vec{B}$-field, $\vec{B}_{DC}(\vec{r})$ via the inverse Primikoff effect. The resulting ectromagnetic fields oscillate at frequency, $\omega_a$, equivalent to the axion mass, $m_a$, with a mixing efficiency of the axion-photon coupling, $g_{a\gamma\gamma}$. The $\vec{E}_{aB}^i(\vec{r},t)$ vector represents the external force per unit charge supplied by the mixing process, which directly oscillates the axion background ``Noether'' charge, with the time derivative equal to the axion displacement current, $\vec{J}_{aB}(\vec{r},t)$. Recently a topological axion was detected as an oscillating charge density wave in a Weyl semimetal topological insulator\cite{Gooth:2019np}. In analogy, we can consider the oscillating ``Noether'' charge density wave described here as the ``actual'' Dark Matter axion.}
\label{Mix}
\end{figure}

The equations for the reacted fields induced by the axion through the inverse Primakoff effect, separated from the external fields may be written as \cite{Younggeun18,TobarModAx19} (see Appendix A for more details);
\begin{align}
&\vec{\nabla}\cdot\vec{D}_a^T =\rho_f,\label{eq:M7}\\
&\vec{\nabla} \times \vec{B}_a-\mu_0\frac{\partial \vec{D}_a^T}{\partial t}=\mu_0\vec{J}_f,\label{eq:M8}\\
&\vec{\nabla} \cdot \vec{B}_a= 0,\label{eq:M9}\\
&\vec{\nabla} \times \vec{E}_a+\frac{\partial \vec{B}_a}{\partial t} = 0\label{eq:M10}
\end{align}
with the following constitutive relationship;
\begin{equation}
\vec{D}_a^T=\epsilon_0\vec{E}_a^T =\epsilon_0(\vec{E}_a+\vec{E}_{aB}^i),
\label{eq:M11}
\end{equation}
where,
\begin{equation}
\vec{E}_{aB}^i(\vec{r},t)=-g_{a\gamma\gamma}a(t)c\vec{B}_{DC}(\vec{r}).
\label{EaB(t)1}
\end{equation}
Here, $\vec{E}_{aB}^i(\vec{r},t)$ \textit{is not an electric field derived from Maxwell's equations}, it is the external axionic/photonic force per unit charge (or electrical action) caused by a non-electromagnetic source in electrodynamics, and is referred to as an impressed electric field (or emf per unit length) \cite{RHbook2012,Balanis2012,TobarElectret}. This electrical action drives and oscillates the ``Noether''  charge due to the mixing of the axion scalar field $a(t)$ with $\vec{B}_{DC}$, with a schematic shown in Fig. \ref{Mix}. This is analogous to the axion in solid state physics, which arises from the formation of a charge-density wave in a Weyl semimetal, and has been recently detected \cite{Gooth:2019np}. Also, $\rho_f$ and $\vec{J}_f$ are any free current and charge that might be in the detection system (not impressed). However, first we will consider the axion environment in vacuum and under DC magnetic field, and in this case $\rho_f$ and $\vec{J}_f$ will be zero. 

By inspection, we can immediately solve for $\vec{E}_{aB}^i(\vec{r},t)$ by substituting equation (\ref{Bo}) into (\ref{EaB(t)1}), to give,
\begin{equation}
\vec{E}_{aB}^i(\vec{r},t)=-g_{a\gamma\gamma}a(t)c\frac {\mu_0}{ 4\pi} \int_{\Omega } \frac {\vec{\nabla} \times\vec{J}_{DC}^i \left( \vec{ r } ^ { \prime } \right) } { \left| \vec{r}-\vec{r} ^ { \prime } \right| } \mathrm{d}^{3}\vec{r }^{\prime},
\label{EaB(t)2}
\end{equation}
From equation (\ref{EaB(t)2}), we may identify the effective impressed magnetic current, $\vec{J}_{ma}^i(\vec{r},t)$, which sources $\vec{E}_{aB}^i(\vec{r},t)$ by the standard left hand rule given by,
\begin{equation}
\vec{E}_{aB}^i(\vec{r},t)=-\frac {1}{ 4\pi} \int_{\Omega } \frac {\vec{\nabla} \times \vec{J}_{ma}^i \left( \vec{ r } ^ { \prime },t \right) } { \left| \vec{r}-\vec{r} ^ { \prime } \right| } \mathrm{d}^{3}\vec{r }^{\prime},
\label{EaB(t)3}
\end{equation}
and deduce that the effective impressed magnetic current is related to the impressed electrical current by,
\begin{equation}
\vec{J}_{ma}^i(\vec{r},t)=g_{a\gamma\gamma}a(t)c\mu_0\vec{J}_{DC}^i(\vec{r}).
\label{MC}
\end{equation}
Thus, we have determined that the impressed DC electrical current converts to a parallel effective impressed magnetic current oscillating at the Compton frequency of the axion \cite{TobarModAx19}. Here, $\vec{J}_{DC}^i(\vec{r})$, acts as a source for $\vec{B}_{DC}(\vec{r})$, $\vec{E}_{aB}^i(\vec{r},t)$ and $\vec{J}_{ma}^i(\vec{r},t)$. 

Furthermore, the modified constitutive relationship connects the external non-electrical force to the electromagnetic forces in the system, similar to relationship in a voltage source \cite{TobarElectret}. Combining equation (\ref{eq:M10}) with the constitutive equation (\ref{eq:M11}) we can also write a modified Faraday equation as (this is similar to an electret model in electrodynamics \cite{TobarElectret}),
\begin{equation}
\vec{\nabla} \times \vec{E}_a^T+\frac{\partial \vec{B}_a}{\partial t} = -\vec{J}_{ma}^i(\vec{r},t),
\label{ModFar}
\end{equation}
or in integral form as,
\begin{equation}
\oint_P\vec{E}_a^T\cdot d\vec{l}+\frac{d}{dt}\int_S\vec{B}_a \cdot d\vec{a}=-\int_S\vec{J}_{ma}^i(\vec{r},t)\cdot d\vec{a}.
\label{ModFarInt}
\end{equation}

The next step in solving for the electromagnetic fields is to apply the quasi-static approximation, which comes from calculating the oscillating axion induced magnetic field, $\vec{B}_{a}(t)$, produced by the time dependence of $\vec{D}^T=\epsilon_0\vec{E}_{aB}^i(\vec{r},t)$ with $\vec{J}_{f}=0$ (vacuum). Thus, from the modified Ampere's law given by equation (\ref{eq:M8}), we obain,
\begin{equation}
\vec{\nabla} \times\vec{B}_a=\mu_0\epsilon_0\frac{\partial \vec{E}_{aB}^i}{\partial t}=\mu_0 \vec{J}_{aB}(\vec{r},t),
\label{amp}
\end{equation}
where
\begin{equation}
 \vec{J}_{aB}(\vec{r},t)=\epsilon_0\frac{\partial \vec{E}_{aB}^i}{\partial t}=-g_{a\gamma\gamma}\sqrt{\frac{\epsilon_0}{\mu_0}}\vec{B}_{DC}(\vec{r})\frac{\partial a(t)}{\partial t}.\label{eq:M2b}
\end{equation}
Equation (\ref{amp}) is consistent with previous work \cite{Ouellet2018,Beutter18,Younggeun18}, but this work does not make the connection that this term is actually generated from the electrical action from the axion mixing with the DC $\vec{B}$-field creating a force per unit charge $\vec{E}_{aB}^i(t)$, as schematically shown in Fig. \ref{Mix}. The calculation of $\vec{E}_{aB}^i(t)$ is in fact part of the first order solution before the quasi-static approximation needs to be applied \cite{TobarModAx19}. Previously, this has been overlooked, as this prior work only calculates electric effects derived from Maxwell's equations, which necessarily has a scalar potential \cite{Ouellet2018,Beutter18,Younggeun18} and ignores the electric effects from the electrical action (or emf) driving the whole process, which has a vector potential. Equation (\ref{amp}) also dictates that the axion induced oscillating magnetic field, $\vec{B}_{a}$ is suppressed by the Compton frequency of the axion with respect to $\vec{E}_{aB}^i(\vec{r},t)$ as it is proportional to the time derivative of the axion field, $\frac{\partial a(t)}{\partial t}$. In this sense we should consider the axion modification as an electromotive force generator represented by an impressed electric field which oscillates ``Noether'' charge and produces the axion displacement current, as shown in Fig. \ref{Mix}. 

Following the procedure of the quasi-static technique, the next term to calculate is $\vec{E}_{a}$ from the time rate of change of $\vec{B}_{a}$ from Faraday's equation (\ref{eq:M10}). 
\begin{equation}
\vec{\nabla} \times \vec{E}_{a}=-\frac{\partial \vec{B}_{a}}{\partial t}.
\label{faraday}
\end{equation}
This equation calculates the generated electric field due to the electric scalar potential, and is further suppressed by another factor of the Compton frequency of the axion, because the solution of $\vec{E}_{a}$ is proportional to $\frac{\partial^2 a(t)}{\partial^2 t}$. 

{\color{red}Fundamentally, it has been standard within the axion dark matter community to argue that equation (\ref{MaxED}) in Appendix A (and hence equations (\ref{eq:M7})-(\ref{eq:M10})) show that as $\omega_a\rightarrow 0$ the scalar axion field cannot enter any physical equations. However, solutions (\ref{EaB(t)2}) and (\ref{MC}) obviously show that this is not the case, and that the axion field is in fact a physical observable, so how could this be possible? The answer, of course, is already known. The equations given by (\ref{MaxED}) does not describe the system without specifying the boundary conditions, which includes the impressed boundary sources. The system under discussion is characterized by nontrivial topological sectors with nontrivial boundary conditions, similar to the well-known Witten's effect, when the axion $\Theta$ parameter becomes a physical observable in the presence of non-trivial boundary conditions as explained in \cite{Cao2017}. However, in this case the non-trivial boundary condition comes from the fact that the process is a non-conservative one with respect to our oscillating photonic degree of freedom, and like a voltage source in standard electrodynamics, there exist an effective impressed magnetic current boundary source \cite{RHbook2012,TobarElectret}, given by equation (\ref{MC}), which is proportional to the same impressed electric DC current, which sources our applied DC magnetic field. 

In the following we show how to make use of this non-trivial boundary condition, to realise a detector sensitivity proportional to the axion scalar field. To do this a conductor winding needs to link this this axion induced magnetic current boundary source. If it does not the experiment with the conductor will just be proportional to the axion time derivative (or the axion mass) in a similar way to the current low-mass axion haloscope detectors, which utilise a DC magnetic field. We show that with the new proposed enhanced topology, experiments may be made sensitive enough to detect the QCD axion at low-mass when the Compton wavelength of the axion is larger than the size of the experiment (the quasi-static regime).}

\section{Conducting Antenna in Spatially Constant DC Magnetic Field}

If a circuit component is small enough and placed within a DC solenoidal electromagnet, both the intensity and direction of the field will be effectively constant with respect to the circuit component. In such a case one can approximate the DC magnetic field as sourced by an infinite solenoid with magnetic field of, $\vec{B}_{DC}=B_{DC} \hat{z}$ (assuming the field is orientated in the laboratory $z$-direction), then $\vec{E}_{aB}^i(t)$ is given by \cite{TobarModAx19},
\begin{equation}
\vec{E}_{aB}^i(t)=-g_{a\gamma\gamma}a_0cB_{DC}\sin{(\omega_at+\phi)}\hat{z}.
\label{EaB}
\end{equation}
{\color{blue} Here the axion field $a(t)$ is considered as a scalar field oscillating at the Compton frequency of the axion, $\omega_a$, its phase, $\phi$ is determined by the production mechanism at the QCD scale in early Universe and has no direct relation to the instrument detection sensitivity, so for simplicity we set it to zero in the analysis of the paper. The same results of sensitivity would be generated if we used a cosine or complex exponential functions or any other phase for this analysis, and the results are independent of this value. Nevertheless, the value does change the detectors transient behaviour as shown in Appendix B. However, we also show that the detectors transient behaviour has no bearing on axion detection sensitivity as we determine that the sensitivity only depends on the steady state response.}

Under constant $\vec{B}$-field the impressed DC current, which creates $\vec{B}_{DC}$ and $\vec{E}_{aB}^i$ (equation (\ref{Bo}) and (\ref{EaB(t)2})) is effectively at infinity. Thus, in the quasi-static limit and under constant $\vec{B}_{DC}$, equation (\ref{ModFarInt}) can be approximated by $\oint_P\vec{E}_a^T\cdot d\vec{l}=0$. 

\begin{figure}[t]
\includegraphics[width=0.8\columnwidth]{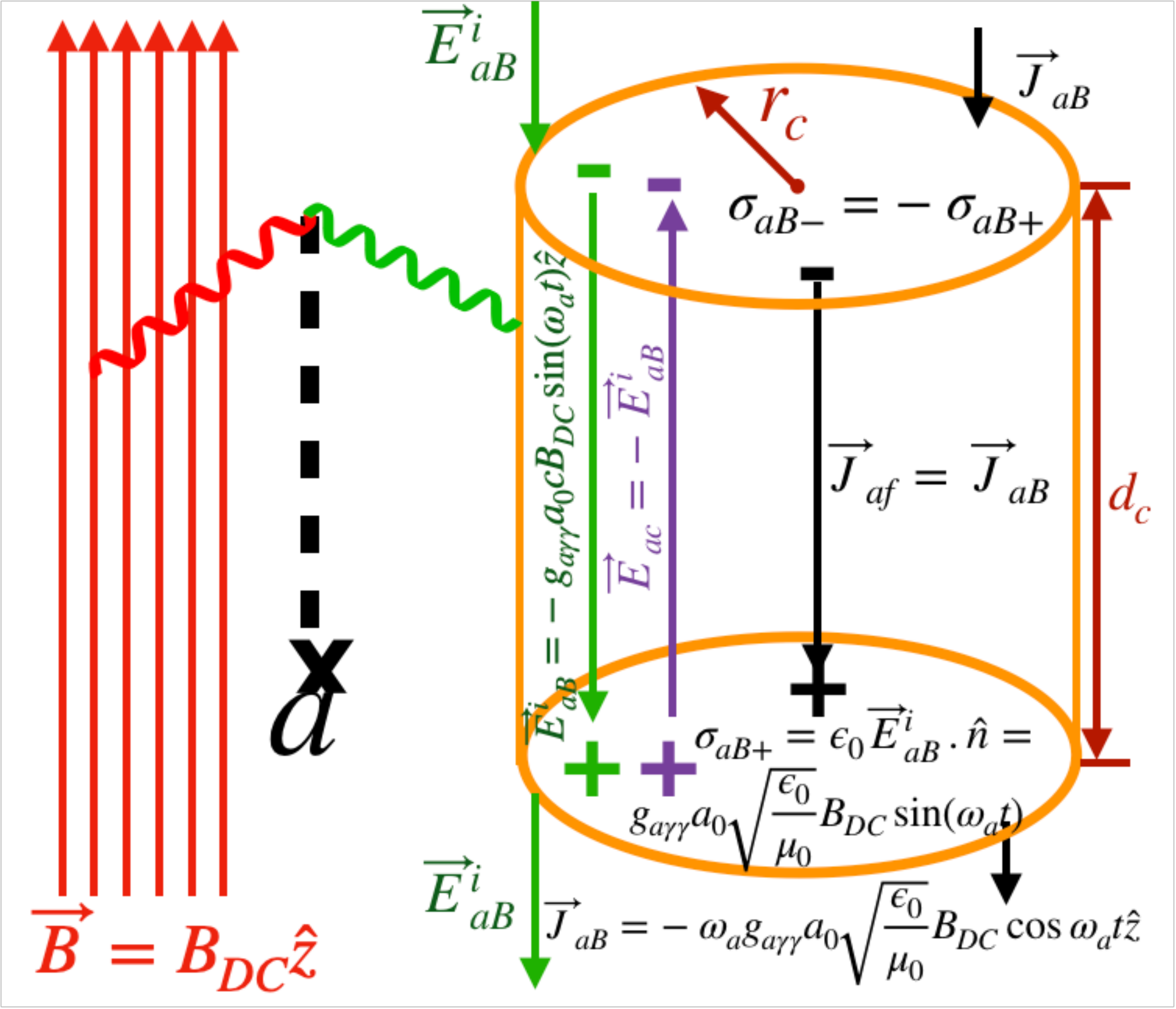}
\caption{Ideal cylindrical normal conductor under high DC magnetic field interacting with the axion scalar field. The induced oscillating impressed electric field, $\vec{E}_{aB}^i$, creates an emf per unit length across the conductor as shown. In the perfect conductor a reverse electrical field is induced, $\vec{E}_{a}=\vec{E}_{ac}$, so that $\vec{E}_{aB}^i+\vec{E}_{ac}=0$ and the voltage across the conductor is zero. This causes an oscillating free current density inside the conductor, $\vec{J}_{af}$, due to the time dependence of the surface charge, $\sigma_{aB\pm}$ that terminates $E_a^T$ outside the conductor.}
\label{perfectcond}
\end{figure}

Here we consider a perfect conducting cylindrical antenna with radius $r_c$, cross sectional area $A_c=\pi r_c^2$ and length $d_c$, which is not polarizable or magnetizable as shown in Fig. \ref{perfectcond}. When $\vec{E}_{aB}^i$ is incident on a conductor the free charges inside the conductor will rearrange themselves until they no longer experience a force. This means the voltage across the perfect conductor $\pm$ terminals must be zero, given by $v_a=\int_-^+\vec{E}_a^T\cdot d\vec{l}=0$. As the free electrons migrate across the conductor, they create an internal electric field, $\vec{E}_{ac}$, to exactly cancel out $\vec{E}_{aB}^i$, so that, $\vec{E}_{ac}=-\vec{E}_{aB}^i$ and $v_a=0$.

To keep the voltage across the conductor zero, an oscillating free current must be generated at a frequency equivalent to the axion mass. The magnitude of this current may be calculated with Gauss' law. The $\vec{E}_{aB}^i$ lines that are perpendicular to the surface will generate surface charges at the boundary as dictated by the integral form of Gauss' Law given by equation (\ref{eq:M7}), giving
\begin{equation}
\sigma_{aB\pm}(t)=\epsilon_0\vec{E}_{aB}^i(t)\cdot\hat{n}=\pm g_{a\gamma\gamma}a_0\sqrt{\frac{\epsilon_0}{\mu_0}}B_{DC}\sin{(\omega_at)}\hat{z},
\label{SigmaaB}
\end{equation}
where the surface charges, $\sigma_{aB\pm}$, are shown in Fig. \ref{perfectcond}. Furthermore, the oscillating free current density inside the conductor may be calculated from the time rate of change of surface charge density to be, 
\begin{equation}
\vec{J}_{af}=-\frac{\partial\sigma_{aB}}{\partial t}\hat{z}=-\omega_ag_{a\gamma\gamma}a_0\sqrt{\frac{\epsilon_0}{\mu_0}}B_{DC}\cos{(\omega_at)}\hat{z}=\vec{J}_{aB}
\label{Current}
\end{equation}
with a total current oscillating in the antenna given by,
\begin{equation}
\vec{i}_{a_{Ant}}(t)=\vec{J}_{af}A_c=-\omega_ag_{a\gamma\gamma}a_0\sqrt{\frac{\epsilon_0}{\mu_0}}A_cB_{DC}\cos{(\omega_at)}\hat{z}
\label{Ia}
\end{equation}
Note, that this solution also satisfies Ampere's law given by equation (\ref{eq:M8}) as $\vec{D}^T$ inside the conductor is zero, and then $\vec{\nabla} \times \vec{B}_a=\mu_0\vec{J}_{af}=\vec{J}_{aB}$, so that the induced $\vec{B}_a$ field will remain continuous inside and outside the conductor.

Of course, we can not have a perfect conductor. If we assume a high conductivity conductor governed by Ohm's law, with a conductivity $\kappa_c$, then there will be a small imbalance between $\vec{E}_{aB}^i$ and $\vec{E}_{ac}$ due to losses, given by $\vec{E}=\vec{E}_{aB}^i-\vec{E}_{ac}$. This imbalance will be $\pi/2$ out of phase with the driving electric field, $\vec{E}_{aB}^i$, representing a resistive power loss or dissipation with an electrical resistance of $R_c= \frac{d_c}{\kappa_c A_c}$ and voltage drop given by $v_a(t)=R_ci_a(t)$ ignoring the skin effect (i.e. assuming the skin depth is larger than the radius of the wire). So from Ohms law and equation (\ref{Ia}) we obtain,
\begin{equation}
v_{a_{Ant}}(t)=-g_{\alpha\gamma\gamma}a_0\sqrt{\frac{\epsilon_0}{\mu_0}}\frac{d_c}{\kappa_c}B_{DC}\omega_a\cos(\omega_a t).
\end{equation}

Now, considering the axions originate from galactic halo dark matter axions, we arrive at
\begin{equation}
v_{a_{Ant}}^{\textrm{RMS}} = g_{a\gamma\gamma}\sqrt{\frac{\epsilon_0}{\mu_0}}\frac{d_c}{\kappa_c}B_{DC}\sqrt{\rho_\textrm{DM}c^3}.\label{eq104}
\end{equation}
\begin{equation}
i_{a_{Ant}}^{\textrm{RMS}} = g_{a\gamma\gamma}\sqrt{\frac{\epsilon_0}{\mu_0}}A_cB_{DC}\sqrt{\rho_\textrm{DM}c^3}.
\label{eq204}
\end{equation}
Equations (\ref{eq104}) and (\ref{eq204}) are the basic expressions for calculating the sensitivity to dark matter axions for an experiment based on a conducting antenna in a uniform DC magnetic field.

In the case that the skin depth is smaller than the radius of the wire, the resistance will increase, as the effective cross sectional area will become tubular like and hence decrease, this can be adjusted by using the effective area instead of $A_c$, decreasing the current for a fixed voltage drop.

\section{Conductor in Spatially Varying DC Magnetic Field}

In actual fact it is impossible to create a constant DC magnetic field throughout free space, and in this section we consider experiments in the quasi-static low-mass regime of the axion, which make full use of the spatial extent of the magnetic field of an electromagnet. Two commonly used electromagnets are the toroid and the solenoid, where the spatial dependence comes from the way the impressed electrical current flows within the solenoidal or toroidal coils that generate the DC magnetic fields. Clearly, in the limit of a thinly wound wire coil, the impressed current creates an effective surface boundary defined by the coil of the electromagnet, which also unequivocally defines the spatial extent of the magnetic field as shown in Fig. \ref{TorSol}, and described mathematically by equation (\ref{Bo}). This in turn generates $\vec{E}_{aB}^i$ under the inverse Primakoff effect, which oscillates at the Compton frequency of the axion, but with a similar spatial dependence and surface boundary condition as the DC magnetic field, as dictated by equation (\ref{EaB(t)2}). Because of this localised surface term, the electric sensing techniques proposed in this paper can be made directly sensitive to $a(t)$. It has been pointed out in reference \cite{Cao2017} that this is indeed possible, depending on the nature of the boundary source and the topology of the system. In this section we extend the topology of our conducting wire to a winding that links the coil of the electromagnet. This means the winding picks up an inductive impedance and generates an output voltage proportional to the linked axion induced effective magnetic current.

\begin{figure}[t]
\includegraphics[width=0.6\columnwidth]{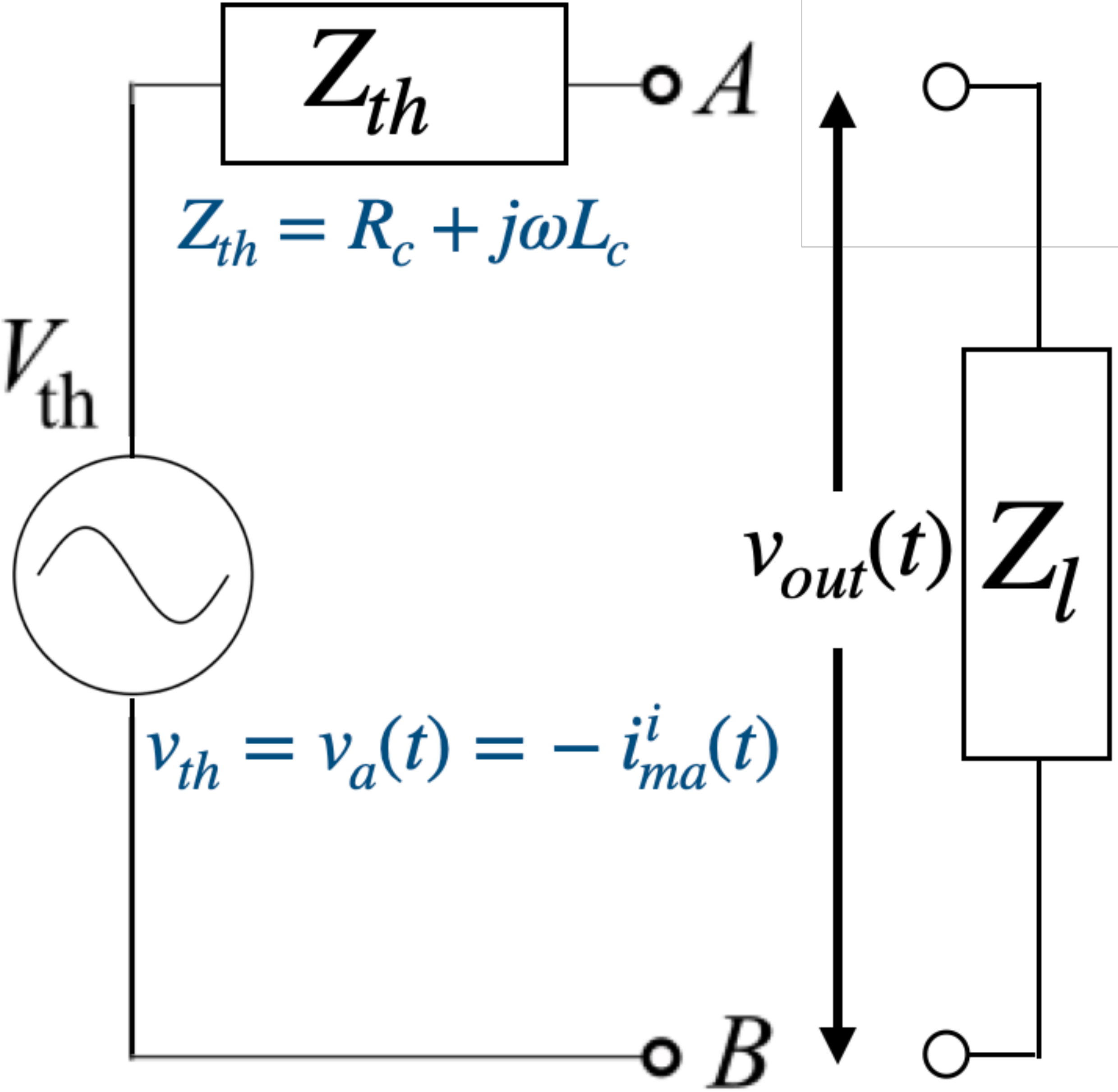}
\caption{Thevenin equivalent circuit of a coil winding, which links the axion induced magnetic current (or coil of the electromagnet). The output ports are labelled as $A$ and $B$, where the first low-noise amplifier will be attached, with the sensitivity depending on its input impedance ($Z_L$) with respect to the source impedance. In general there should a small resistance in the coil, which is represented by $R_c$.}
\label{WireThNo}
\end{figure}

To come up with a complete circuit model of the effective voltage source, we must utilise the integral form of the modified Faraday's law given in equation (\ref{ModFarInt}). Thus, the emf, $\mathcal{E}_a(t)$, produced by the non-conservative process can be calculated by \cite{TobarModAx19},
\begin{equation}
\mathcal{E}_a(t)=\oint_P\vec{E}_a^T\cdot d\vec{l}=-\frac{d}{dt}\int_S\vec{B}_a \cdot d\vec{a}-\int_S\vec{J}_{ma}^i\cdot d\vec{a}.
\label{emf}
\end{equation}
The axion induced magnetic current, $i_{{ma}}^i(t)$, enclosed by the winding, can be calculated from the enclosed magnetic current density as,
\begin{equation}
i_{{ma}}^i(t)=\int_S\vec{J}_{ma}^i\cdot d\vec{a}=g_{a\gamma\gamma}a_0\cos(\omega_a t)c\mu_0I_{{DC}_{enc}}^i,
\label{emfMagC}
\end{equation}
where
\begin{equation}
I_{{DC}_{enc}}^i=\int_S\vec{J}_{DC}^i\cdot d\vec{a}.
\end{equation}
Following this, we can determine the magnetic flux linking a single winding as,
\begin{equation}
\Phi_a(t)=\int_S\vec{B}_a \cdot d\vec{a}.
\end{equation}
More generally if the enclosed winding has more than one number of turns, $N_c$, then the emf induced will be given by the rate of change of flux linkage, where the flux linkage is defined by $\lambda=N_c\Phi$. Thus, the emf induced around a closed path of integration is related to the enclosed magnetic flux and magnetic current by,
\begin{equation}
\mathcal{E}_a(t)=-i_{{ma}}^i(t)-N_c\frac{d\Phi_a(t)}{dt},
\end{equation}
This means a sensitive experiment may be realised by using a conducting winding or coil to link the axion induced magnetic current to read out a voltage or current. In this example the winding or coil will produce a time rate of change of magnetic flux (due to its self-inductance). Assuming a current of  $i_{a}(t)$ is induced in the winding, with a self-inductance of $L_{c}$, then $N_c\frac{d\Phi_a(t)}{dt}=L_c\frac{di_{a}}{dt}$, so that $\mathcal{E}_a(t)=-i_{{ma}}^i(t)-L_c\frac{dI_{a}(t)}{dt}$. Then the effective source voltage may be recognised as, $v_{a}(t)=-i_{{ma}}^i(t)$, with an effective output voltage of $v_{out}(t)=\mathcal{E}_a(t)-i_{a}(t)R_c$, so that;
\begin{equation}
v_{a}(t)=-i_{{ma}}^i(t)=v_{out}(t)+i_{a}(t)R_c+L_c\frac{di_{a}(t)}{dt}, 
\end{equation}
which becomes a standard equation for a voltage source with an $LR$ source impedance, with the Thevenin equivalent circuit shown in Fig. \ref{WireThNo}. Note for an open circuit load ($i_{a}=0$), $v_{a}=v_{out}$, and for a short circuit load ($v_{out}=0$), $v_{a}=i_{a}R_c+L_c\frac{di_{a}}{dt}$. In the quasi-static regime, dynamical transient effects are of fundamental importance and will be governed by the circuit time constant. In the case of the short circuit current the time constant is of order $\tau_c=\frac{L_c}{R_c}$. However, inevitably the response of the readout circuit will also depend on the input impedance of the first amplifier, which will act as a load with impedance $Z_L$. More detailed calculations of these responses are given in Appendix B, with calculation of circuit parameters given in Appendix C.

\subsection{Toroidal Magnet}

\begin{figure}[t]
\includegraphics[width=0.65\columnwidth]{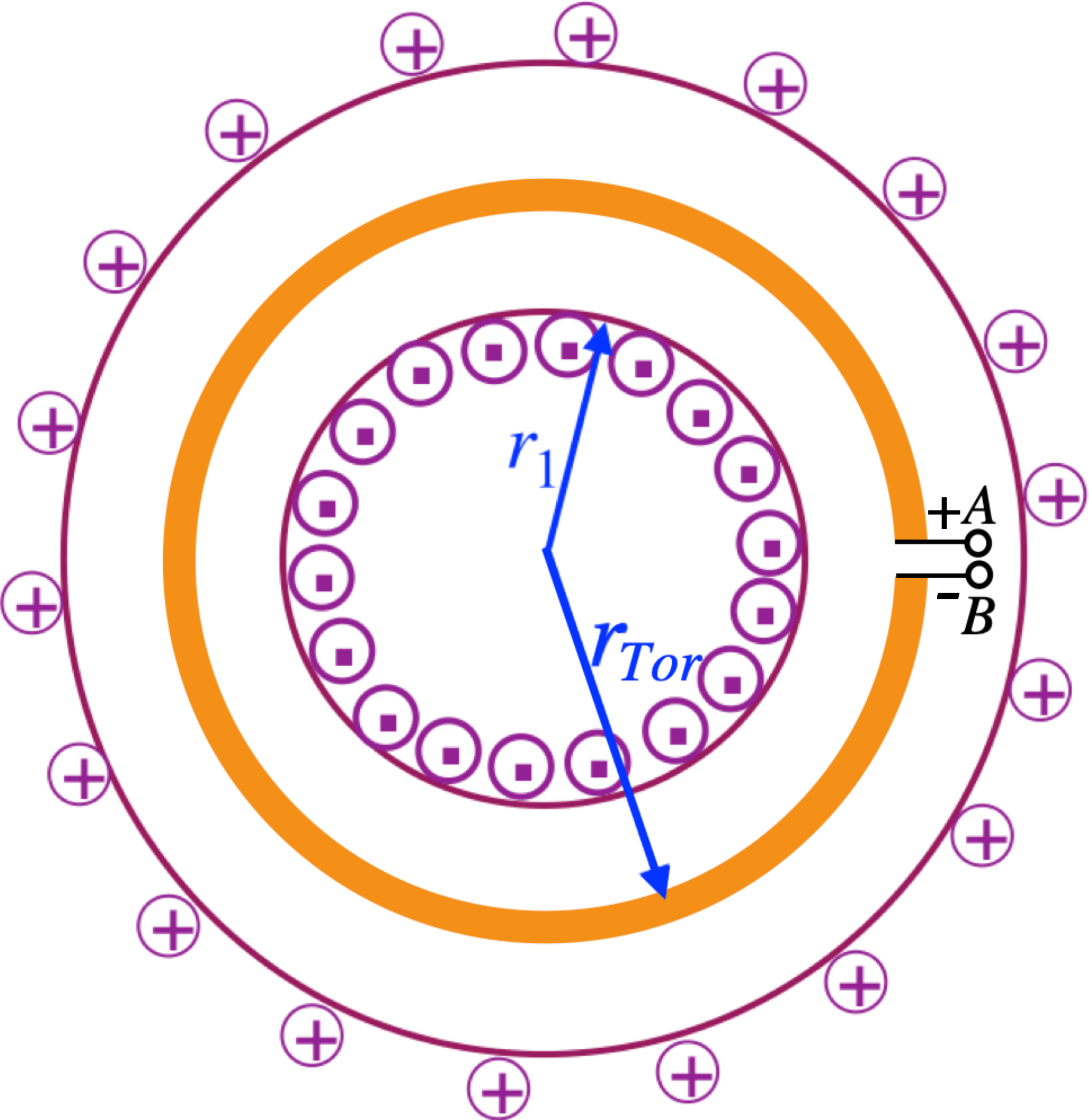}
\caption{Schematic of a low-mass axion experiment, which utilises electric sensing by linking the main coil of a toroidal magnet with a conducting wire loop (or winding), with the output ports as defined in Fig. \ref{WireThNo} shown.}
\label{Tor}
\end{figure}

To utilise the full extent of the spatial dependence of a toroidal magnet as shown in Fig. \ref{TorSol}, a circuit which fully links the current of the toroid should be made. An example of a possible experiment, which winds a coil through the centre of the toroid is shown in Fig. \ref{Tor}. 

Setting the $DC$ magnetic field at the centre of a toroid (at radius $r_{Tor}$) to, $\vec{B}=B_{DC}\hat{\theta}$, and then by implementing Ampere's law, the solution of the magnitude of the magnetic field at $r_{Tor}$, inside the toroid is well-known (see Fig. \ref{Tor}) and can be calculated to be,
\begin{equation}
B_{DC}=\mu_0N_lI_{{DC}_{Tor}}^i
\label{BoTor}
\end{equation}
Here $N_l$ is the number of turns per unit length of the toroid and $I_{{DC}_{Tor}}^i$ is the impressed current in the toroid coil. As shown in Fig. \ref{Tor} the enclosed electrical current along the path of the wire is,
\begin{equation}
I_{{DC}_{enc}}^i=N_{Tor}I_{DC_{Tor}}^i=2\pi r_{Tor}\frac{B_{DC}}{\mu_0}
\label{Ienc}
\end{equation}
where $N_{Tor}=2\pi r_{Tor}N_l$ is the total number of turns for the toroidal coil. Then by substituting equation (\ref{Ienc}) in (\ref{emfMagC}) the open circuit voltage at the input terminal to the load amplifier shown in Fig. \ref{Tor} can be determined to be,
\begin{equation}
v_a(t)_{Tor}=g_{a\gamma\gamma}a_0cB_{DC}2\pi r_{Tor}N_{c}\sin{(\omega_at)},
\label{vat}
\end{equation}
or in terms of dark matter density,
\begin{equation}
v_{{a}_{Tor}}^{\textrm{RMS}} = g_{a\gamma\gamma}2\pi r_{Tor}N_{c}\Big(\frac{c}{\omega_a}\Big)B_{DC} \sqrt{\rho_\textrm{DM}c^3},
\label{VaRMStor}
\end{equation}
The conductor within the toroid is essentially a pick up coil, and for every number of turns (or windings), $N_{c}$, the voltage may be enhanced proportionally and is included in the equation. Unfortunately, one could not increase the number of windings indefinitely, as this will increase the impedance of the effective voltage source, with the resistance of the wire, $R_c$ being proportional to the length and hence $N_{c}$ and the inductance of the coil, $L_c$, proportional to $N_{c}^2$, limiting the voltage across the load. To measure the corresponding voltage output, $v_{out}(t)$, with minimal degradation with respect to $v_a(t)$, we need an amplifier with high input impedance. Inevitably, the finite input impedance of the amplifier will reduce the signal, and for a good design this should be minimised.

If we short circuit the terminals $A$-$B$ we essential generate a ring of current in the conducting loop, with details on how this is calculated given in Appendix B. Assuming we are focused on measuring the signal current at $\omega_a$ we can ignore all transient effects, with the following current generated,
\begin{equation}
i_{a}(t)_{Tor}=\frac{g_{a\gamma\gamma}a_0cB_{DC}2\pi r_{Tor}N_{c}}{\sqrt{R_c^{2}+L_c^2\omega_a^{2}}}\sin\left(\omega_a t-\tan ^{-1}(\omega_a \tau_c)\right),
\label{SCtor}
\end{equation}
and in terms of dark matter density at $\omega_a$ the RMS current will be,
\begin{equation}
i_{{a}_{Tor}}^{\textrm{RMS}} = \frac{g_{a\gamma\gamma}2\pi r_{Tor}N_{c}cB_{DC}}{\omega_a\sqrt{R_c^{2}+L_c^2\omega_a^{2}}}\sqrt{\rho_\textrm{DM}c^3}.
\label{IaRMStor}
\end{equation}
To measure this current with minimal degradation, we need a current amplifier with a low input impedance. 

By comparing equation (\ref{IaRMStor}) with equation (\ref{eq204}), we can estimate how much sensitivity we have gained by extending the topology from a small conductor in a constant field, to a winding that links the impressed current of the electromagnet, assuming $N_c=1$ we obtain,
\begin{equation}
\frac{i_{{a}_{Tor}}^{\textrm{RMS}}}{i_{a_{Ant}}^{\textrm{RMS}}}= \frac{r_{Tor}\lambda_a}{A_c}\frac{{Z_0}}{|Z_{Tor}|}.
\label{IaComp}
\end{equation}
Here, $\lambda_a=\frac{2\pi c}{\omega_a}$ is the Compton wavelength of the axion, $Z_0=\sqrt{\frac{\mu_0}{\epsilon_0}}$ is the impedance of free space, and $Z_{Tor}$ is the total impedance of the coil winding plus load, as shown in Fig. \ref{WireThNo}. Note, increasing $N_c$ also increases the impedance of the winding and hence $Z_{Tor}$, so increasing the number of coil windings does not increase the sensitivity for experiments that couple to the short circuit current in a conductor generated by the axion.

Another way to measure the current is indirectly through coupling to the magnetic flux, and since $N_c\Phi_{a}(t)_{Tor}=L_ci_{a}(t)_{Tor}$, we can write the AC component of magnetic flux as,
\begin{equation}
\Phi_{a}(t)_{Tor}=\frac{g_{a\gamma\gamma}a_0cB_{DC}2\pi r_{Tor}\tau_c}{\sqrt{1+(\omega_a\tau_c)^{2}}}\sin\left(\omega_a t-\tan ^{-1}(\omega_a \tau_c)\right).
\end{equation}
or in terms of dark matter density,
\begin{equation}
\Phi_{{a}_{Tor}}^{\textrm{RMS}} = g_{a\gamma\gamma}\Big(\frac{cB_{DC}2\pi r_{Tor}\tau_c}{\omega_a\sqrt{1+(\omega_a\tau_c)^{2}}}\Big) \sqrt{\rho_\textrm{DM}c^3}.
\label{PhiaRMStor}
\end{equation}
These equations do not include the effect of the load of the readout amplifier, but represent the best we could achieve in a properly designed system. However, adding the load just adds to the impedance and is not hard to include, but just modifies the circuit parameters.

\subsection{Solenoidal Magnet}

\begin{figure}[t]
\includegraphics[width=0.45\columnwidth]{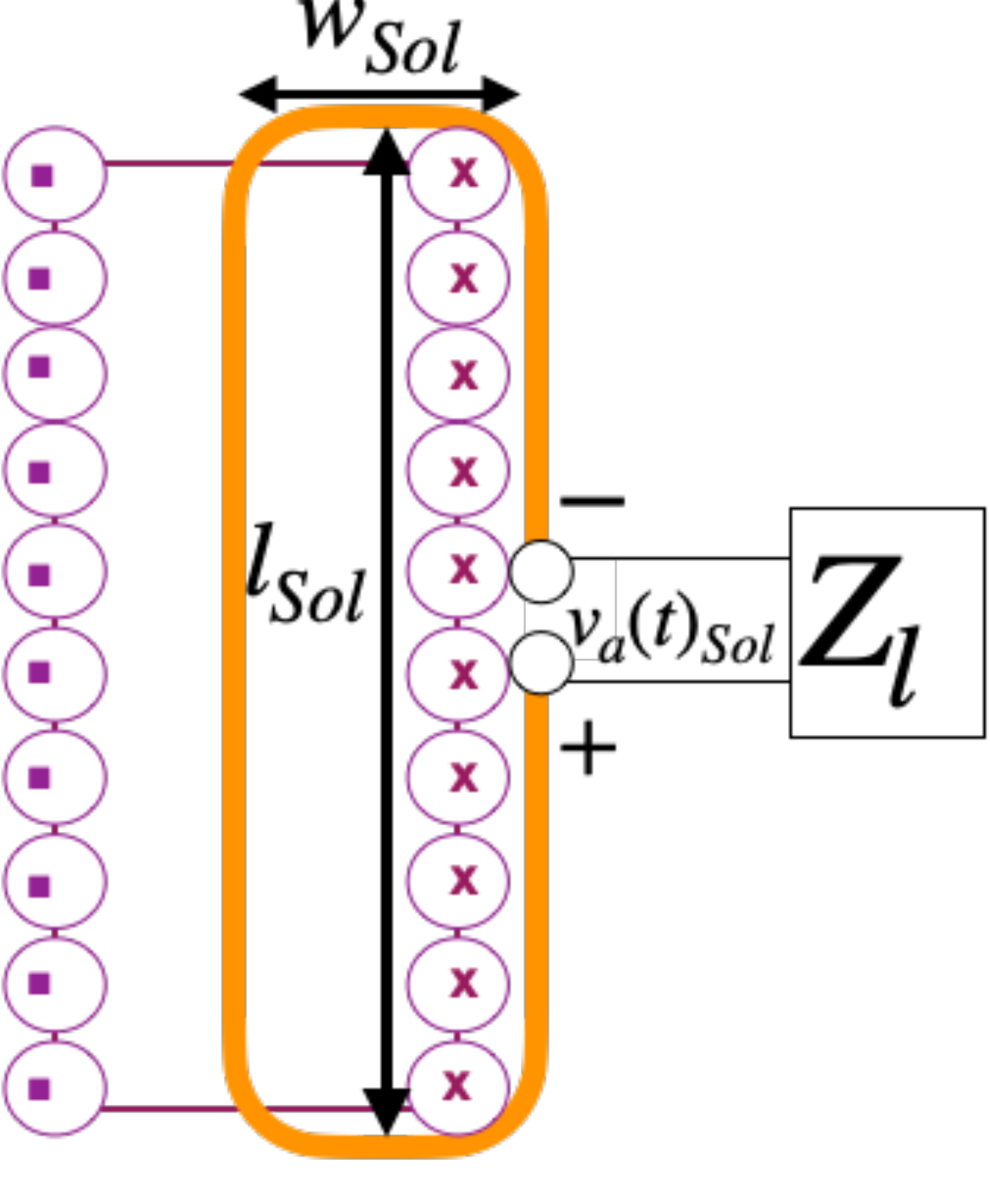}
\caption{Schematic of a low-mass axion experiments, which utilises electric sensing by linking the main coil of a solenoidal magnet with a single conducting winding.}
\label{Sol}
\end{figure}

In a similar way to the toroid example, to utilize the full extent of the spatial dependence of a solenoidal magnet (see Fig. \ref{TorSol}), a circuit which fully links the current of the solenoid should be made as shown in Fig. \ref{Sol}. Assuming that the magnetic field produced by the solenoid, $\vec{B}(r,z)$, has a maximum value of magnetic field in the centre of the solenoid of $\vec{B}(0,0)=B_{DC}\hat{z}$. By implementing Ampere's law we can show that,
\begin{equation}
B_{DC}=\mu_0N_lI_{DC_{Sol}}^i
\label{BoSol}
\end{equation}
Here $N_l$ is the number of turns per unit length of the solenoidal coil and $I_{DC_{Sol}}^i$ is the impressed current in the coil. As shown in Fig. \ref{Sol} the enclosed electrical current along the path of the wire is,
\begin{equation}
I_{{DC}_{enc}}^i=NI_{DC_{Sol}}^i=l_{Sol}\frac{B_{DC}}{\mu_0}
\label{Ienc1}
\end{equation}
where $N=l_{Sol}N_l$ is the total number of turns of the solenoidal coil of length, $l_{Sol}$, as indicated in Fig. \ref{Sol}. 

Considering a conducting wire which links the axion induced magnetic current as in Fig. \ref{Sol}, the voltage at the output terminals supplied to the load impedance $Z_l$ may be found by combining equation. (\ref{Ienc1}) with (\ref{emfMagC}) and can be determined to be,
\begin{equation}
v_a(t)_{Sol}=-g_{a\gamma\gamma}a_0cB_{DC}l_{Sol}N_{c}\sin{(\omega_at)},
\end{equation}
or in terms of dark matter density,
\begin{equation}
v_{a_{Sol}}^{\textrm{RMS}} = g_{a\gamma\gamma} l_{Sol}N_{c}\Big(\frac{c}{\omega_a}\Big)B_{DC} \sqrt{\rho_\textrm{DM}c^3}.
\label{VaRMSsol}
\end{equation}
which is similar to the toroid calculation with the length of the solenoid, $l_{Sol}$ replacing the length of the toroid, $2\pi r_{Tor}$, from the previous subsection. The conductor within the solenoid is essentially becomes a pick up coil with a small inductance, this means, like in the toroid calculation, the voltage may be enhance by increasing the number of turns, $N_{c}$.

The short circuit current between $A$-$B$ for a perfect conductor, and the magnetic flux produced by the inductive read out winding may be calculated in a similar way to equations (\ref{SCtor})-(\ref{PhiaRMStor}) but with the length of the solenoid, $l_{Sol}$ replacing the length of the toroid, $2\pi r_{Tor}$.

\section{Sensitivity To Low-Mass Axions}

The sinusoidal current or voltage under measurement will not be completely coherent, and depends on the distribution of frequencies in the signal. The dark matter axion is expected to have a spread of frequencies due to random dispersion within the Dark Matter Halo following a Maxwell Boltzmann distribution. This means the signal really takes the form of a narrow band noise process centred around a frequency of $\omega_a/2\pi$. Thus to measure the required $RMS$ current or voltage, an integration with respect to the axion line width must be performed. This is usually considered to be a part in $10^6$ so that the coherence time of the axion depends on the axion mass, (and hence $\omega_a$) given approximately by, $\tau_a\approx\frac{2\pi\times10^6}{\omega_a}$, where the bandwidth in Hz is given by $\Delta f_a\approx\frac{1}{\tau_a}$. This means as the axion mass decreases, it takes longer to average at the required bandwidth. Thus, if we are to reach the calculated sensitivity of the lowest frequency of interest, we need to sample at least with a measurement time equal to the inverse bandwidth of the lowest frequency of interest. If the lowest frequency of measurement is equivalent to an axion mass of $10^{-12} eV$, then $\frac{\omega_a}{2\pi}=242 Hz$ and $\tau_{a_{max}}=1$ hour and $9$ minutes. Taking data for longer periods only increases the sensitivity by $t^{\frac{1}{4}}$, for example if we measure for a year we only gain a factor of 9 in sensitivity, even though we have increased the time of measurement by a factor of $7.6\times10^3$. 

\begin{figure}[t]
\includegraphics[width=1.0\columnwidth]{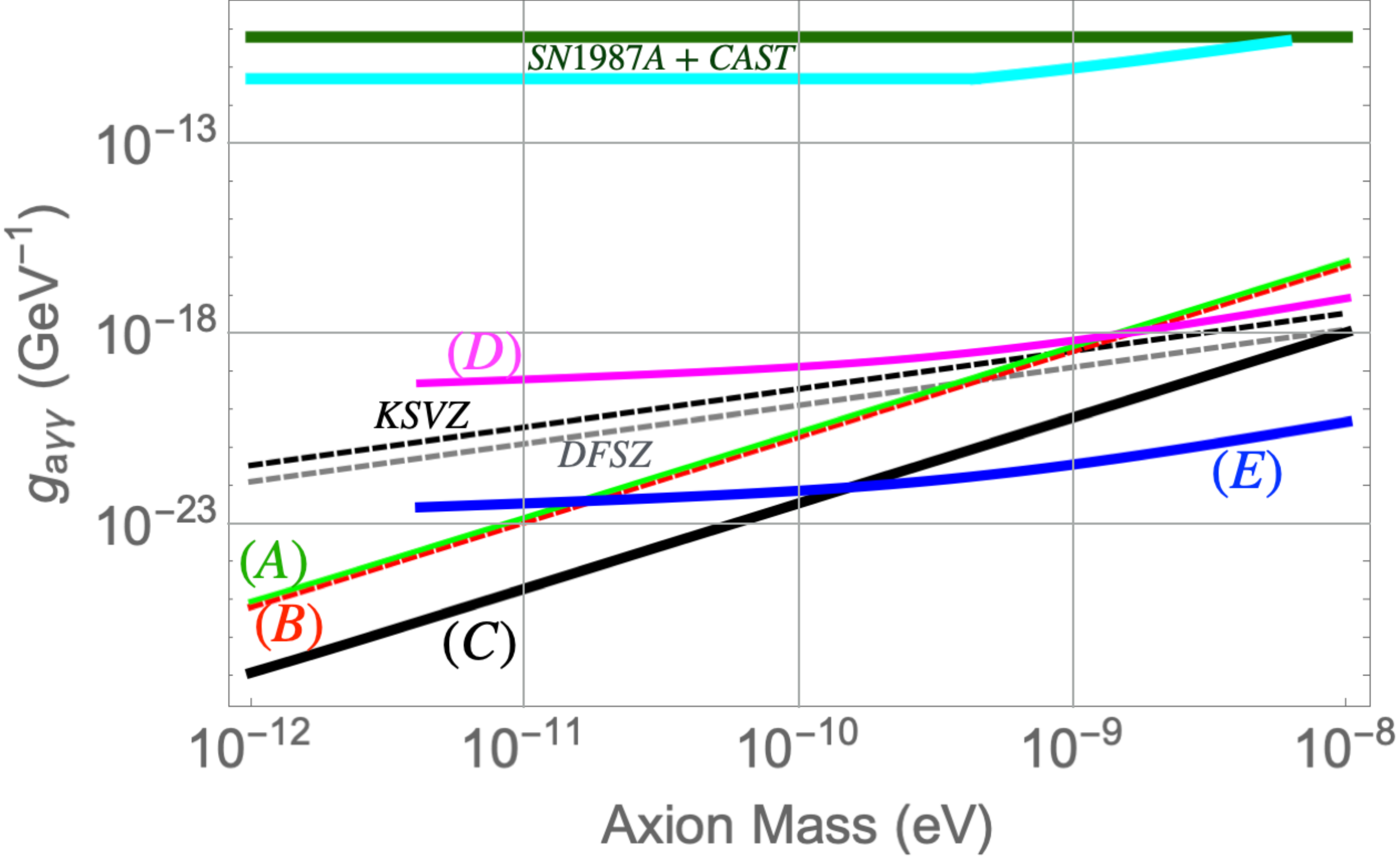}
\caption{Sensitivity estimates of the Broadband Electrical Action Sensing Technique, compared to KSVZ and DFSZ axions and supernova and CAST limits. (A) Solenoid and (B) toroid magnets of order $10cm$ in size, configured with a single coil winding to read out the short circuit current generated by the axion, with one hour and nine minutes of data and equivalent parameters. (C) Increasing the sensitivity of (A) by implementing the ORGAN 14 Tesla solenoid magnet \cite{ORGAN} and collecting data for 1.5 months. In this case we can achieve the QCD axion limit for the whole axion masses range of $10^{-12}$ to $10^{-8}~eV$. (D) Solenoid magnet of (A) configured with a multiple coil toroid winding readout, to directly measure the emf generated by the axion. (E) Increasing the sensitivity of (D) by implementing the ORGAN 14 Tesla solenoid magnet \cite{ORGAN} and collecting data for 1.5 months. This setup is better than the short circuit single winding readout for axion masses above $10^{-10} eV$.}
\label{Sens}
\end{figure}

Since averaging is a slow improver of sensitivity, it is important to investigate other ways to improve measurements to enable axion searches at the QCD limit. Importantly, the new designs presented in this work are inherently more sensitive than other known techniques in the low-mass band, allowing the QCD axion sensitivity to be achieved in a relative small amount of time, as shown in Fig. \ref{Sens}.

To understand the best ways to improve the sensitivity of the experiment, we analyse the value of the signal to noise ratio. The signal to noise ratio of the discussed experiments can be determined with respect to the flux noise or the effective current noise of the SQUID amplifier (which are related) or the equivalent voltage noise of the high impedance amplifier. Calculating the signal to noise ratio and setting it to unity allows us to set the order of magnitude that our detectors are capable of achieving.

Typically, the $RMS$ magnetic flux fluctuations of the SQUID amplifier are of order, $\sqrt{S_\Phi}=1.2\times10^{-6}~\Phi_0/\sqrt{Hz}=2.5\times10^{-21}~Wb/\sqrt{Hz}$, with a flicker corner on the order of $1$ to $10 Hz$ \cite{FlickSquid}, which we have verified in the laboratory \cite{SQUIDQuartz}. The effective current noise is calculated through the SQUID mutual inductance, $M_{in}$, which is $4.2nH$ for the SQUID in \cite{SQUIDQuartz} and $2.5nH$ for the SQUID in \cite{Oue19}, so assuming $M_{in}=2.5nH$ then the effective spectral density of current noise will be $\sqrt{S_{I}}=\frac{\sqrt{S_\Phi}}{M_{in}}=1.0pA/\sqrt{Hz}$. The signal to noise ratio can be written in terms of either current or flux noise by the following equations,
\begin{equation}
\begin{aligned}
&SNR=\frac{i_{{a}_{Tor}}^{\textrm{RMS}}(\tau_a t)^{\frac{1}{4}}}{\sqrt{S_{I}}}= \frac{g_{a\gamma\gamma}2\pi r_{Tor}cB_{DC}(\tau_a t)^{\frac{1}{4}}\sqrt{\rho_\textrm{DM}c^3}}{\omega_a\sqrt{R_c^{2}+L_T^2\omega_a^{2}}\sqrt{S_{I}}} \\
&=\frac{\Phi_{{a}_{Tor}}^{\textrm{RMS}}(\tau_a t)^{\frac{1}{4}}}{\sqrt{S_{\Phi}}}= \frac{M_{in}}{L_T}\frac{g_{a\gamma\gamma}2\pi r_{Tor}cB_{DC}\tau_T(\tau_a t)^{\frac{1}{4}}\sqrt{\rho_\textrm{DM}c^3}}{\omega_a\sqrt{1+(\omega_a\tau_T)^{2}}\sqrt{S_{\Phi}}}
\label{SNRI}
\end{aligned}
\end{equation}
Here, we assume that the integration time, $t$, exceeds or is equal to the axion coherence time, $\tau_a$. 

For the voltage HIA readout, the white noise level is of order $0.3 nV/\sqrt{Hz}$ up to 2.4 MHz in Fourier frequency. However, the flicker corner is much higher than the SQUID amplifier and is close to 200 kHz. The model of voltage noise we use is $\sqrt{S_V}=\frac{3.55\times10^{-5}}{f}+0.29\times 10^{-9} V/\sqrt{Hz}$. Given that $v_{out}(t)\approx v_{a}(t)$, the signal to noise ratio for these experiments can be determined to be,
\begin{equation}
SNR=\frac{v_{{a}_{Sol}}^{\textrm{RMS}}(\tau_a t)^{\frac{1}{4}}}{\sqrt{S_{V}}}=\frac{ g_{a\gamma\gamma} l_{Sol}N_{c}cB_{DC} \sqrt{\rho_\textrm{DM}c^3}(\tau_a t)^{\frac{1}{4}}}{\omega_a\sqrt{S_V}},
\label{SNRV}
\end{equation}
assuming that the integration time, $t$, exceeds or is equal to the axion coherence time, $\tau_a$. 

To compare the sensitivity of different detection techniques we can plot the expected sensitivity limit we would set on $g_{a\gamma\gamma}$ with a SNR=1, as presented in Fig. \ref{Sens}. However, this plot makes some assumptions, which must be highlighted. 1) The calculation assumes we know the waveform of detection, which we expect to be a narrow band noise process of coherence time $\tau_a$ and therefore represents our knowledge of the signal shape. 2) The time, $t$, is how long we average for, which may be set differently, so to really compare different detectors it should be set to the same amount. 3) The signal size of the detection process (RMS value of the signal) and 4) the noise in the detector (square root spectral density) are major design parameters, which determine the detectors sensitivity but may vary from proposal to proposal and are apparent in equations. (\ref{SNRI}) and (\ref{SNRV}). 5) Also tied in with our assumptions is the local density of dark matter, $\rho_{DM}$, which we take to be $0.45 GeV/cm^3$. Note in the noise calculations, we just include the noise supplied by the first cryogenic readout amplifier. This is because the gain of the amplifier usually renders noise processes up the chain as irrelevant. However, there may be systematic effects and other environmental random noise sources. For example fluctuations in magnetic field, pressure in the system, external electromagnetic interference etc. There will also be Nyquist noise in the readout itself, however this will just effectively add to the noise temperature of the amplifier, but at these low frequencies and low temperatures of 4 K and below, the Nyquist noise is much smaller than the amplifier noise.

The sensitivity of a variety of configurations are plotted in Fig. \ref{Sens}, which reveals some interesting points. For example, the sensitivity of the short circuit current configuration is only weakly dependent on size. This is because as we increase the magnet size to increase the generated emf, the size of the coil also increases in line with the inductance and resistance, limiting the value of the short circuit current. Thus, the best ways to increase sensitivity is to either apply a larger magnetic field, or if the frequency pole is too high, install a larger conductor radius to reduce the resistance and inductance of the coil. For the read out amplifier an optimally designed SQUID will lower the effective current noise when referred to the short circuit current and could significantly improve the sensitivity. In contrast the multiple loop winding, which directly detects the axion generated voltage source, $v_a(t)$, is increased with every turn of winding added to the readout. However the effective source inductance increases as $N_c^2$ as indicated by equation. (\ref{MTor}), so there is a limit to how many windings can be added before the effective voltage divider starts to reduce the sensitivity. This effective voltage divider depends on the HIA input impedance, so when designing such an experiment we must balance these effects. From Fig. \ref{Sens} we can see that experiments of size order $10cm$ can set limit on a significant portion of the low-mass axion band at QCD axion sensitivity with only one hour and 9 minutes of integration. If we use a 14 Tesla solenoid, similar to the ORGAN magnet, with one and a half months of data the whole QCD axion range can be searched. It is apparent that the short circuit current technique is more sensitive between axion masses of $10^{-12} - 10^{-10} GeV$, while the multiple coil voltage readout is more sensitive between $10^{-10} - 10^{-8} GeV$. This is mainly due to the excess flicker noise in the high impedance voltage amplifier \cite{HIA}.

\begin{figure}[t]
\includegraphics[width=1.0\columnwidth]{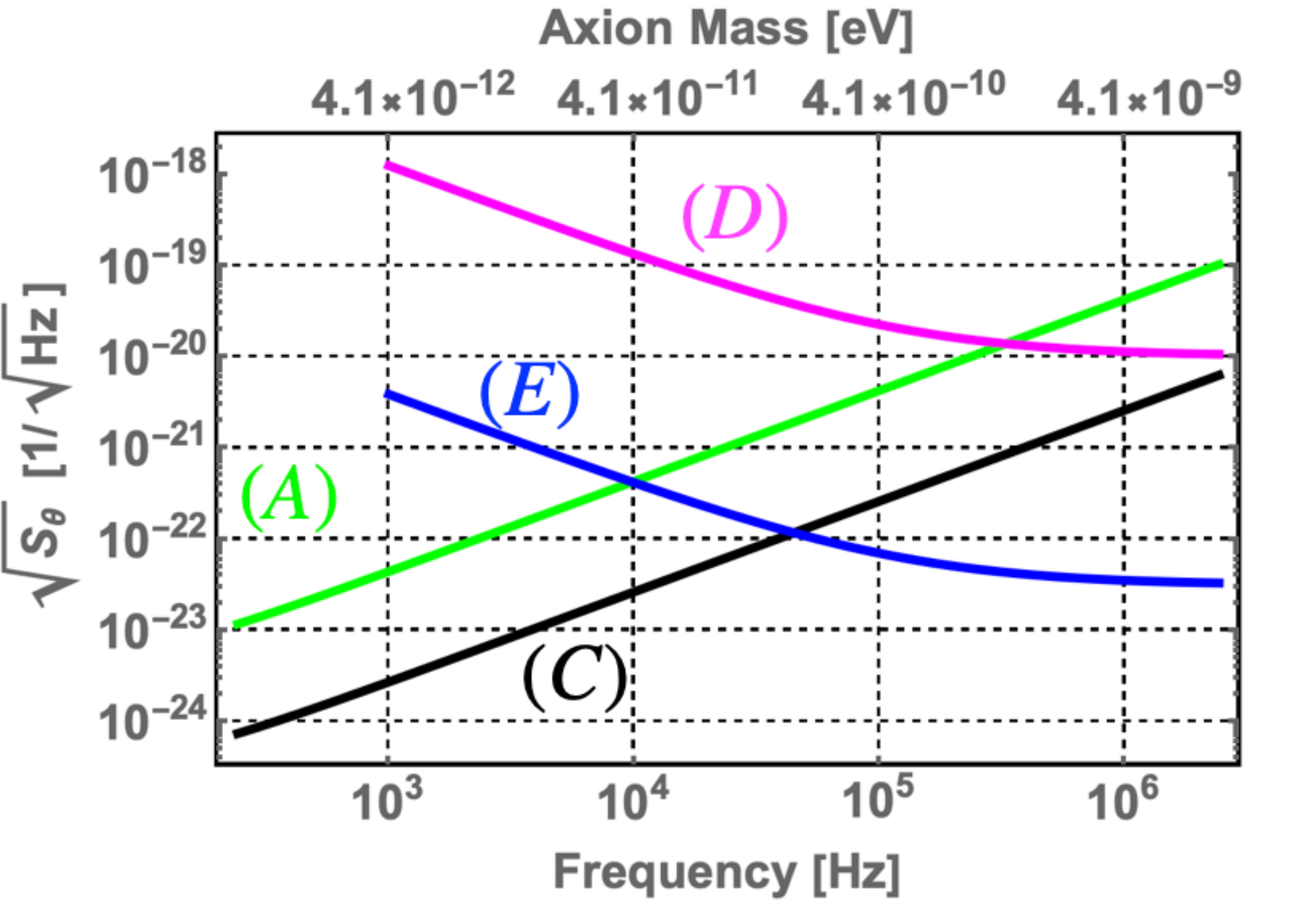}
\caption{Square root spectral density of axion-photon theta angle noise, $\sqrt{S_{\theta}}$ in the masses range of $10^{-12}$-$10^{-8}~eV$ the configurations presented in Fig. \ref{Sens} (same labels).}
\label{SpecSens}
\end{figure}

To compare different types of detectors with less assumptions, it may be useful to characterise detectors in a similar vain to the gravitational wave detector community, which use the calculated spectral strain sensitivity per square root Hz. Thus, we introduce a similar spectral density of noise that assumes nothing about the signal or the way we detect it (including the value of the dark matter density), and only considers the efficiency of detection and the noise in the detector itself. To do this we characterise the noise within the detector with respect to the mean square of the axion-photon theta angle noise $\langle\theta^2\rangle=g_{a\gamma\gamma}^2\langle a^2\rangle$, as a spectral density, $S_{\theta}/Hz$, which are related by $\langle\theta^2\rangle=\int_{f_1}^{f_2}S_{\theta}df$. This requires a conversion of the current and voltage noise spectral densities to spectral densities of $\sqrt{S_{\theta}}/\sqrt{Hz}$ using the relations between voltage and current as a function of $\theta=g_{a\gamma\gamma}a$. For example, given by equations (\ref{vat}) and (\ref{SCtor}) for the toroid configuration. Thus, the values of $\sqrt{S_v}$ volts$/\sqrt{Hz}$ and $\sqrt{S_I}$ amps$/\sqrt{Hz}$ may be converted to $\sqrt{S_{\theta}}/\sqrt{Hz}$ and are plotted in Fig. \ref{SpecSens}. This plot highlights our conclusion that between $10^{-10}$-$10^{-8} eV$ the multiple coils is inherently a more sensitive technique, but below $10^{-10} eV$ the short circuit current readout is more sensitive.
 
Of course to further improve sensitivity one could design a capacitive load, which would make the readout resonant, similar to a haloscope experiment, requiring scanning and would no longer be broad band. However, this new techniques is sensitive enough to reach the QCD axion limit as a broad band detector. In the advent that it is necessary to search beyond these values in the future, this modification would be possible. This may occur if it is shown that axions exist in this mass range, but are not $100\%$ of the dark matter, or maybe new calculations suggest a new axion model and that more sensitive measurements are necessary.

\subsection{Prospects of Searching for Ultra-Light Axion Dark Matter}

Models of light scalar DM have recently gained attention in the scientific community, in these models the scalar field couples to standard-model fields leading to the violation of the Einstein equivalence principle \cite{Damour1990,Damour1994}. The scalar-matter coupling in these models depend on fundamental constants \cite{Flambaum2004,Stadnik2015}, and means the local dark matter density can manifest as an oscillating fundamental constant of nature. There are several ways to search for variations in fundamental constants or scalar dark matter, which includes frequency modulations of atomic and classical oscillators \cite{Turneaure1983,Tobar2009,Tilburg2015,Hees2018,Arvanitaki2015}, torsion balances \cite{Schlamminger2008,Smith1999} and accelerometers in space \cite{Berge2018}, which typically search for masses between $10^{-24}$ to $10^{-8}eV$, referred to as Ultra-Light Axion Dark Matter (ULDM). Such experiments must be maintained for multiple years to be able to search for such low-masses, for example a particle mass of $10^{-22}eV$ corresponds to a frequency of $24.2 nHz$ with a period of 1.3 years. This technique relies on comparing two systems with different dependence on fundamental constants.

\begin{figure}[t]
\includegraphics[width=0.8\columnwidth]{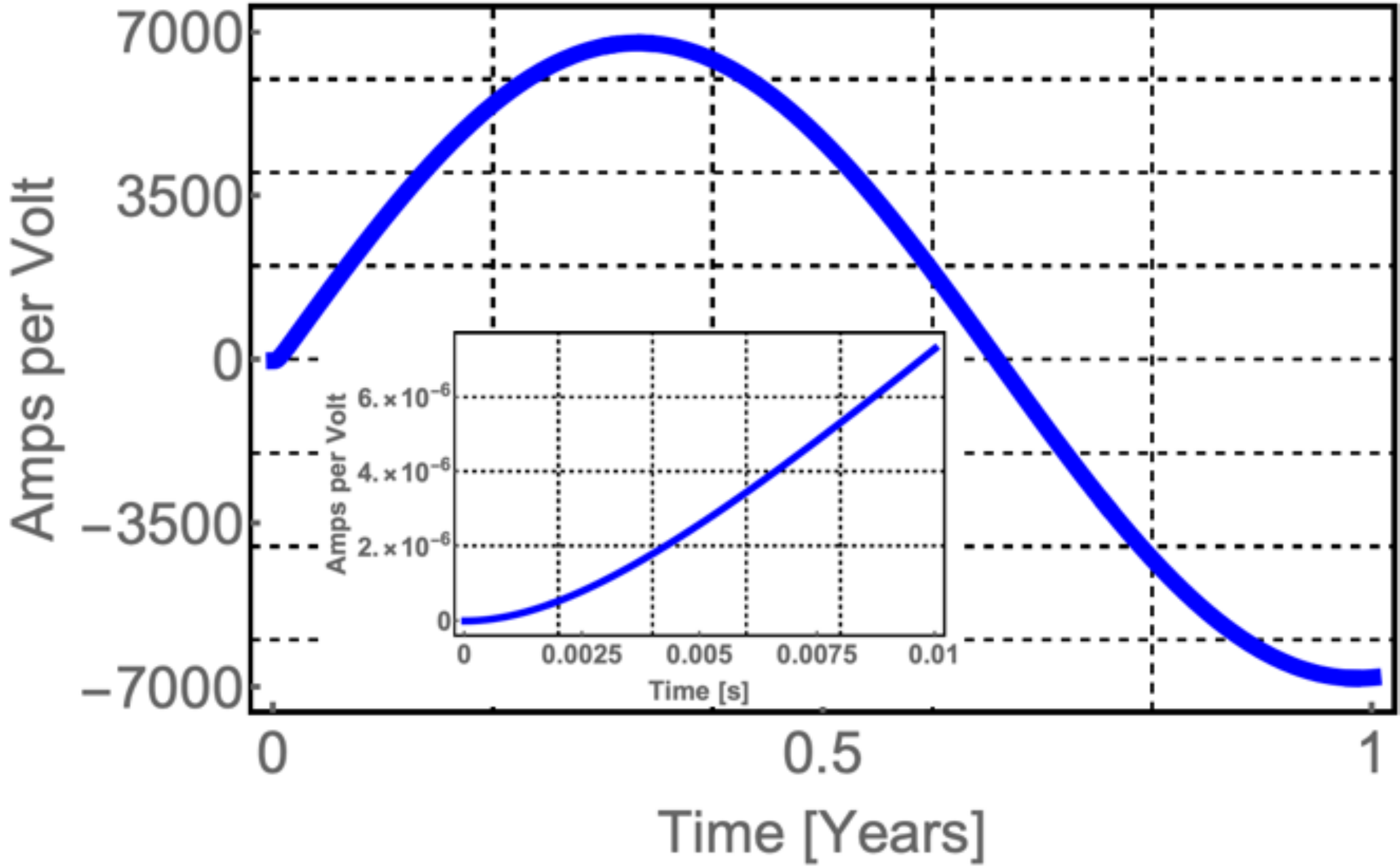}
\caption{Blue curve, $i_a(t)/v_0$ as a function of time as given by equation. (\ref{Curr}) in Appendix B, for the single winding circuit linking the ORGAN $14 T$ magnet at 24.2 nHz in frequency. Inset, the transient response for the first one tenth of a second.}
\label{FuzzyAx}
\end{figure}

Likewise, it is possible to devise experiments to look for ULDM axions \cite{ULACMB2017,Marsh17,Zhang_2018,Fedderke19}. For example, a recently developed frequency technique which utilises two real photons for the two photon degrees of freedom can work in a similar way to the ULDM frequency comparison experiments if they are made degenerate with non-zero $\int \vec{E}\cdot\vec{B}dv$ between the two photonic modes \cite{freqmetrology,Cat19}. It is also apparent in this work that the short circuit current technique with one winding is suited to ULDM experiments. This is because of the low flicker noise in SQUID amplifiers, and the fact that the experiment is only weakly dependent on size, so in principle can remain compact.

\begin{figure}[t]
\includegraphics[width=1.0\columnwidth]{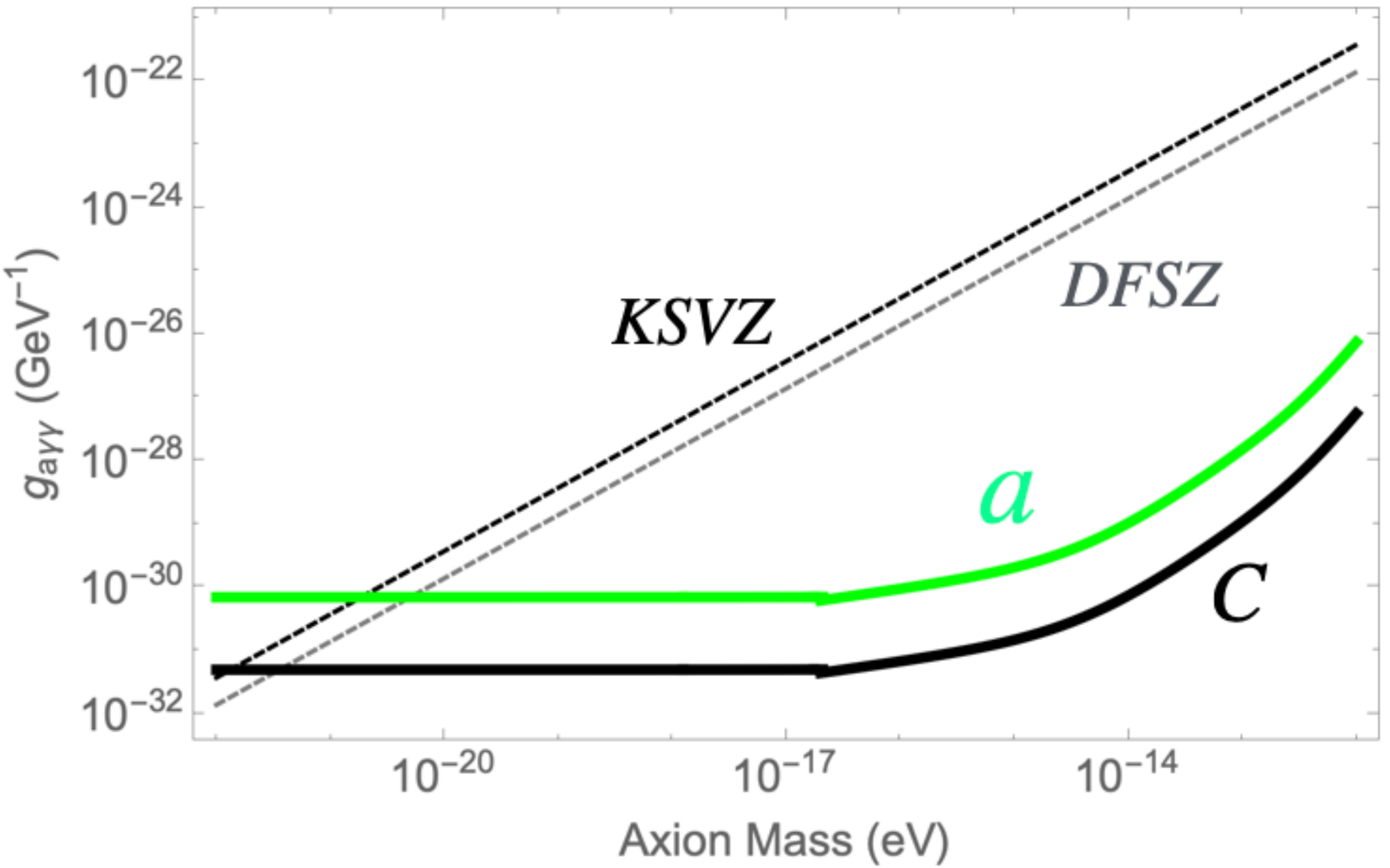}
\caption{Extrapolation of the sensitivity of the short circuit current technique to ULDM axions assuming 4 years of data. Below an axion mass of $3.3\times10^{-17}eV$ the measurement time is less than the inverse of the axion bandwidth.}
\label{ULSens}
\end{figure}

The expected short circuit current per volt is shown in Fig. \ref{FuzzyAx} for a axion mass of $10^{-22} eV$. Note, that the experiment can be still very sensitive at these low-masses (see Fig. \ref{ULSens}) even if the sensitivity $\rightarrow 0$ as the axion mass, $m_a\rightarrow 0$ (depends on the axion initial phase). This is because there will always be a significant signal at any oscillating axion frequency independent of this limit, so they are not related. The calculations presented in Fig. \ref{ULSens} only give an indication of the best sensitivity possible, because at such low frequencies other processes not considered can be present. For instance, temperature fluctuations if not controlled properly can give extra noise at low frequencies, and a range of other possible systematics need to be monitored and considered, in a similar way to the scalar ULDM and Lorentz Invariance experiments, which need to be operated for greater than a year. In our calculations we only extrapolate the SQUID flicker noise, as an ultimate level of sensitivity that in principle could be obtained. However, this limit shows the potential of this setup to test ULDM axions at DFSZ and KSVZ levels of sensitivity.

\section{Acknowledgements}
This work was funded by Australian Research Council grant numbers DP190100071 and CE170100009. The authors thank Ian McArthur for his critical analysis of our work and taking the time to understand the new concepts put forward. They also thank John Clarke and Eugene Ivanov for information about the low noise amplifiers discussed in this work, and Joerg Jaeckel, Javier Redondo, Matt Dolan and Ariel Zhitnitsky for interesting discussions at the workshop on Axion Cosmology at the Munich Institute for Astro- and Particle Physics (MIAPP), which was funded by the Deutsche Forschungsgemeinschaft (DFG, German Research Foundation) under Germany's Excellence Strategy EXC-2094 390783311.

\appendix

\section{Appendix A: Separating the External Applied Fields from the Axion Induced Reacted Fields}

This appendix discusses in more detail how we obtained equations (\ref{ExStatic}) and (\ref{eq:M7})-(\ref{EaB(t)1}) in the main body of the paper. The axion two-photon coupling to the electromagnetic field consists of three degrees of freedom, one axion and two photon, with a schematic shown in Fig. \ref{Mix}. The axion modification to Maxwell electrodynamics leads to the following set of equations,
\begin{equation}
\begin{aligned}
&\vec{\nabla} \cdot\left(\vec{E}-cg_{a \gamma \gamma} a \vec{B}\right)=\frac{\rho_{f}}{\varepsilon_0}\\
&\vec{\nabla} \times\left(\vec{B}+\frac{1}{c}g_{a \gamma \gamma} a \vec{E}\right)-\frac{1}{c^2} \frac{\partial}{\partial t}\left(\vec{E}-cg_{a \gamma \gamma} a \vec{B}\right)=\mu_0 \vec{J}_{f}\\
&\vec{\nabla} \cdot \vec{B}=0\\
&\vec{\nabla} \times \vec{E}+\frac{\partial \vec{B}}{\partial t}=0,
\end{aligned}
\label{ModAxED}
\end{equation}
which describe the two photonic degrees of freedom. Here $g_{a \gamma \gamma}$ is the two-photon coupling to an axion field, $a$, is the amplitude of the axion field, $\rho_f$, is the free volume charge density and $\vec{J}_{f}$ is the free volume current density. 

{\color{red} Explicitly, from this set of equations alone, a scalar value of $a$ cannot become a physical observable. However, this representation has no source terms to create electric or magnetic fields, and in a source free medium there are no currents or charges inputing energy to the system. Thus, the free currents, $\vec{J}_{f}$, and charges, $\rho_f$, in the above equations either propagate without loss due to the interaction with the electromagnetic fields, or can describe a dissipative (or resistive) system where electromagnetic energy is lost, usually by conversion to heat (in this case the electric and magnetic field phasors can become complex). What is not so obvious from these equations, is the photonic terms described by the $\vec{E}$ and $\vec{B}$ fields represent two degrees of freedom and the scalar value of $a$ can become a physical observable when a boundary source term capable of generating magnetic and/or electric fields is added as a forcing function to the equations.}

Trying to better understand the two photonic degrees of freedom in axion electrodynamics, recently Kim et.al. \cite{Younggeun18} came up with a new unambiguous way of describing inverse Primakoff effect of the axion-photon interaction when one of the photonic degrees of freedom is an externally applied field. They did this by applying an effective first order approximation to axion modified electrodynamics in vacuum, which allowed the separation of the two photonic degrees of freedom of axion electrodynamics to a set of integro-differential equations that represent the external applied EM fields, $\vec{E}_{0}(\vec{r},t)$ and $\vec{B}_{0}(\vec{r},t)$, from the set of integro-differential equations that describe the axion induced reacted EM fields, $\vec{E}_{a}(\vec{r},t)$ and $\vec{B}_{a}(\vec{r},t)$, where $\vec{E}=\vec{E}_{0}(\vec{r},t)+\vec{E}_{a}(\vec{r},t)$ and $\vec{B}=\vec{B}_{0}(\vec{r},t)+\vec{B}_{a}(\vec{r},t)$. {\color{red} Here we deviate slightly from Kim et.al. \cite{Younggeun18} and consider the free charge and current terms on the right hand side of equation (\ref{ModAxED}) in a more general way by including impressed source terms. To do this we make the following substitutions, $\vec{J}_f\rightarrow\vec{J}_e^i+\vec{J}_f$ and $\rho_f\rightarrow\rho_e^i+\rho_f$. Here, $\vec{J}_e^i$ and $\rho_e^i$ are impressed source terms that can create a magnetic field with a vector potential and an electric field with a scalar potential respectively. In contrast, $\vec{J}_f$ and $\rho_f$ are any free current and charge that might be in the detection system (not impressed) and will react to the fields induced in the photonic degrees of freedom. It is important to make the distinction between source and free currents and charges, because the former adds energy to the system while the later does not.}

Following the method of Kim et.al. \cite{Younggeun18}, the integro-differential equations that describe the external applied fields can be shown to be given by the normal Maxwell's equations with impressed sources as forcing functions,
\begin{equation}
\begin{aligned}
&\vec{\nabla} \cdot \vec{E}_{0}(\vec{r},t) =\rho_{e}^i / \epsilon_{0} \\
&\vec{\nabla} \times \vec{B}_{0}(\vec{r},t)-\frac{1}{c^2} \frac{\partial}{\partial t} \vec{E}_{0}(\vec{r},t)=\mu_{0} \vec{J}_{e}^i\\
&\vec{\nabla} \cdot \vec{B}_{0}(\vec{r},t)=0 \\
&\vec{\nabla} \times \vec{E}_{0}(\vec{r},t)+\frac{\partial}{\partial t} \vec{B}_{0}(\vec{r},t) =0,
\end{aligned}
\label{MaxED}
\end{equation}
and the set of integro-differential equations to solve for the axion induced reacted EM fields can be shown to be given by,
\begin{equation}
\begin{aligned}
&\vec{\nabla} \cdot\left(\vec{E}_{a}(\vec{r},t)-c g_{a \gamma \gamma} a(\vec{r},t) \vec{B}_{0}(\vec{r},t)\right)=\rho_f\\
&\vec{\nabla} \times\left(\vec{B}_{a}(\vec{r},t)+\frac{1}{c} g_{a \gamma \gamma} a(\vec{r},t) \vec{E}_{0}(\vec{r},t)\right)\\ 
&-\frac{1}{c^{2}} \frac{\partial}{\partial t}\left(\vec{E}_{a}(\vec{r},t)-c g_{a \gamma \gamma} a(\vec{r},t) \vec{B}_{0}(\vec{r},t)\right)=\mu_0\vec{J}_f\\
&\vec{\nabla} \cdot \vec{B}_{a}(\vec{r},t)=0\\
&\vec{\nabla} \times \vec{E}_{a}(\vec{r},t)+\frac{\partial \vec{B}_{a}(\vec{r},t)}{\partial t}=0.
\end{aligned}
\label{Reacted}
\end{equation}
This way of representing modified axion electrodynamics is extremely useful. It means for an arbitrary experiment we can calculate $\vec{E}_{0}(\vec{r},t)$ and $\vec{B}_{0}(\vec{r},t)$ from Maxwell's equations given by eqn (\ref{MaxED}), and if they then interact with a putative axion, then the generated photon EM fields may be calculated from eqn (\ref{Reacted}), which are inevitably at a different frequency when compared to the applied fields given by eqn (\ref{MaxED}). 

Now taking this a step further we can rewrite the reacted equations in a similar form to Maxwell electrodynamics, given by (\ref{Reacted}), as.
\begin{equation}
\begin{aligned}
&\vec{\nabla}\cdot\vec{D}_a^T =\rho_f\\
&\vec{\nabla} \times \vec{H}_a^T-\frac{\partial \vec{D}_a^T}{\partial t}=\vec{J}_f\\
&\vec{\nabla} \cdot \vec{B}_a= 0\\
&\vec{\nabla} \times \vec{E}_a+\frac{\partial \vec{B}_a}{\partial t} = 0
\end{aligned}
\label{axed1}
\end{equation}
with the following constitutive relationships;
\begin{equation}
\begin{aligned}
&\vec{D}_a^T=\epsilon_0\vec{E}_a^T =\epsilon_0(\vec{E}_a+\vec{E}_{aB}^i)\\
&\mu_0\vec{H}_a^T=\vec{B}_a^T =\vec{B}_a+\vec{B}_{aE}^i,
\end{aligned}
\label{const1}
\end{equation}
where,
\begin{equation}
\begin{aligned}
&\vec{E}_{aB}^i(\vec{r},t)=-g_{a\gamma\gamma}a(\vec{r},t)c\vec{B}_{0}(\vec{r},t)\\
&c\vec{B}_{aE}^i(\vec{r},t)=g_{a\gamma\gamma}a(\vec{r},t)\vec{E}_{0}(\vec{r},t).
\end{aligned}
\label{aximp1}
\end{equation}

Kim et al. \cite{Younggeun18} remark that the two sets of equations, which represent the two different photonic degrees of freedom, are decoupled. However, it is probably more correct to just call them separated. This is because the EM fields calculated from eqn (\ref{Reacted}) depend on $\vec{E}_{0}(\vec{r},t) $ and $\vec{B}_{0}(\vec{r},t) $, so it is not generally correct to solve the reacted equations from eqn (\ref{Reacted}) independently form eqn (\ref{MaxED}), as the results from eqn (\ref{MaxED}) must be substituted back into eqn (\ref{Reacted}). Furthermore, the equations given by (\ref{Reacted}) are now four equations with four variables $(\vec{E}_{a},\vec{B}_{a},\vec{E}_{0},\vec{B}_{0})$. While eqn (\ref{MaxED}) are also four equations, but with only two variables, $(\vec{E}_{0},\vec{B}_{0})$. Inevitably the information from eqn (\ref{MaxED}) must be used to solve for the EM fields given by eqn (\ref{Reacted}), and if any of the information is missing it may mean the most general form of the solution may be missed. 

To derive equations (\ref{ExStatic}) and (\ref{eq:M7})-(\ref{EaB(t)1}) in the main text, we assume the external applied field consists of only a DC $\vec{B}$-field, so that $\vec{B}_{0}(\vec{r},t)=\vec{B}_{DC}(\vec{r})$ and $\vec{E}_{0}=0$, sourced only by an impressed $DC$ electrical current, $\vec{J}^i_{DC}$, so equations (\ref{MaxED}) become,
\begin{equation}
\begin{aligned}
&\vec{\nabla} \cdot \vec{B}_{DC}(\vec{r})=0 \\
&\vec{\nabla} \times \vec{B}_{DC}(\vec{r})=\mu_{0} \vec{J}^i_{DC},
\end{aligned}
\end{equation}
equivalent to eqn (\ref{ExStatic}) in the text. Following this eqns. (\ref{axed1})-(\ref{aximp1}) become,
\begin{align}
&\vec{\nabla}\cdot\vec{D}_a^T =\rho_f,\label{eeq:M7}\\
&\vec{\nabla} \times \vec{B}_a-\mu_0\frac{\partial \vec{D}_a^T}{\partial t}=\mu_0\vec{J}_f,\label{eeq:M8}\\
&\vec{\nabla} \cdot \vec{B}_a= 0,\label{eeq:M9}\\
&\vec{\nabla} \times \vec{E}_a+\frac{\partial \vec{B}_a}{\partial t} = 0\label{eeq:M10}
\end{align}
with the following constitutive relationship;
\begin{equation}
\vec{D}_a^T=\epsilon_0\vec{E}_a^T =\epsilon_0(\vec{E}_a+\vec{E}_{aB}^i),
\end{equation}
where,
\begin{equation}
\vec{E}_{aB}^i(\vec{r},t)=-g_{a\gamma\gamma}a(t)c\vec{B}_{DC}(\vec{r}).
\end{equation}
equivalent to eqn (\ref{eq:M7})-(\ref{EaB(t)1}) in the text.

\section{Appendix B: Detection Circuit Properties and Response}

\subsubsection{Short Circuit Response}

\begin{figure}[t]
\includegraphics[width=0.95\columnwidth]{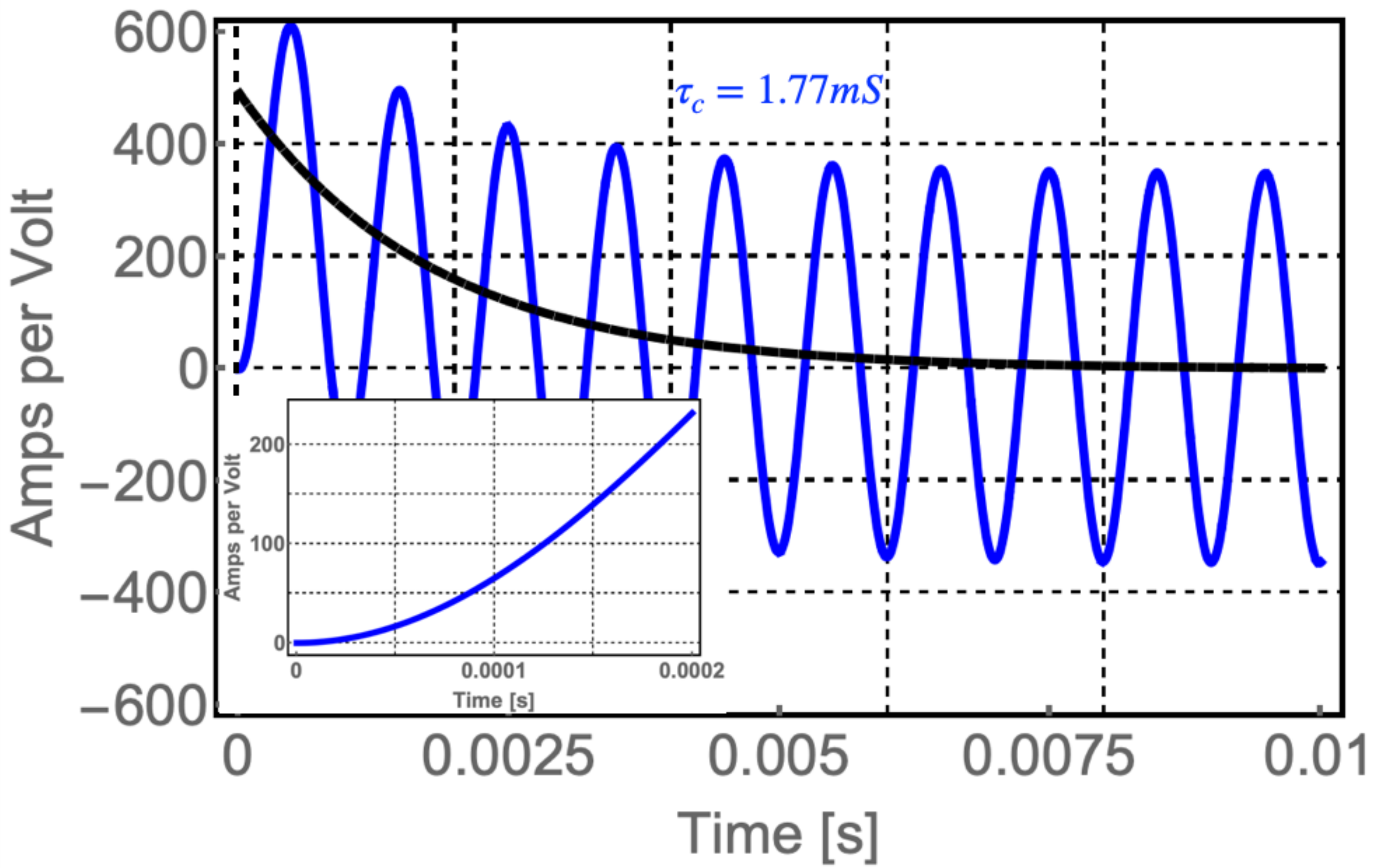}
\caption{Blue curve, $i_a(t)/v_0$ response of a single winding read out, as a function of time and at 1 kHz in frequency as given by eqn (\ref{Curr}) for the first $10 ms$ when $\phi$=0. Here the circuit parameters are given in appendix C for a toroid of order $10cm$ in size. Black curve, the transient response, which is the first term in eqn (\ref{Curr}). {\color{blue}In general the transient response will vary depending on the value of $\phi$, however note that the amplitude at the Fourier frequency equivalent to $\omega_a$ remains constant, even throughout the transient process, which is independent of the value of $\phi$.}}
\label{Transient}
\end{figure}

If we short circuit the Thevenin equivalent circuit across terminals $A-B$ as shown in Fig. \ref{WireThNo}, we will have a standard $LR$ circuit response. In the quasi static low-loss limit, it is important to investigate transient effects. For example, if our readout winding is a perfect conductor, $R_c=0$, then the circuit time constant, $\tau_c=\frac{Lc}{Rc}$ will be infinite and will not reach the steady state response. Assuming a sinusoidal voltage of the form $v_{a}(t)=v_0\sin(\omega_a t+\phi)$ driving the circuit, the text book way of solving this problem is by using Laplace transforms. In the Laplace domain the short circuit current is given by,
\begin{equation}
I_a(s)=\frac{V_a(s)}{R_c+sL_c}=\frac{v_0}{R_c+sL_c}\frac{s \sin (\phi)+\omega_a\cos (\phi )}{s^2+\omega_a^2}.
\label{Laplace}
\end{equation}
Here $V_a(s)$ is the Laplace transform of the input voltage $v_{a}(t)$. Then by taking the inverse Laplace transform of eqn (\ref{Laplace}) and assuming zero initial current, we can calculate the time domain response to be,
\begin{equation}
\begin{aligned}
&i_a(t)=\frac{v_0}{\sqrt{R_c^2+L_c^2 \omega_a ^2}}\Bigg[\cos\left(\phi+\tan^{-1}\left(\frac{1}{\omega_a\tau_c}\right)\right)e^{-\frac{t}{\tau_c}}+ \\
&\cos\phi\sin \left(\omega_a t-\tan ^{-1}(\omega_a \tau_c)\right)+ \\
&\sin\phi\sin \left(\omega_a t+\tan^{-1}\left(\frac{1}{\omega_a\tau_c}\right)\right)\Bigg],
\label{Curr}
\end{aligned}
\end{equation}
with an example of the time dependence shown in Fig. \ref{Transient} for when $\phi=0$. The first term in eqn (\ref{Curr}) is the transient response and the second and third term represent the steady state response.

When $\omega_a>\tau_c^{-1}$, the inductance of the winding dominates, and when $\omega_a<\tau_c^{-1}$  the resistance of the winding dominates. The value of $\tau_c^{-1}$ determines the pole frequency of the circuit. The sensitivity of a well designed experiment sets the pole of the circuit below all frequencies of interest. In reality there will be some resistance, especially as we need to use a normal conductor, as a superconductor will expel the magnetic field from the wire and will not be sensitive to axions. Furthermore, even if $\omega_a>\tau_c^{-1}$, the DC component given by the transient response will decay, with the steady state response reached in a few time constants, $\tau_c$ as shown in Fig. \ref{Curr}. In this limit the solution can be approximated by (setting $R_c\approx0$),
\begin{equation}
i_a(t)\approx-\left(1-e^{-\frac{t}{\tau_c}}\right)\frac{v_0}{L_c\omega_a}\cos(\omega_at+\phi),
\label{TC}
\end{equation}
Since the sensitivity of the experiment governed by the steady state  response, in this case when $\omega_a>\tau_c^{-1}$
\begin{equation}
i_a(\omega_at)\approx-\frac{v_0}{L_c\omega_a}\cos(\omega_at+\phi),
\label{TC}
\end{equation}
Independent of the transient effect the sinusoidal signal at $\omega_a$ has a constant amplitude of $\frac{v_0}{Lc\omega_a}$ (as indicated in Fig. \ref{Transient}) so in principle the transient signal does not effect the AC measurement at $\omega_a$ in frequency. This regime dominates in our calculation for axion masses between $10^{-12}$ to $10^{-8}$ eV.

In the limit when $\omega_a<\tau_c^{-1}$ the steady state response will be dominated by,
\begin{equation}
i_a(\omega_at)\approx\frac{v_0}{R_c} \sin \left(\omega_a t+\phi\right).
\label{SSCurr}
\end{equation}
This approximation will dominant when searching for ultra light axions.

\subsubsection{Impedance of a Single Circular Winding}

To analyse the current induced in a single circular winding readout of radius $r$ with a wire cross section of radius of $r_c$, we must calculate the inductance, which is well-known and in the limit $\frac{r_c}{r}<1$, is given by,
\begin{equation}
L_{c}=\mu_{0} r\left[\ln \left(\frac{8 r}{r_c}\right)-2+\frac{1}{4} Y\right]
\label{LoopInd}
\end{equation}
where
\begin{equation}
Y \approx \frac{1}{1+r_c \sqrt{\frac{1}{8} \mu_0 \kappa_c \omega_a}}
\end{equation}
Here, $Y$ is a value between 0 and 1 that depends on the distribution of the current in the wire, $Y = 0$ when the current flows only on the surface of the wire (complete skin effect like in a superconductor), Y = 1 when the current is DC and thus evenly spread over the cross-section of the wire \cite{Rosa1908}.

To model the resistance of the winding, at high frequencies there will be an influence of the skin depth, which will increase the resistance from its DC value. Empirically we can model the resistance in the wire as a function of skin depth, and hence frequency by the following formula,
\begin{equation}
R_c(\delta)=\frac{l_c}{\kappa_c}\left(\frac{1}{\pi r_c^2}+\frac{1}{2\pi r_c\delta}\right).
\label{rescoil}
\end{equation}
Here, $l_c=2\pi r$ is the length of the wire coil and $\delta$ is the skin depth, given by,
\begin{equation}
\delta=\sqrt{\frac{2 }{\omega_a \mu_0\kappa_c}} \sqrt{\sqrt{1+\left(\frac{\omega_a \epsilon_0}{\kappa_c}\right)^{2}}+\frac{\omega_a \epsilon_0}{\kappa_c}}.
\end{equation}

In principle it is best to design the pole frequency to be below the lowest frequency of interest. This is equivalent to $10^{-12}$ eV or 242 Hz. Note that both the resistance and the inductance of the coil can be modified by varying $r_c$, so an optimum design would minimise the inductance keeping the pole below 242 Hz owing to the fact that we are searching for dark matter axions. However, in practice we also need to consider the finite impedance of the readout amplifier. If we set the search for the axion mass between $10^{-12}$ to $10^{-8}$ eV we need to search between 242 Hz to 2.42 MHz in frequency. For these frequencies, we can calculate the short circuit current and flux induced in the coil for the QCD axion or ALP dark matter using equations (\ref{IaRMStor}) and (\ref{PhiaRMStor}). 

\subsubsection{Impedance of a Single Rectangular Winding}

A single winding read out for solenoid can be configured as shown in Fig. \ref{Sol}. To compare with a toroid we consider the case when the value of $v_a(t)$ in both systems is equal, i.e. when $2\pi r_{Tor}=l_{Sol}$. In comparison the length of coil required to link the current for the solenoid is, $l_c=2(l_{Sol}+w_{Sol})$, which is at least as twice as long the coil required in the equivalent toroid. This makes the toroid a more efficient experiment to measure the axion induced short circuit current. Nevertheless the solenoid is a more common and cheaper magnetic structure, so we follow through with the calculation and compare it to the toroid.

The single winding inductance for the readout coil shown in Fig. \ref{Sol} can be shown to be given by \cite{Grover46},
\begin{equation}
\begin{aligned}
\begin{aligned}
&L_c=\frac{\mu_{0}}{\pi}\left[l_{Sol} \ln \left(\frac{2 l_{Sol}}{r_c}\right)+w_{Sol} \ln \left(\frac{2 w_{Sol}}{r_c}\right)\right.\\
&\left.+2 \sqrt{l_{Sol}^{2}+w_{Sol}^{2}}-l_{Sol} \sinh ^{-1}\left(\frac{l_{Sol}}{w_{Sol}}\right)\right.\\
&\left.-w_{Sol} \sinh ^{-1}\left(\frac{w_{Sol}}{l_{Sol}}\right)-\left(2-\frac{1}{4} Y\right)(l_{Sol}+w_{Sol})\right],
\label{SloopInd}
\end{aligned}
\end{aligned}
\end{equation}
where the resistance of the coil can be calculated by eqn (\ref{rescoil}). 

\subsubsection{Impedance of a Multiple Winding Inductive Voltage Readout}

\begin{figure}[t]
\includegraphics[width=0.7\columnwidth]{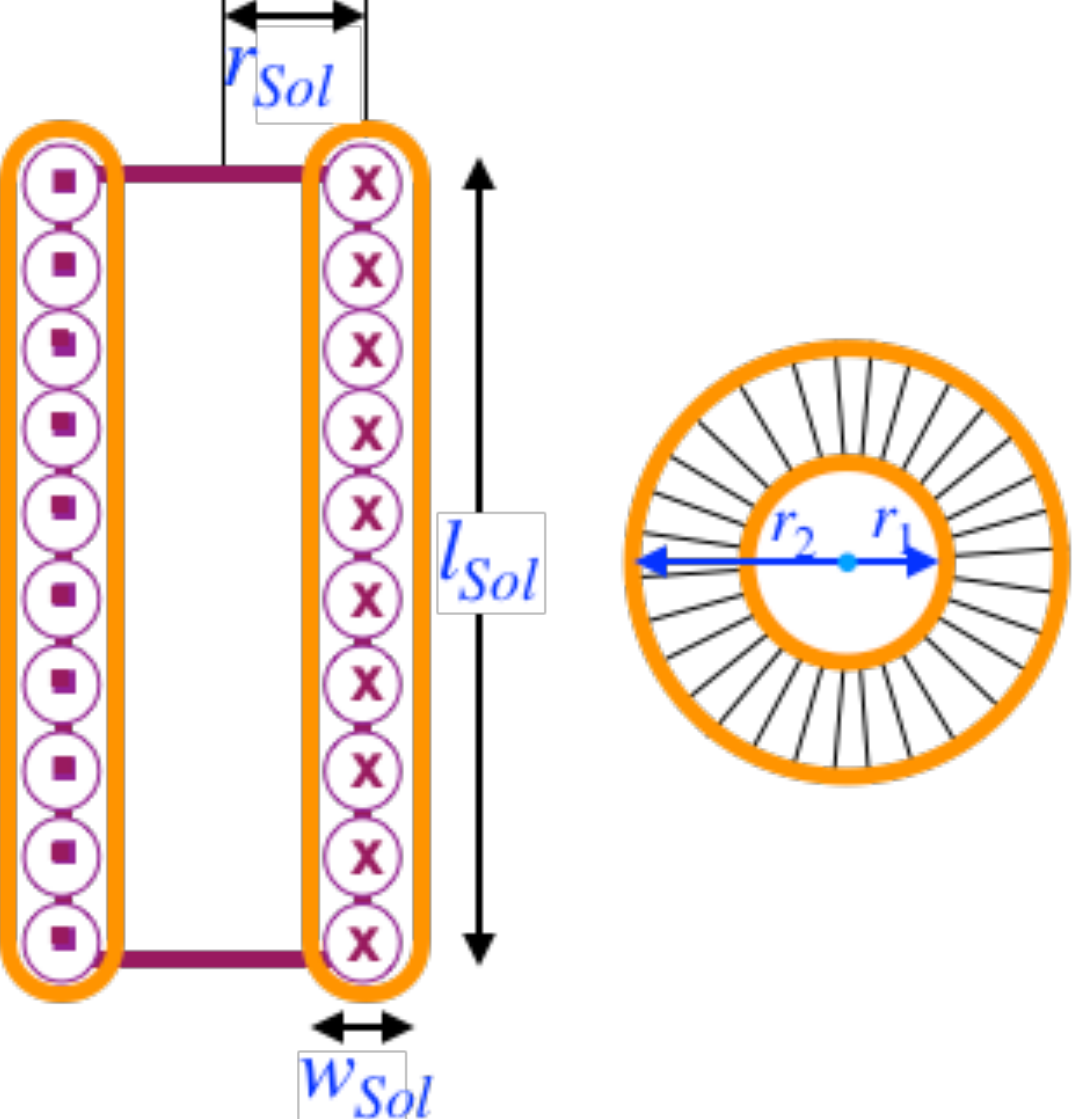}
\caption{A multi winding inductive readout, which consists of a rectangular cross section toroid wound around a superconducting magnet. Left, side view cross section and right, top view. The superconducting solenoid has a length of, $l_{Sol}$, and radius, $r_{Sol}$, while the readout toroidal coil has an inner radius of $r_1$ and an outer radius of $r_2$.}
\label{Multi}
\end{figure}
Another way to make a sensitive measurement is to make a readout with multiple windings ($N_c>0$) to increase the effective axion voltage source as given by eqns. (\ref{VaRMStor}) and (\ref{VaRMSsol}). However, this will not increase the short circuit current, as the impedance of the readout will increase proportionally. In fact we can show that the short circuit current readout is optimum for a single winding readout when coupled to a SQUID amplifier.

In contrast, a multiple coil readout requires a high impedance voltage amplifier to measure the axion induced voltage source. There are two ways we could do this, 1) wrap a solenoidal coil inside a toroidal coil, or 2) wrap a toroidal coil around a solenoidal coil as shown in Fig. \ref{Multi}. Here, we only chose to analyse the latter, as it would be more practical to wind normal conducting toroid windings around a superconducting magnet of solenoidal structure, and both systems would achieve similar sensitivity for the equivalent systems.

The effective inductance of the toroid read out is given by,
\begin{equation}
L_{RO}=\frac{\mu_{0} N_c^{2} l_{sol}}{2 \pi} \ln \left(\frac{r_2}{r_1}\right)
\label{MTor}
\end{equation}
Assuming a tightly wound coil, $r_1=r_{sol}-\frac{w_{sol}}{2}$ and $r_2=r_{sol}+\frac{w_{sol}}{2}$ with $100\%$ packing efficiency, then the single layer number of windings will be $N_c=\frac{\pi r_1}{r_c}$. Again the resistance of the coil can be calculated by eqn (\ref{rescoil}).

\section{Appendix C: Calculation of Circuit Parameters}

In this appendix we present the detailed calculations of the possible experimental circuit parameters for toroidal and solenoidal electromagnets to allow the sensitivity to QCD axions to be estimated in the main text.

\subsubsection{Toroidal Electro-Magnet with Single Winding Current Readout} 

A leading example of a low-mass toroidal experiment is ABRACADABRA-$10cm$ \cite{ABRACADABRA,FirstAbra,Oue19}, which is named as such because the dimensional size is of order $10cm$, with an inner toroidal volume of $V_{abra}=890cm^3$. This experiment implements a pick up coil in the central region outside the magnets coil winding, and detects the magnetic flux generated by the axion displacement current given by eqn (\ref{eq:M2b}). Since the operation of this experiment is well documented, it provides a good comparison to our new way of searching for low-mass axions documented in this paper.
 
The new technique described here, requires the detection coil (or pick up coil) to be placed inside the toroid as shown in Fig. \ref{Tor}. Such a superconducting coil exists in ABRACADABRA-$10cm$, which was primarily added for calibration purposes \cite{Oue19}. Here we assume a similar coil made from oxygen free copper to allow the DC magnetic field to penetrate. 
 \begin{figure}[t]
\includegraphics[width=1.0\columnwidth]{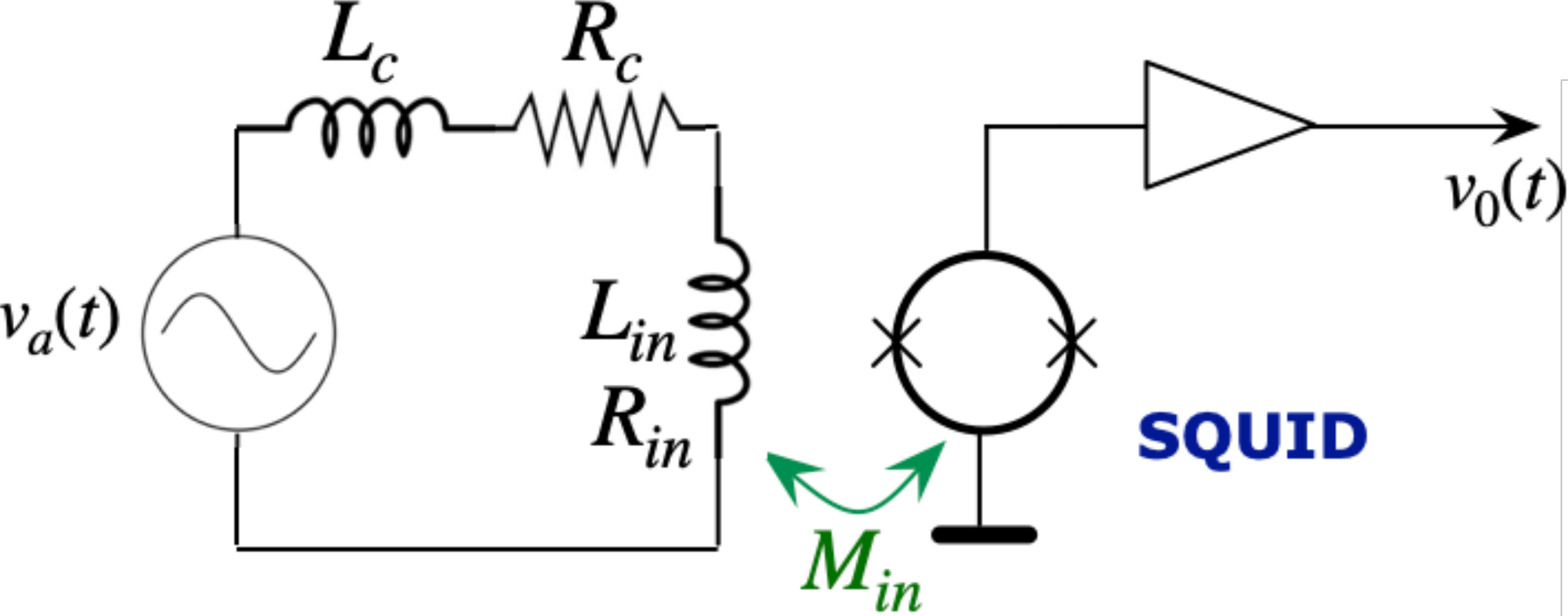}
\caption{Equivalent circuit of the single winding detection coil, configured as a short circuit current loop, coupled to a low impedance SQUID amplifier. Here, $v_a(t)$ is the induced axion voltage, shown in Fig. \ref{WireThNo}, with the impedance in the circuit governed by the small inductance, $L_c$, and resistance, $R_c$ of the readout coil, along with the small input inductance, $L_{in}$ and input resistance, $R_{in}$of the SQUID amplifier input coil. Since the response is close to a short circuit, an amplified current is obtained.}
\label{sqcct}
\end{figure}

The coil for this experiment has dimensions, $r_{Tor}=45mm$ and $r_c=0.25mm$, which from eqn (\ref{LoopInd}), gives a low frequency inductance of $0.31~\mu H$. If the coil is assumed to be made from oxygen free copper, at low temperatures it will exhibit a resistivity of order $10^{-2}\mu\Omega~cm$ \cite{NBS1053} equivalent to a conductivity of $\kappa_c=10^{10}\Omega^{-1}m^{-1}$, so that the DC resistance is equal to $R_c =0.144 m\Omega$. One of the proven ways to implement a sensitive readout is to implement a SQUID amplifier, with a schematic of the experiment shown in Fig. \ref{sqcct}. The input coil of the SQUID will load the circuit as indicated in Fig. \ref{WireThNo}. Here we use typical values of a Magnicon SQUID amplifier, which have previously been used for similar measurements of this kind \cite{Oue19,BEAST}. Given the SQUID has a superconducting input coil (so $R_{in}$ is negligible), with inductance of $L_{in}=150 nH$ \cite{Oue19} the total circuit will exhibit a DC resistance of, $R_T =R_c+R_{in}=0.144 m\Omega$ and a low frequency inductance of $L_T=L_c+L_{in}=0.46~\mu H$ resulting in a low frequency time constant of $\tau_T=\frac{L_T}{R_c}$ of about $3 ms$ and a pole frequency of about $50 Hz$. These properties are used to calculate the short circuit current created by axion dark matter under DC magnetic field, and estimate the sensitivity of such an experiment.

\subsubsection{Solenoidal Electro-Magnet with Single Winding Current Readout} 

In this subsection and the next we use the experimental properties of the equivalent solenoid of equivalent size to the toroid electromagnet discussed in the previous subsection To do this we assume a solenoid of the same magnetic field, $B_{DC}=1$ Tesla, with a length of, $l_{Sol}=2\pi\times45mm=282.7mm$, and an equivalent volume of $Vol_{Tor}=890 cm^3$, so the radius of the equivalent solenoid is given by $r_{Sol}=\sqrt{\frac{Vol_{Tor}}{\pi l_{sol}}}=31.7mm$.

The configuration of this solenoid, with a single winding detection coil is shown in Fig. \ref{Sol}, with the dimensions above and a coil winding radius of $r_c=0.25mm$ to be consistent with the previous example. Then assuming $w_{Sol}\approx 1cm$, from eqn (\ref{SloopInd}) we obtain a low frequency inductance of $0.46~\mu H$. If the coil is to be made from oxygen free copper, at low temperatures the DC resistance will be equal to $R_c=0.30 m\Omega$. Assuming a SQUID readout as shown in Fig. \ref{sqcct}, with a superconducting coil input inductance of $L_{in}=150 nH$ \cite{Oue19} the total circuit will exhibit a DC resistance of, $R_T=R_c=0.30 m\Omega$ and a low frequency inductance of $L_T=L_c+L_{in}=0.61~\mu H$ resulting in a low frequency time constant of $\tau_T=\frac{L_T}{R_c}$ of about $2 ms$ and a pole frequency of about $78 Hz$. To compute the response for both the Solenoid and Toroid magnets, we take into account the full frequency dependence as outline in the main text, including extra resistive skin effects and changes in inductance.

\subsubsection{Solenoidal Electro-Magnet with Multiple Winding Voltage Readout} 
 
 \begin{figure}[t]
\includegraphics[width=0.7\columnwidth]{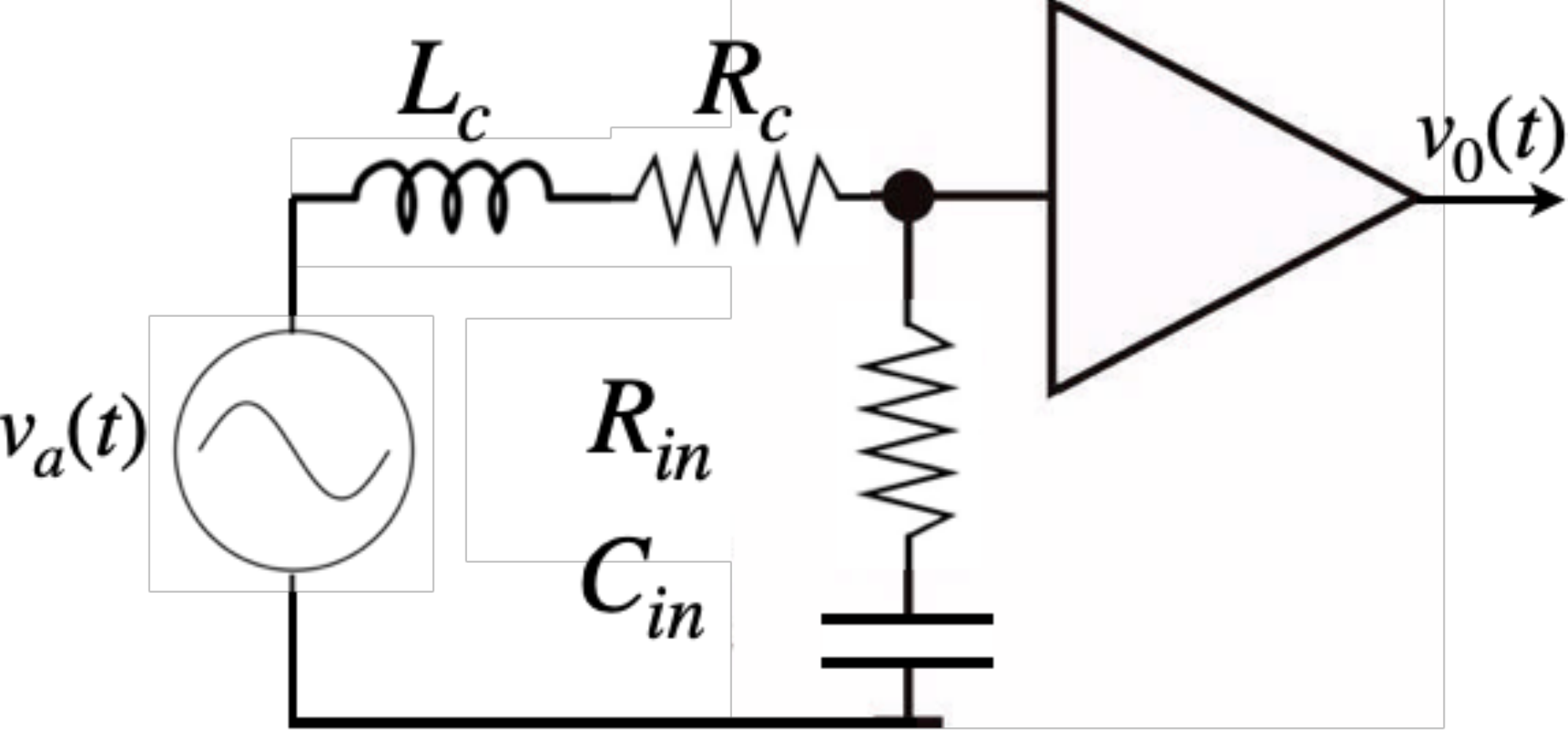}
\caption{Schematic of a High Impedance voltage Amplifier (HIA) \cite{HIA}. The amplifier exhibits a large input impedance of greater than $15 M\Omega$ with an input capacitance to ground of $4.2 pF$ and can supply a voltage gain between $30$ to $70$ dB above 1 kHz in frequency. Below this frequency the gain is not sufficient.}
\label{Vamp}
\end{figure}

Assuming the same solenoid as defined in the prior subsection, we can estimate the properties of the multiple winding toroidal readout coil as indicated from Fig. \ref{Multi} and equation (\ref{MTor}). Here, we assume the following dimensions, $r_1=r_{sol}-\frac{w_{sol}}{2}=26.7mm$, $r_2=r_{sol}+\frac{w_{sol}}{2}=36.7mm$ and $N_c=\frac{\pi r_1}{r_c}=335$ so that $L_{RO}=2mH$. Assuming the coil is made from oxygen free copper, at low temperatures the DC resistance will be equal to $R_c=100 m\Omega$. To compare to the other circuits discussed previously, we can calculate the short circuit current time constant and frequency pole to be $20ms$ and $21 Hz$ respectively. 

However, the multiple winding readout is not efficient for reading out the short circuit current, and is better suited to a voltage readout. The readout amplifier for this experiment is a CX-4 ``Cryogenic Super Low Noise Amplifier'', from stahl-electronics with high input impedance \cite{HIA}, with a schematic shown in Fig. \ref{Vamp}. The amplifier has an input capacitance of $C_{in}=4.2 pF$ and an overall input impedance $15 M\Omega$ below $100 kHz$. The net effect is that the input impedance is much higher than the equivalent source impedance combined of the readout inductance, $L_{RO}=2mH$ and the coil resistance, $R_c=100 m\Omega$, over all frequencies of interest. The frequencies of interest are also limited by the amplifier bandwidth, equivalent to an axion mass of $4.137\times10^{-12} eV$ or frequency of 1 kHz, up to an axion mass of $10^{-8} eV$ or $2.42 MHz$. So effectively in these experiments the output voltage of the equivalent circuit (see Fig. \ref{WireThNo}), gives, $v_{out}(t)\approx v_{a}(t)$. In this case we ignore the frequency dependance of the inductance due to the skin effect, as it is a small variation and has little impact on the conclusions above, and hence sensitivity estimations.


\begin{thebibliography}{76}%
\makeatletter
\providecommand \@ifxundefined [1]{%
 \@ifx{#1\undefined}
}%
\providecommand \@ifnum [1]{%
 \ifnum #1\expandafter \@firstoftwo
 \else \expandafter \@secondoftwo
 \fi
}%
\providecommand \@ifx [1]{%
 \ifx #1\expandafter \@firstoftwo
 \else \expandafter \@secondoftwo
 \fi
}%
\providecommand \natexlab [1]{#1}%
\providecommand \enquote  [1]{``#1''}%
\providecommand \bibnamefont  [1]{#1}%
\providecommand \bibfnamefont [1]{#1}%
\providecommand \citenamefont [1]{#1}%
\providecommand \href@noop [0]{\@secondoftwo}%
\providecommand \href [0]{\begingroup \@sanitize@url \@href}%
\providecommand \@href[1]{\@@startlink{#1}\@@href}%
\providecommand \@@href[1]{\endgroup#1\@@endlink}%
\providecommand \@sanitize@url [0]{\catcode `\\12\catcode `\$12\catcode
  `\&12\catcode `\#12\catcode `\^12\catcode `\_12\catcode `\%12\relax}%
\providecommand \@@startlink[1]{}%
\providecommand \@@endlink[0]{}%
\providecommand \url  [0]{\begingroup\@sanitize@url \@url }%
\providecommand \@url [1]{\endgroup\@href {#1}{\urlprefix }}%
\providecommand \urlprefix  [0]{URL }%
\providecommand \Eprint [0]{\href }%
\providecommand \doibase [0]{http://dx.doi.org/}%
\providecommand \selectlanguage [0]{\@gobble}%
\providecommand \bibinfo  [0]{\@secondoftwo}%
\providecommand \bibfield  [0]{\@secondoftwo}%
\providecommand \translation [1]{[#1]}%
\providecommand \BibitemOpen [0]{}%
\providecommand \bibitemStop [0]{}%
\providecommand \bibitemNoStop [0]{.\EOS\space}%
\providecommand \EOS [0]{\spacefactor3000\relax}%
\providecommand \BibitemShut  [1]{\csname bibitem#1\endcsname}%
\let\auto@bib@innerbib\@empty
%</preamble>
\bibitem [{\citenamefont {Peccei}\ and\ \citenamefont {Quinn}(1977)}]{PQ1977}%
  \BibitemOpen
  \bibfield  {author} {\bibinfo {author} {\bibfnamefont {R.~D.}\ \bibnamefont
  {Peccei}}\ and\ \bibinfo {author} {\bibfnamefont {H.~R.}\ \bibnamefont
  {Quinn}},\ }\href {\doibase 10.1103/PhysRevLett.38.1440} {\bibfield
  {journal} {\bibinfo  {journal} {Phys. Rev. Lett.}\ }\textbf {\bibinfo
  {volume} {38}},\ \bibinfo {pages} {1440} (\bibinfo {year}
  {1977})}\BibitemShut {NoStop}%
\bibitem [{\citenamefont {Wilczek}(1978)}]{Wilczek1978}%
  \BibitemOpen
  \bibfield  {author} {\bibinfo {author} {\bibfnamefont {F.}~\bibnamefont
  {Wilczek}},\ }\href {\doibase 10.1103/PhysRevLett.40.279} {\bibfield
  {journal} {\bibinfo  {journal} {Phys. Rev. Lett.}\ }\textbf {\bibinfo
  {volume} {40}},\ \bibinfo {pages} {279} (\bibinfo {year} {1978})}\BibitemShut
  {NoStop}%
\bibitem [{\citenamefont {Weinberg}(1978)}]{Weinberg1978}%
  \BibitemOpen
  \bibfield  {author} {\bibinfo {author} {\bibfnamefont {S.}~\bibnamefont
  {Weinberg}},\ }\href {\doibase 10.1103/PhysRevLett.40.223} {\bibfield
  {journal} {\bibinfo  {journal} {Phys. Rev. Lett.}\ }\textbf {\bibinfo
  {volume} {40}},\ \bibinfo {pages} {223} (\bibinfo {year} {1978})}\BibitemShut
  {NoStop}%
\bibitem [{\citenamefont {Jaeckel}\ and\ \citenamefont
  {Ringwald}(2010)}]{wisps}%
  \BibitemOpen
  \bibfield  {author} {\bibinfo {author} {\bibfnamefont {J.}~\bibnamefont
  {Jaeckel}}\ and\ \bibinfo {author} {\bibfnamefont {A.}~\bibnamefont
  {Ringwald}},\ }\href {\doibase 10.1146/annurev.nucl.012809.104433} {\bibfield
   {journal} {\bibinfo  {journal} {Annual Review of Nuclear and Particle
  Science}\ }\textbf {\bibinfo {volume} {60}},\ \bibinfo {pages} {405}
  (\bibinfo {year} {2010})}\BibitemShut {NoStop}%
\bibitem [{\citenamefont {Kim}(1979)}]{K79}%
  \BibitemOpen
  \bibfield  {author} {\bibinfo {author} {\bibfnamefont {J.~E.}\ \bibnamefont
  {Kim}},\ }\href {\doibase 10.1103/PhysRevLett.43.103} {\bibfield  {journal}
  {\bibinfo  {journal} {Phys. Rev. Lett.}\ }\textbf {\bibinfo {volume} {43}},\
  \bibinfo {pages} {103} (\bibinfo {year} {1979})}\BibitemShut {NoStop}%
\bibitem [{\citenamefont {Kim}\ and\ \citenamefont {Carosi}(2010)}]{Kim2010}%
  \BibitemOpen
  \bibfield  {author} {\bibinfo {author} {\bibfnamefont {J.~E.}\ \bibnamefont
  {Kim}}\ and\ \bibinfo {author} {\bibfnamefont {G.}~\bibnamefont {Carosi}},\
  }\href {\doibase 10.1103/RevModPhys.82.557} {\bibfield  {journal} {\bibinfo
  {journal} {Rev. Mod. Phys.}\ }\textbf {\bibinfo {volume} {82}},\ \bibinfo
  {pages} {557} (\bibinfo {year} {2010})}\BibitemShut {NoStop}%
\bibitem [{\citenamefont {Zhitnitsky}(1980)}]{Zhitnitsky:1980tq}%
  \BibitemOpen
  \bibfield  {author} {\bibinfo {author} {\bibfnamefont {A.~R.}\ \bibnamefont
  {Zhitnitsky}},\ }\href@noop {} {\bibfield  {journal} {\bibinfo  {journal}
  {Sov. J. Nucl. Phys.}\ }\textbf {\bibinfo {volume} {31}},\ \bibinfo {pages}
  {260} (\bibinfo {year} {1980})}\BibitemShut {NoStop}%
\bibitem [{\citenamefont {Dine}\ \emph {et~al.}(1981)\citenamefont {Dine},
  \citenamefont {Fischler},\ and\ \citenamefont {Srednicki}}]{DFS81}%
  \BibitemOpen
  \bibfield  {author} {\bibinfo {author} {\bibfnamefont {M.}~\bibnamefont
  {Dine}}, \bibinfo {author} {\bibfnamefont {W.}~\bibnamefont {Fischler}}, \
  and\ \bibinfo {author} {\bibfnamefont {M.}~\bibnamefont {Srednicki}},\ }\href
  {\doibase http://dx.doi.org/10.1016/0370-2693(81)90590-6} {\bibfield
  {journal} {\bibinfo  {journal} {Physics Letters B}\ }\textbf {\bibinfo
  {volume} {104}},\ \bibinfo {pages} {199 } (\bibinfo {year}
  {1981})}\BibitemShut {NoStop}%
\bibitem [{\citenamefont {Shifman}\ \emph {et~al.}(1980)\citenamefont
  {Shifman}, \citenamefont {Vainshtein},\ and\ \citenamefont
  {Zakharov}}]{SVZ80}%
  \BibitemOpen
  \bibfield  {author} {\bibinfo {author} {\bibfnamefont {M.}~\bibnamefont
  {Shifman}}, \bibinfo {author} {\bibfnamefont {A.}~\bibnamefont {Vainshtein}},
  \ and\ \bibinfo {author} {\bibfnamefont {V.}~\bibnamefont {Zakharov}},\
  }\href {\doibase http://dx.doi.org/10.1016/0550-3213(80)90209-6} {\bibfield
  {journal} {\bibinfo  {journal} {Nuclear Physics B}\ }\textbf {\bibinfo
  {volume} {166}},\ \bibinfo {pages} {493 } (\bibinfo {year}
  {1980})}\BibitemShut {NoStop}%
\bibitem [{\citenamefont {Dine}\ and\ \citenamefont
  {Fischler}(1983)}]{Dine1983}%
  \BibitemOpen
  \bibfield  {author} {\bibinfo {author} {\bibfnamefont {M.}~\bibnamefont
  {Dine}}\ and\ \bibinfo {author} {\bibfnamefont {W.}~\bibnamefont
  {Fischler}},\ }\href {\doibase
  http://dx.doi.org/10.1016/0370-2693(83)90639-1} {\bibfield  {journal}
  {\bibinfo  {journal} {Physics Letters B}\ }\textbf {\bibinfo {volume}
  {120}},\ \bibinfo {pages} {137 } (\bibinfo {year} {1983})}\BibitemShut
  {NoStop}%
\bibitem [{\citenamefont {Preskill}\ \emph {et~al.}(1983)\citenamefont
  {Preskill}, \citenamefont {Wise},\ and\ \citenamefont
  {Wilczek}}]{Preskill1983}%
  \BibitemOpen
  \bibfield  {author} {\bibinfo {author} {\bibfnamefont {J.}~\bibnamefont
  {Preskill}}, \bibinfo {author} {\bibfnamefont {M.~B.}\ \bibnamefont {Wise}},
  \ and\ \bibinfo {author} {\bibfnamefont {F.}~\bibnamefont {Wilczek}},\ }\href
  {\doibase http://dx.doi.org/10.1016/0370-2693(83)90637-8} {\bibfield
  {journal} {\bibinfo  {journal} {Physics Letters B}\ }\textbf {\bibinfo
  {volume} {120}},\ \bibinfo {pages} {127 } (\bibinfo {year}
  {1983})}\BibitemShut {NoStop}%
\bibitem [{\citenamefont {Abbott}\ and\ \citenamefont
  {Sikivie}(1983)}]{Sikivie1983}%
  \BibitemOpen
  \bibfield  {author} {\bibinfo {author} {\bibfnamefont {L.}~\bibnamefont
  {Abbott}}\ and\ \bibinfo {author} {\bibfnamefont {P.}~\bibnamefont
  {Sikivie}},\ }\href {\doibase http://dx.doi.org/10.1016/0370-2693(83)90638-X}
  {\bibfield  {journal} {\bibinfo  {journal} {Physics Letters B}\ }\textbf
  {\bibinfo {volume} {120}},\ \bibinfo {pages} {133 } (\bibinfo {year}
  {1983})}\BibitemShut {NoStop}%
\bibitem [{\citenamefont {Ipser}\ and\ \citenamefont
  {Sikivie}(1983)}]{Sikivie1983b}%
  \BibitemOpen
  \bibfield  {author} {\bibinfo {author} {\bibfnamefont {J.}~\bibnamefont
  {Ipser}}\ and\ \bibinfo {author} {\bibfnamefont {P.}~\bibnamefont
  {Sikivie}},\ }\href {\doibase 10.1103/PhysRevLett.50.925} {\bibfield
  {journal} {\bibinfo  {journal} {Phys. Rev. Lett.}\ }\textbf {\bibinfo
  {volume} {50}},\ \bibinfo {pages} {925} (\bibinfo {year} {1983})}\BibitemShut
  {NoStop}%
\bibitem [{\citenamefont {Wilczek}(1987)}]{Wilczek:1987aa}%
  \BibitemOpen
  \bibfield  {author} {\bibinfo {author} {\bibfnamefont {F.}~\bibnamefont
  {Wilczek}},\ }\href {http://link.aps.org/doi/10.1103/PhysRevLett.58.1799}
  {\bibfield  {journal} {\bibinfo  {journal} {Physical Review Letters}\
  }\textbf {\bibinfo {volume} {58}},\ \bibinfo {pages} {1799} (\bibinfo {year}
  {1987})}\BibitemShut {NoStop}%
\bibitem [{\citenamefont {Rodr{\'\i}guez-Tzompantzi}(2020)}]{ORT20}%
  \BibitemOpen
  \bibfield  {author} {\bibinfo {author} {\bibfnamefont {O.}~\bibnamefont
  {Rodr{\'\i}guez-Tzompantzi}},\ }\href@noop {} {\bibfield  {journal} {\bibinfo
   {journal} {arXiv:2001.07101 [hep-th]}\ } (\bibinfo {year}
  {2020})}\BibitemShut {NoStop}%
\bibitem [{\citenamefont {Tobar}\ \emph
  {et~al.}(2019{\natexlab{a}})\citenamefont {Tobar}, \citenamefont
  {McAllister},\ and\ \citenamefont {Goryachev}}]{TobarModAx19}%
  \BibitemOpen
  \bibfield  {author} {\bibinfo {author} {\bibfnamefont {M.~E.}\ \bibnamefont
  {Tobar}}, \bibinfo {author} {\bibfnamefont {B.~T.}\ \bibnamefont
  {McAllister}}, \ and\ \bibinfo {author} {\bibfnamefont {M.}~\bibnamefont
  {Goryachev}},\ }\href {\doibase https://doi.org/10.1016/j.dark.2019.100339}
  {\bibfield  {journal} {\bibinfo  {journal} {Physics of the Dark Universe}\
  }\textbf {\bibinfo {volume} {26}},\ \bibinfo {pages} {100339} (\bibinfo
  {year} {2019}{\natexlab{a}})}\BibitemShut {NoStop}%
\bibitem [{\citenamefont {Wuensch}\ \emph {et~al.}(1989)\citenamefont
  {Wuensch}, \citenamefont {De~Panfilis-Wuensch}, \citenamefont {Semertzidis},
  \citenamefont {Rogers}, \citenamefont {Melissinos}, \citenamefont {Halama},
  \citenamefont {Moskowitz}, \citenamefont {Prodell}, \citenamefont {Fowler},\
  and\ \citenamefont {Nezrick}}]{Wuensch}%
  \BibitemOpen
  \bibfield  {author} {\bibinfo {author} {\bibfnamefont {W.~U.}\ \bibnamefont
  {Wuensch}}, \bibinfo {author} {\bibfnamefont {S.}~\bibnamefont
  {De~Panfilis-Wuensch}}, \bibinfo {author} {\bibfnamefont {Y.~K.}\
  \bibnamefont {Semertzidis}}, \bibinfo {author} {\bibfnamefont {J.~T.}\
  \bibnamefont {Rogers}}, \bibinfo {author} {\bibfnamefont {A.~C.}\
  \bibnamefont {Melissinos}}, \bibinfo {author} {\bibfnamefont {H.~J.}\
  \bibnamefont {Halama}}, \bibinfo {author} {\bibfnamefont {B.~E.}\
  \bibnamefont {Moskowitz}}, \bibinfo {author} {\bibfnamefont {A.~G.}\
  \bibnamefont {Prodell}}, \bibinfo {author} {\bibfnamefont {W.~B.}\
  \bibnamefont {Fowler}}, \ and\ \bibinfo {author} {\bibfnamefont {F.~A.}\
  \bibnamefont {Nezrick}},\ }\href {\doibase 10.1103/PhysRevD.40.3153}
  {\bibfield  {journal} {\bibinfo  {journal} {Phys. Rev. D}\ }\textbf {\bibinfo
  {volume} {40}},\ \bibinfo {pages} {3153} (\bibinfo {year}
  {1989})}\BibitemShut {NoStop}%
\bibitem [{\citenamefont {Hagmann}\ \emph {et~al.}(1990)\citenamefont
  {Hagmann}, \citenamefont {Sikivie}, \citenamefont {Sullivan}, \citenamefont
  {Tanner},\ and\ \citenamefont {Cho}}]{hagmann1990}%
  \BibitemOpen
  \bibfield  {author} {\bibinfo {author} {\bibfnamefont {C.}~\bibnamefont
  {Hagmann}}, \bibinfo {author} {\bibfnamefont {P.}~\bibnamefont {Sikivie}},
  \bibinfo {author} {\bibfnamefont {N.}~\bibnamefont {Sullivan}}, \bibinfo
  {author} {\bibfnamefont {D.~B.}\ \bibnamefont {Tanner}}, \ and\ \bibinfo
  {author} {\bibfnamefont {S.-I.}\ \bibnamefont {Cho}},\ }\href {\doibase
  10.1063/1.1141427} {\bibfield  {journal} {\bibinfo  {journal} {Review of
  Scientific Instruments}\ }\textbf {\bibinfo {volume} {61}},\ \bibinfo {pages}
  {1076} (\bibinfo {year} {1990})}\BibitemShut {NoStop}%
\bibitem [{\citenamefont {Bradley}\ \emph {et~al.}(2003)\citenamefont
  {Bradley}, \citenamefont {Clarke}, \citenamefont {Kinion}, \citenamefont
  {Rosenberg}, \citenamefont {van Bibber}, \citenamefont {Matsuki},
  \citenamefont {M\"uck},\ and\ \citenamefont {Sikivie}}]{Bradley2003}%
  \BibitemOpen
  \bibfield  {author} {\bibinfo {author} {\bibfnamefont {R.}~\bibnamefont
  {Bradley}}, \bibinfo {author} {\bibfnamefont {J.}~\bibnamefont {Clarke}},
  \bibinfo {author} {\bibfnamefont {D.}~\bibnamefont {Kinion}}, \bibinfo
  {author} {\bibfnamefont {L.~J.}\ \bibnamefont {Rosenberg}}, \bibinfo {author}
  {\bibfnamefont {K.}~\bibnamefont {van Bibber}}, \bibinfo {author}
  {\bibfnamefont {S.}~\bibnamefont {Matsuki}}, \bibinfo {author} {\bibfnamefont
  {M.}~\bibnamefont {M\"uck}}, \ and\ \bibinfo {author} {\bibfnamefont
  {P.}~\bibnamefont {Sikivie}},\ }\href {\doibase 10.1103/RevModPhys.75.777}
  {\bibfield  {journal} {\bibinfo  {journal} {Rev. Mod. Phys.}\ }\textbf
  {\bibinfo {volume} {75}},\ \bibinfo {pages} {777} (\bibinfo {year}
  {2003})}\BibitemShut {NoStop}%
\bibitem [{\citenamefont {Asztalos}\ \emph {et~al.}(2010)\citenamefont
  {Asztalos}, \citenamefont {Carosi}, \citenamefont {Hagmann}, \citenamefont
  {Kinion}, \citenamefont {van Bibber}, \citenamefont {Hotz}, \citenamefont
  {Rosenberg}, \citenamefont {Rybka}, \citenamefont {Hoskins}, \citenamefont
  {Hwang}, \citenamefont {Sikivie}, \citenamefont {Tanner}, \citenamefont
  {Bradley},\ and\ \citenamefont {Clarke}}]{ADMXaxions2010}%
  \BibitemOpen
  \bibfield  {author} {\bibinfo {author} {\bibfnamefont {S.~J.}\ \bibnamefont
  {Asztalos}}, \bibinfo {author} {\bibfnamefont {G.}~\bibnamefont {Carosi}},
  \bibinfo {author} {\bibfnamefont {C.}~\bibnamefont {Hagmann}}, \bibinfo
  {author} {\bibfnamefont {D.}~\bibnamefont {Kinion}}, \bibinfo {author}
  {\bibfnamefont {K.}~\bibnamefont {van Bibber}}, \bibinfo {author}
  {\bibfnamefont {M.}~\bibnamefont {Hotz}}, \bibinfo {author} {\bibfnamefont
  {L.~J.}\ \bibnamefont {Rosenberg}}, \bibinfo {author} {\bibfnamefont
  {G.}~\bibnamefont {Rybka}}, \bibinfo {author} {\bibfnamefont
  {J.}~\bibnamefont {Hoskins}}, \bibinfo {author} {\bibfnamefont
  {J.}~\bibnamefont {Hwang}}, \bibinfo {author} {\bibfnamefont
  {P.}~\bibnamefont {Sikivie}}, \bibinfo {author} {\bibfnamefont {D.~B.}\
  \bibnamefont {Tanner}}, \bibinfo {author} {\bibfnamefont {R.}~\bibnamefont
  {Bradley}}, \ and\ \bibinfo {author} {\bibfnamefont {J.}~\bibnamefont
  {Clarke}},\ }\href {\doibase 10.1103/PhysRevLett.104.041301} {\bibfield
  {journal} {\bibinfo  {journal} {Phys. Rev. Lett.}\ }\textbf {\bibinfo
  {volume} {104}},\ \bibinfo {pages} {041301} (\bibinfo {year}
  {2010})}\BibitemShut {NoStop}%
\bibitem [{\citenamefont {Hoskins}\ \emph {et~al.}(2011)\citenamefont
  {Hoskins}, \citenamefont {Hwang}, \citenamefont {Martin}, \citenamefont
  {Sikivie}, \citenamefont {Sullivan}, \citenamefont {Tanner}, \citenamefont
  {Hotz}, \citenamefont {Rosenberg}, \citenamefont {Rybka}, \citenamefont
  {Wagner}, \citenamefont {Asztalos}, \citenamefont {Carosi}, \citenamefont
  {Hagmann}, \citenamefont {Kinion}, \citenamefont {van Bibber}, \citenamefont
  {Bradley},\ and\ \citenamefont {Clarke}}]{ADMX2011}%
  \BibitemOpen
  \bibfield  {author} {\bibinfo {author} {\bibfnamefont {J.}~\bibnamefont
  {Hoskins}}, \bibinfo {author} {\bibfnamefont {J.}~\bibnamefont {Hwang}},
  \bibinfo {author} {\bibfnamefont {C.}~\bibnamefont {Martin}}, \bibinfo
  {author} {\bibfnamefont {P.}~\bibnamefont {Sikivie}}, \bibinfo {author}
  {\bibfnamefont {N.~S.}\ \bibnamefont {Sullivan}}, \bibinfo {author}
  {\bibfnamefont {D.~B.}\ \bibnamefont {Tanner}}, \bibinfo {author}
  {\bibfnamefont {M.}~\bibnamefont {Hotz}}, \bibinfo {author} {\bibfnamefont
  {L.~J.}\ \bibnamefont {Rosenberg}}, \bibinfo {author} {\bibfnamefont
  {G.}~\bibnamefont {Rybka}}, \bibinfo {author} {\bibfnamefont
  {A.}~\bibnamefont {Wagner}}, \bibinfo {author} {\bibfnamefont {S.~J.}\
  \bibnamefont {Asztalos}}, \bibinfo {author} {\bibfnamefont {G.}~\bibnamefont
  {Carosi}}, \bibinfo {author} {\bibfnamefont {C.}~\bibnamefont {Hagmann}},
  \bibinfo {author} {\bibfnamefont {D.}~\bibnamefont {Kinion}}, \bibinfo
  {author} {\bibfnamefont {K.}~\bibnamefont {van Bibber}}, \bibinfo {author}
  {\bibfnamefont {R.}~\bibnamefont {Bradley}}, \ and\ \bibinfo {author}
  {\bibfnamefont {J.}~\bibnamefont {Clarke}},\ }\href {\doibase
  10.1103/PhysRevD.84.121302} {\bibfield  {journal} {\bibinfo  {journal} {Phys.
  Rev. D}\ }\textbf {\bibinfo {volume} {84}},\ \bibinfo {pages} {121302}
  (\bibinfo {year} {2011})}\BibitemShut {NoStop}%
\bibitem [{\citenamefont {Sikivie}\ \emph {et~al.}(2014)\citenamefont
  {Sikivie}, \citenamefont {Sullivan},\ and\ \citenamefont
  {Tanner}}]{Sikivie2014a}%
  \BibitemOpen
  \bibfield  {author} {\bibinfo {author} {\bibfnamefont {P.}~\bibnamefont
  {Sikivie}}, \bibinfo {author} {\bibfnamefont {N.}~\bibnamefont {Sullivan}}, \
  and\ \bibinfo {author} {\bibfnamefont {D.~B.}\ \bibnamefont {Tanner}},\
  }\href {\doibase 10.1103/PhysRevLett.112.131301} {\bibfield  {journal}
  {\bibinfo  {journal} {Phys. Rev. Lett.}\ }\textbf {\bibinfo {volume} {112}},\
  \bibinfo {pages} {131301} (\bibinfo {year} {2014})}\BibitemShut {NoStop}%
\bibitem [{\citenamefont {Gupta}\ \emph {et~al.}(2016)\citenamefont {Gupta},
  \citenamefont {Anerella}, \citenamefont {Ghosh}, \citenamefont {Sampson},
  \citenamefont {Schmalzle}, \citenamefont {Konikowska}, \citenamefont
  {Semertzidis},\ and\ \citenamefont {Shin}}]{7390193}%
  \BibitemOpen
  \bibfield  {author} {\bibinfo {author} {\bibfnamefont {R.}~\bibnamefont
  {Gupta}}, \bibinfo {author} {\bibfnamefont {M.}~\bibnamefont {Anerella}},
  \bibinfo {author} {\bibfnamefont {A.}~\bibnamefont {Ghosh}}, \bibinfo
  {author} {\bibfnamefont {W.}~\bibnamefont {Sampson}}, \bibinfo {author}
  {\bibfnamefont {J.}~\bibnamefont {Schmalzle}}, \bibinfo {author}
  {\bibfnamefont {D.}~\bibnamefont {Konikowska}}, \bibinfo {author}
  {\bibfnamefont {Y.~K.}\ \bibnamefont {Semertzidis}}, \ and\ \bibinfo {author}
  {\bibfnamefont {Y.}~\bibnamefont {Shin}},\ }\href {\doibase
  10.1109/TASC.2016.2518240} {\bibfield  {journal} {\bibinfo  {journal} {IEEE
  Transactions on Applied Superconductivity}\ }\textbf {\bibinfo {volume}
  {26}},\ \bibinfo {pages} {1} (\bibinfo {year} {2016})}\BibitemShut {NoStop}%
\bibitem [{\citenamefont {McAllister}\ \emph
  {et~al.}(2016{\natexlab{a}})\citenamefont {McAllister}, \citenamefont
  {Parker},\ and\ \citenamefont {Tobar}}]{McAllisterFormFactor}%
  \BibitemOpen
  \bibfield  {author} {\bibinfo {author} {\bibfnamefont {B.~T.}\ \bibnamefont
  {McAllister}}, \bibinfo {author} {\bibfnamefont {S.~R.}\ \bibnamefont
  {Parker}}, \ and\ \bibinfo {author} {\bibfnamefont {M.~E.}\ \bibnamefont
  {Tobar}},\ }\href {\doibase 10.1103/PhysRevLett.117.159901,
  10.1103/PhysRevLett.116.161804} {\bibfield  {journal} {\bibinfo  {journal}
  {Phys. Rev. Lett.}\ }\textbf {\bibinfo {volume} {116}},\ \bibinfo {pages}
  {161804} (\bibinfo {year} {2016}{\natexlab{a}})},\ \bibinfo {note} {[Erratum:
  Phys. Rev. Lett.117,no.15,159901(2016)]},\ \Eprint
  {http://arxiv.org/abs/1607.01928} {arXiv:1607.01928 [hep-ph]} \BibitemShut
  {NoStop}%
%%CITATION = ARXIV:1607.01928;%%
\bibitem [{\citenamefont {McAllister}\ \emph
  {et~al.}(2016{\natexlab{b}})\citenamefont {McAllister}, \citenamefont
  {Parker},\ and\ \citenamefont {Tobar}}]{McAllister:2016fux}%
  \BibitemOpen
  \bibfield  {author} {\bibinfo {author} {\bibfnamefont {B.~T.}\ \bibnamefont
  {McAllister}}, \bibinfo {author} {\bibfnamefont {S.~R.}\ \bibnamefont
  {Parker}}, \ and\ \bibinfo {author} {\bibfnamefont {M.~E.}\ \bibnamefont
  {Tobar}},\ }\href {\doibase 10.1103/PhysRevD.94.042001} {\bibfield  {journal}
  {\bibinfo  {journal} {Phys. Rev.}\ }\textbf {\bibinfo {volume} {D94}},\
  \bibinfo {pages} {042001} (\bibinfo {year} {2016}{\natexlab{b}})},\ \Eprint
  {http://arxiv.org/abs/1605.05427} {arXiv:1605.05427 [physics.ins-det]}
  \BibitemShut {NoStop}%
%%CITATION = ARXIV:1605.05427;%%
\bibitem [{\citenamefont {Alesini}\ \emph {et~al.}(2017)\citenamefont
  {Alesini}, \citenamefont {Babusci}, \citenamefont {Gioacchino}, \citenamefont
  {Gatti}, \citenamefont {Lamanna},\ and\ \citenamefont {Ligi}}]{Klash}%
  \BibitemOpen
  \bibfield  {author} {\bibinfo {author} {\bibfnamefont {D.}~\bibnamefont
  {Alesini}}, \bibinfo {author} {\bibfnamefont {D.}~\bibnamefont {Babusci}},
  \bibinfo {author} {\bibfnamefont {D.~D.}\ \bibnamefont {Gioacchino}},
  \bibinfo {author} {\bibfnamefont {C.}~\bibnamefont {Gatti}}, \bibinfo
  {author} {\bibfnamefont {G.}~\bibnamefont {Lamanna}}, \ and\ \bibinfo
  {author} {\bibfnamefont {C.}~\bibnamefont {Ligi}},\ }\href@noop {} {\bibfield
   {journal} {\bibinfo  {journal} {arXiv:1707.06010 [physics.ins-det]}\ }
  (\bibinfo {year} {2017})}\BibitemShut {NoStop}%
\bibitem [{\citenamefont {Jeong}\ \emph {et~al.}(2018)\citenamefont {Jeong},
  \citenamefont {Youn}, \citenamefont {Ahn}, \citenamefont {Kim},\ and\
  \citenamefont {Semertzidis}}]{JEONG2018412}%
  \BibitemOpen
  \bibfield  {author} {\bibinfo {author} {\bibfnamefont {J.}~\bibnamefont
  {Jeong}}, \bibinfo {author} {\bibfnamefont {S.}~\bibnamefont {Youn}},
  \bibinfo {author} {\bibfnamefont {S.}~\bibnamefont {Ahn}}, \bibinfo {author}
  {\bibfnamefont {J.~E.}\ \bibnamefont {Kim}}, \ and\ \bibinfo {author}
  {\bibfnamefont {Y.~K.}\ \bibnamefont {Semertzidis}},\ }\href {\doibase
  https://doi.org/10.1016/j.physletb.2017.12.066} {\bibfield  {journal}
  {\bibinfo  {journal} {Physics Letters B}\ }\textbf {\bibinfo {volume}
  {777}},\ \bibinfo {pages} {412 } (\bibinfo {year} {2018})}\BibitemShut
  {NoStop}%
\bibitem [{\citenamefont {Baker}\ \emph {et~al.}(2012)\citenamefont {Baker},
  \citenamefont {Betz}, \citenamefont {Caspers}, \citenamefont {Jaeckel},
  \citenamefont {Lindner}, \citenamefont {Ringwald}, \citenamefont
  {Semertzidis}, \citenamefont {Sikivie},\ and\ \citenamefont
  {Zioutas}}]{Baker2012}%
  \BibitemOpen
  \bibfield  {author} {\bibinfo {author} {\bibfnamefont {O.~K.}\ \bibnamefont
  {Baker}}, \bibinfo {author} {\bibfnamefont {M.}~\bibnamefont {Betz}},
  \bibinfo {author} {\bibfnamefont {F.}~\bibnamefont {Caspers}}, \bibinfo
  {author} {\bibfnamefont {J.}~\bibnamefont {Jaeckel}}, \bibinfo {author}
  {\bibfnamefont {A.}~\bibnamefont {Lindner}}, \bibinfo {author} {\bibfnamefont
  {A.}~\bibnamefont {Ringwald}}, \bibinfo {author} {\bibfnamefont
  {Y.}~\bibnamefont {Semertzidis}}, \bibinfo {author} {\bibfnamefont
  {P.}~\bibnamefont {Sikivie}}, \ and\ \bibinfo {author} {\bibfnamefont
  {K.}~\bibnamefont {Zioutas}},\ }\href {\doibase 10.1103/PhysRevD.85.035018}
  {\bibfield  {journal} {\bibinfo  {journal} {Phys. Rev. D}\ }\textbf {\bibinfo
  {volume} {85}},\ \bibinfo {pages} {035018} (\bibinfo {year}
  {2012})}\BibitemShut {NoStop}%
\bibitem [{\citenamefont {Kahn}\ \emph {et~al.}(2016)\citenamefont {Kahn},
  \citenamefont {Safdi},\ and\ \citenamefont {Thaler}}]{ABRACADABRA}%
  \BibitemOpen
  \bibfield  {author} {\bibinfo {author} {\bibfnamefont {Y.}~\bibnamefont
  {Kahn}}, \bibinfo {author} {\bibfnamefont {B.~R.}\ \bibnamefont {Safdi}}, \
  and\ \bibinfo {author} {\bibfnamefont {J.}~\bibnamefont {Thaler}},\ }\href
  {\doibase 10.1103/PhysRevLett.117.141801} {\bibfield  {journal} {\bibinfo
  {journal} {Phys. Rev. Lett.}\ }\textbf {\bibinfo {volume} {117}},\ \bibinfo
  {pages} {141801} (\bibinfo {year} {2016})},\ \Eprint
  {http://arxiv.org/abs/1602.01086} {arXiv:1602.01086 [hep-ph]} \BibitemShut
  {NoStop}%
%%CITATION = ARXIV:1602.01086;%%
\bibitem [{\citenamefont {{Silva-Feaver}}\ \emph {et~al.}(2017)\citenamefont
  {{Silva-Feaver}}, \citenamefont {{Chaudhuri}}, \citenamefont {{Cho}},
  \citenamefont {{Dawson}}, \citenamefont {{Graham}}, \citenamefont {{Irwin}},
  \citenamefont {{Kuenstner}}, \citenamefont {{Li}}, \citenamefont {{Mardon}},
  \citenamefont {{Moseley}}, \citenamefont {{Mule}}, \citenamefont {{Phipps}},
  \citenamefont {{Rajendran}}, \citenamefont {{Steffen}},\ and\ \citenamefont
  {{Young}}}]{DMRadio17}%
  \BibitemOpen
  \bibfield  {author} {\bibinfo {author} {\bibfnamefont {M.}~\bibnamefont
  {{Silva-Feaver}}}, \bibinfo {author} {\bibfnamefont {S.}~\bibnamefont
  {{Chaudhuri}}}, \bibinfo {author} {\bibfnamefont {H.}~\bibnamefont {{Cho}}},
  \bibinfo {author} {\bibfnamefont {C.}~\bibnamefont {{Dawson}}}, \bibinfo
  {author} {\bibfnamefont {P.}~\bibnamefont {{Graham}}}, \bibinfo {author}
  {\bibfnamefont {K.}~\bibnamefont {{Irwin}}}, \bibinfo {author} {\bibfnamefont
  {S.}~\bibnamefont {{Kuenstner}}}, \bibinfo {author} {\bibfnamefont
  {D.}~\bibnamefont {{Li}}}, \bibinfo {author} {\bibfnamefont {J.}~\bibnamefont
  {{Mardon}}}, \bibinfo {author} {\bibfnamefont {H.}~\bibnamefont {{Moseley}}},
  \bibinfo {author} {\bibfnamefont {R.}~\bibnamefont {{Mule}}}, \bibinfo
  {author} {\bibfnamefont {A.}~\bibnamefont {{Phipps}}}, \bibinfo {author}
  {\bibfnamefont {S.}~\bibnamefont {{Rajendran}}}, \bibinfo {author}
  {\bibfnamefont {Z.}~\bibnamefont {{Steffen}}}, \ and\ \bibinfo {author}
  {\bibfnamefont {B.}~\bibnamefont {{Young}}},\ }\href {\doibase
  10.1109/TASC.2016.2631425} {\bibfield  {journal} {\bibinfo  {journal} {IEEE
  Transactions on Applied Superconductivity}\ }\textbf {\bibinfo {volume}
  {27}},\ \bibinfo {pages} {1} (\bibinfo {year} {2017})}\BibitemShut {NoStop}%
\bibitem [{\citenamefont {Hoang}\ \emph {et~al.}(2017)\citenamefont {Hoang},
  \citenamefont {Jeong}, \citenamefont {Cao}, \citenamefont {Shin},
  \citenamefont {Ko},\ and\ \citenamefont {Semertzidis}}]{HOANG2017}%
  \BibitemOpen
  \bibfield  {author} {\bibinfo {author} {\bibfnamefont {P.~L.}\ \bibnamefont
  {Hoang}}, \bibinfo {author} {\bibfnamefont {J.}~\bibnamefont {Jeong}},
  \bibinfo {author} {\bibfnamefont {B.~X.}\ \bibnamefont {Cao}}, \bibinfo
  {author} {\bibfnamefont {Y.}~\bibnamefont {Shin}}, \bibinfo {author}
  {\bibfnamefont {B.}~\bibnamefont {Ko}}, \ and\ \bibinfo {author}
  {\bibfnamefont {Y.}~\bibnamefont {Semertzidis}},\ }\href {\doibase
  https://doi.org/10.1016/j.dark.2017.04.004} {\bibfield  {journal} {\bibinfo
  {journal} {Physics of the Dark Universe}\ } (\bibinfo {year} {2017}),\
  https://doi.org/10.1016/j.dark.2017.04.004}\BibitemShut {NoStop}%
\bibitem [{\citenamefont {Choi}\ \emph {et~al.}(2017)\citenamefont {Choi},
  \citenamefont {Themann}, \citenamefont {Lee}, \citenamefont {Ko},\ and\
  \citenamefont {Semertzidis}}]{PhysRevD.96.061102}%
  \BibitemOpen
  \bibfield  {author} {\bibinfo {author} {\bibfnamefont {J.}~\bibnamefont
  {Choi}}, \bibinfo {author} {\bibfnamefont {H.}~\bibnamefont {Themann}},
  \bibinfo {author} {\bibfnamefont {M.~J.}\ \bibnamefont {Lee}}, \bibinfo
  {author} {\bibfnamefont {B.~R.}\ \bibnamefont {Ko}}, \ and\ \bibinfo {author}
  {\bibfnamefont {Y.~K.}\ \bibnamefont {Semertzidis}},\ }\href {\doibase
  10.1103/PhysRevD.96.061102} {\bibfield  {journal} {\bibinfo  {journal} {Phys.
  Rev. D}\ }\textbf {\bibinfo {volume} {96}},\ \bibinfo {pages} {061102}
  (\bibinfo {year} {2017})}\BibitemShut {NoStop}%
\bibitem [{\citenamefont {Ouellet}\ \emph
  {et~al.}(2019{\natexlab{a}})\citenamefont {Ouellet}, \citenamefont {Salemi},
  \citenamefont {Foster}, \citenamefont {Henning}, \citenamefont {Bogorad},
  \citenamefont {Conrad}, \citenamefont {Formaggio}, \citenamefont {Kahn},
  \citenamefont {Minervini}, \citenamefont {Radovinsky}, \citenamefont {Rodd},
  \citenamefont {Safdi}, \citenamefont {Thaler}, \citenamefont {Winklehner},\
  and\ \citenamefont {Winslow}}]{FirstAbra}%
  \BibitemOpen
  \bibfield  {author} {\bibinfo {author} {\bibfnamefont {J.~L.}\ \bibnamefont
  {Ouellet}}, \bibinfo {author} {\bibfnamefont {C.~P.}\ \bibnamefont {Salemi}},
  \bibinfo {author} {\bibfnamefont {J.~W.}\ \bibnamefont {Foster}}, \bibinfo
  {author} {\bibfnamefont {R.}~\bibnamefont {Henning}}, \bibinfo {author}
  {\bibfnamefont {Z.}~\bibnamefont {Bogorad}}, \bibinfo {author} {\bibfnamefont
  {J.~M.}\ \bibnamefont {Conrad}}, \bibinfo {author} {\bibfnamefont {J.~A.}\
  \bibnamefont {Formaggio}}, \bibinfo {author} {\bibfnamefont {Y.}~\bibnamefont
  {Kahn}}, \bibinfo {author} {\bibfnamefont {J.}~\bibnamefont {Minervini}},
  \bibinfo {author} {\bibfnamefont {A.}~\bibnamefont {Radovinsky}}, \bibinfo
  {author} {\bibfnamefont {N.~L.}\ \bibnamefont {Rodd}}, \bibinfo {author}
  {\bibfnamefont {B.~R.}\ \bibnamefont {Safdi}}, \bibinfo {author}
  {\bibfnamefont {J.}~\bibnamefont {Thaler}}, \bibinfo {author} {\bibfnamefont
  {D.}~\bibnamefont {Winklehner}}, \ and\ \bibinfo {author} {\bibfnamefont
  {L.}~\bibnamefont {Winslow}},\ }\href {\doibase
  10.1103/PhysRevLett.122.121802} {\bibfield  {journal} {\bibinfo  {journal}
  {Phys. Rev. Lett.}\ }\textbf {\bibinfo {volume} {122}},\ \bibinfo {pages}
  {121802} (\bibinfo {year} {2019}{\natexlab{a}})}\BibitemShut {NoStop}%
\bibitem [{\citenamefont {Harrington}(2012)}]{RHbook2012}%
  \BibitemOpen
  \bibfield  {author} {\bibinfo {author} {\bibfnamefont {R.~E.}\ \bibnamefont
  {Harrington}},\ }\href@noop {} {\emph {\bibinfo {title} {Introduction to
  Electromagnetic Engineering}}},\ \bibinfo {edition} {2nd}\ ed.\ (\bibinfo
  {publisher} {Dover Publications, Inc.},\ \bibinfo {address} {31 East 2nd
  Street, Mineola, NY 11501},\ \bibinfo {year} {2012})\BibitemShut {NoStop}%
\bibitem [{\citenamefont {Balanis}(2012)}]{Balanis2012}%
  \BibitemOpen
  \bibfield  {author} {\bibinfo {author} {\bibfnamefont {C.~A.}\ \bibnamefont
  {Balanis}},\ }\href@noop {} {\emph {\bibinfo {title} {Advanced Engineering
  Electromagnetics}}}\ (\bibinfo  {publisher} {John Wiley},\ \bibinfo {year}
  {2012})\BibitemShut {NoStop}%
\bibitem [{\citenamefont {Tobar}\ \emph
  {et~al.}(2019{\natexlab{b}})\citenamefont {Tobar}, \citenamefont
  {McAllister},\ and\ \citenamefont {Goryachev}}]{TobarElectret}%
  \BibitemOpen
  \bibfield  {author} {\bibinfo {author} {\bibfnamefont {M.~E.}\ \bibnamefont
  {Tobar}}, \bibinfo {author} {\bibfnamefont {B.~T.}\ \bibnamefont
  {McAllister}}, \ and\ \bibinfo {author} {\bibfnamefont {M.}~\bibnamefont
  {Goryachev}},\ }\href@noop {} {\bibfield  {journal} {\bibinfo  {journal}
  {arXiv:1904.05774 [physics.class-ph]}\ } (\bibinfo {year}
  {2019}{\natexlab{b}})}\BibitemShut {NoStop}%
\bibitem [{\citenamefont {Gratus}\ \emph {et~al.}(2020)\citenamefont {Gratus},
  \citenamefont {McCall},\ and\ \citenamefont {Kinsler}}]{Kinsler20}%
  \BibitemOpen
  \bibfield  {author} {\bibinfo {author} {\bibfnamefont {J.}~\bibnamefont
  {Gratus}}, \bibinfo {author} {\bibfnamefont {M.~W.}\ \bibnamefont {McCall}},
  \ and\ \bibinfo {author} {\bibfnamefont {P.}~\bibnamefont {Kinsler}},\ }\href
  {\doibase 10.1103/PhysRevA.101.043804} {\bibfield  {journal} {\bibinfo
  {journal} {Phys. Rev. A}\ }\textbf {\bibinfo {volume} {101}},\ \bibinfo
  {pages} {043804} (\bibinfo {year} {2020})}\BibitemShut {NoStop}%
\bibitem [{\citenamefont {Visinelli}(2013)}]{VISINELLI}%
  \BibitemOpen
  \bibfield  {author} {\bibinfo {author} {\bibfnamefont {L.}~\bibnamefont
  {Visinelli}},\ }\href {\doibase 10.1142/S0217732313501629} {\bibfield
  {journal} {\bibinfo  {journal} {Modern Physics Letters A}\ }\textbf {\bibinfo
  {volume} {28}},\ \bibinfo {pages} {1350162} (\bibinfo {year} {2013})},\
  \Eprint {http://arxiv.org/abs/https://doi.org/10.1142/S0217732313501629}
  {https://doi.org/10.1142/S0217732313501629} \BibitemShut {NoStop}%
\bibitem [{\citenamefont {Goryachev}\ \emph {et~al.}(2019)\citenamefont
  {Goryachev}, \citenamefont {McAllister},\ and\ \citenamefont
  {Tobar}}]{freqmetrology}%
  \BibitemOpen
  \bibfield  {author} {\bibinfo {author} {\bibfnamefont {M.}~\bibnamefont
  {Goryachev}}, \bibinfo {author} {\bibfnamefont {B.}~\bibnamefont
  {McAllister}}, \ and\ \bibinfo {author} {\bibfnamefont {M.}~\bibnamefont
  {Tobar}},\ }\href@noop {} {\bibfield  {journal} {\bibinfo  {journal} {Physics
  of the Dark Universe}\ }\textbf {\bibinfo {volume} {26}},\ \bibinfo {pages}
  {100345} (\bibinfo {year} {2019})}\BibitemShut {NoStop}%
\bibitem [{\citenamefont {Crisosto}\ \emph {et~al.}(2019)\citenamefont
  {Crisosto}, \citenamefont {Rybka}, \citenamefont {Sikivie}, \citenamefont
  {Sullivan}, \citenamefont {Tanner},\ and\ \citenamefont {Yang}}]{Nicol19}%
  \BibitemOpen
  \bibfield  {author} {\bibinfo {author} {\bibfnamefont {N.}~\bibnamefont
  {Crisosto}}, \bibinfo {author} {\bibfnamefont {G.}~\bibnamefont {Rybka}},
  \bibinfo {author} {\bibfnamefont {P.}~\bibnamefont {Sikivie}}, \bibinfo
  {author} {\bibfnamefont {N.~S.}\ \bibnamefont {Sullivan}}, \bibinfo {author}
  {\bibfnamefont {D.~B.}\ \bibnamefont {Tanner}}, \ and\ \bibinfo {author}
  {\bibfnamefont {J.}~\bibnamefont {Yang}},\ }\href@noop {} {\emph {\bibinfo
  {title} {ADMX SLIC: Results from a Superconducting LC Circuit Investigating
  Cold Axions}}}\ (\bibinfo  {publisher} {arXiv:1911.05772 [astro-ph.CO]},\
  \bibinfo {year} {2019})\BibitemShut {NoStop}%
\bibitem [{\citenamefont {Thomson}\ \emph {et~al.}(2019)\citenamefont
  {Thomson}, \citenamefont {McAllister}, \citenamefont {Goryachev},
  \citenamefont {Ivanov},\ and\ \citenamefont {Tobar}}]{Cat19}%
  \BibitemOpen
  \bibfield  {author} {\bibinfo {author} {\bibfnamefont {C.~A.}\ \bibnamefont
  {Thomson}}, \bibinfo {author} {\bibfnamefont {B.~T.}\ \bibnamefont
  {McAllister}}, \bibinfo {author} {\bibfnamefont {M.}~\bibnamefont
  {Goryachev}}, \bibinfo {author} {\bibfnamefont {E.~N.}\ \bibnamefont
  {Ivanov}}, \ and\ \bibinfo {author} {\bibfnamefont {M.~E.}\ \bibnamefont
  {Tobar}},\ }\href@noop {} {\bibfield  {journal} {\bibinfo  {journal}
  {arXiv:1912.07751 [hep-ex]}\ } (\bibinfo {year} {2019})}\BibitemShut
  {NoStop}%
\bibitem [{\citenamefont {Ouellet}\ and\ \citenamefont
  {Bogorad}(2019)}]{Ouellet2018}%
  \BibitemOpen
  \bibfield  {author} {\bibinfo {author} {\bibfnamefont {J.}~\bibnamefont
  {Ouellet}}\ and\ \bibinfo {author} {\bibfnamefont {Z.}~\bibnamefont
  {Bogorad}},\ }\href {\doibase 10.1103/PhysRevD.99.055010} {\bibfield
  {journal} {\bibinfo  {journal} {Phys. Rev. D}\ }\textbf {\bibinfo {volume}
  {99}},\ \bibinfo {pages} {055010} (\bibinfo {year} {2019})}\BibitemShut
  {NoStop}%
\bibitem [{\citenamefont {Beutter}\ \emph {et~al.}(2019)\citenamefont
  {Beutter}, \citenamefont {Pargner}, \citenamefont {Schwetz},\ and\
  \citenamefont {Todarello}}]{Beutter18}%
  \BibitemOpen
  \bibfield  {author} {\bibinfo {author} {\bibfnamefont {M.}~\bibnamefont
  {Beutter}}, \bibinfo {author} {\bibfnamefont {A.}~\bibnamefont {Pargner}},
  \bibinfo {author} {\bibfnamefont {T.}~\bibnamefont {Schwetz}}, \ and\
  \bibinfo {author} {\bibfnamefont {E.}~\bibnamefont {Todarello}},\ }\href@noop
  {} {\bibfield  {journal} {\bibinfo  {journal} {JCAP}\ }\textbf {\bibinfo
  {volume} {02}},\ \bibinfo {pages} {026} (\bibinfo {year} {2019})}\BibitemShut
  {NoStop}%
\bibitem [{\citenamefont {Kim}\ \emph {et~al.}(2019)\citenamefont {Kim},
  \citenamefont {Kim}, \citenamefont {Jeong}, \citenamefont {Kim},
  \citenamefont {Shin},\ and\ \citenamefont {Semertzidis}}]{Younggeun18}%
  \BibitemOpen
  \bibfield  {author} {\bibinfo {author} {\bibfnamefont {Y.}~\bibnamefont
  {Kim}}, \bibinfo {author} {\bibfnamefont {D.}~\bibnamefont {Kim}}, \bibinfo
  {author} {\bibfnamefont {J.}~\bibnamefont {Jeong}}, \bibinfo {author}
  {\bibfnamefont {J.}~\bibnamefont {Kim}}, \bibinfo {author} {\bibfnamefont
  {Y.~C.}\ \bibnamefont {Shin}}, \ and\ \bibinfo {author} {\bibfnamefont
  {Y.~K.}\ \bibnamefont {Semertzidis}},\ }\href {\doibase
  https://doi.org/10.1016/j.dark.2019.100362} {\bibfield  {journal} {\bibinfo
  {journal} {Physics of the Dark Universe}\ }\textbf {\bibinfo {volume} {26}},\
  \bibinfo {pages} {100362} (\bibinfo {year} {2019})}\BibitemShut {NoStop}%
\bibitem [{\citenamefont {Graham}\ and\ \citenamefont
  {Rajendran}(2011)}]{Graham11}%
  \BibitemOpen
  \bibfield  {author} {\bibinfo {author} {\bibfnamefont {P.~W.}\ \bibnamefont
  {Graham}}\ and\ \bibinfo {author} {\bibfnamefont {S.}~\bibnamefont
  {Rajendran}},\ }\href {\doibase 10.1103/PhysRevD.84.055013} {\bibfield
  {journal} {\bibinfo  {journal} {Phys. Rev. D}\ }\textbf {\bibinfo {volume}
  {84}},\ \bibinfo {pages} {055013} (\bibinfo {year} {2011})}\BibitemShut
  {NoStop}%
\bibitem [{\citenamefont {Graham}\ and\ \citenamefont
  {Rajendran}(2013)}]{altmethods2013}%
  \BibitemOpen
  \bibfield  {author} {\bibinfo {author} {\bibfnamefont {P.~W.}\ \bibnamefont
  {Graham}}\ and\ \bibinfo {author} {\bibfnamefont {S.}~\bibnamefont
  {Rajendran}},\ }\href {\doibase 10.1103/PhysRevD.88.035023} {\bibfield
  {journal} {\bibinfo  {journal} {Phys. Rev. D}\ }\textbf {\bibinfo {volume}
  {88}},\ \bibinfo {pages} {035023} (\bibinfo {year} {2013})}\BibitemShut
  {NoStop}%
\bibitem [{\citenamefont {Budker}\ \emph {et~al.}(2014)\citenamefont {Budker},
  \citenamefont {Graham}, \citenamefont {Ledbetter}, \citenamefont
  {Rajendran},\ and\ \citenamefont {Sushkov}}]{BudkerPRX}%
  \BibitemOpen
  \bibfield  {author} {\bibinfo {author} {\bibfnamefont {D.}~\bibnamefont
  {Budker}}, \bibinfo {author} {\bibfnamefont {P.~W.}\ \bibnamefont {Graham}},
  \bibinfo {author} {\bibfnamefont {M.}~\bibnamefont {Ledbetter}}, \bibinfo
  {author} {\bibfnamefont {S.}~\bibnamefont {Rajendran}}, \ and\ \bibinfo
  {author} {\bibfnamefont {A.~O.}\ \bibnamefont {Sushkov}},\ }\href {\doibase
  10.1103/PhysRevX.4.021030} {\bibfield  {journal} {\bibinfo  {journal} {Phys.
  Rev. X}\ }\textbf {\bibinfo {volume} {4}},\ \bibinfo {pages} {021030}
  (\bibinfo {year} {2014})}\BibitemShut {NoStop}%
\bibitem [{\citenamefont {Wu}\ \emph {et~al.}(2019)\citenamefont {Wu},
  \citenamefont {Blanchard}, \citenamefont {Centers}, \citenamefont {Figueroa},
  \citenamefont {Garcon}, \citenamefont {Graham}, \citenamefont {Kimball},
  \citenamefont {Rajendran}, \citenamefont {Stadnik}, \citenamefont {Sushkov},
  \citenamefont {Wickenbrock},\ and\ \citenamefont {Budker}}]{Budker19}%
  \BibitemOpen
  \bibfield  {author} {\bibinfo {author} {\bibfnamefont {T.}~\bibnamefont
  {Wu}}, \bibinfo {author} {\bibfnamefont {J.~W.}\ \bibnamefont {Blanchard}},
  \bibinfo {author} {\bibfnamefont {G.~P.}\ \bibnamefont {Centers}}, \bibinfo
  {author} {\bibfnamefont {N.~L.}\ \bibnamefont {Figueroa}}, \bibinfo {author}
  {\bibfnamefont {A.}~\bibnamefont {Garcon}}, \bibinfo {author} {\bibfnamefont
  {P.~W.}\ \bibnamefont {Graham}}, \bibinfo {author} {\bibfnamefont {D.~F.~J.}\
  \bibnamefont {Kimball}}, \bibinfo {author} {\bibfnamefont {S.}~\bibnamefont
  {Rajendran}}, \bibinfo {author} {\bibfnamefont {Y.~V.}\ \bibnamefont
  {Stadnik}}, \bibinfo {author} {\bibfnamefont {A.~O.}\ \bibnamefont
  {Sushkov}}, \bibinfo {author} {\bibfnamefont {A.}~\bibnamefont
  {Wickenbrock}}, \ and\ \bibinfo {author} {\bibfnamefont {D.}~\bibnamefont
  {Budker}},\ }\href {\doibase 10.1103/PhysRevLett.122.191302} {\bibfield
  {journal} {\bibinfo  {journal} {Phys. Rev. Lett.}\ }\textbf {\bibinfo
  {volume} {122}},\ \bibinfo {pages} {191302} (\bibinfo {year}
  {2019})}\BibitemShut {NoStop}%
\bibitem [{\citenamefont {Abel}\ \emph {et~al.}(2017)\citenamefont {Abel},
  \citenamefont {Ayres}, \citenamefont {Ban}, \citenamefont {Bison},
  \citenamefont {Bodek}, \citenamefont {Bondar}, \citenamefont {Daum},
  \citenamefont {Fairbairn}, \citenamefont {Flambaum}, \citenamefont
  {Geltenbort}, \citenamefont {Green}, \citenamefont {Griffith}, \citenamefont
  {van~der Grinten}, \citenamefont {Grujic}, \citenamefont {Harris},
  \citenamefont {Hild}, \citenamefont {Iaydjiev}, \citenamefont {Ivanov},
  \citenamefont {Kasprzak}, \citenamefont {Kermaidic}, \citenamefont {Kirch},
  \citenamefont {Koch}, \citenamefont {Komposch}, \citenamefont {Koss},
  \citenamefont {Kozela}, \citenamefont {Krempel}, \citenamefont {Lauss},
  \citenamefont {Lefort}, \citenamefont {Lemi\`ere}, \citenamefont {Marsh},
  \citenamefont {Mohanmurthy}, \citenamefont {Mtchedlishvili}, \citenamefont
  {Musgrave}, \citenamefont {Piegsa}, \citenamefont {Pignol}, \citenamefont
  {Rawlik}, \citenamefont {Rebreyend}, \citenamefont {Ries}, \citenamefont
  {Roccia}, \citenamefont {Rozpedzik}, \citenamefont {Schmidt-Wellenburg},
  \citenamefont {Severijns}, \citenamefont {Shiers}, \citenamefont {Stadnik},
  \citenamefont {Weis}, \citenamefont {Wursten}, \citenamefont {Zejma},\ and\
  \citenamefont {Zsigmond}}]{RALnEDM17}%
  \BibitemOpen
  \bibfield  {author} {\bibinfo {author} {\bibfnamefont {C.}~\bibnamefont
  {Abel}}, \bibinfo {author} {\bibfnamefont {N.~J.}\ \bibnamefont {Ayres}},
  \bibinfo {author} {\bibfnamefont {G.}~\bibnamefont {Ban}}, \bibinfo {author}
  {\bibfnamefont {G.}~\bibnamefont {Bison}}, \bibinfo {author} {\bibfnamefont
  {K.}~\bibnamefont {Bodek}}, \bibinfo {author} {\bibfnamefont
  {V.}~\bibnamefont {Bondar}}, \bibinfo {author} {\bibfnamefont
  {M.}~\bibnamefont {Daum}}, \bibinfo {author} {\bibfnamefont {M.}~\bibnamefont
  {Fairbairn}}, \bibinfo {author} {\bibfnamefont {V.~V.}\ \bibnamefont
  {Flambaum}}, \bibinfo {author} {\bibfnamefont {P.}~\bibnamefont
  {Geltenbort}}, \bibinfo {author} {\bibfnamefont {K.}~\bibnamefont {Green}},
  \bibinfo {author} {\bibfnamefont {W.~C.}\ \bibnamefont {Griffith}}, \bibinfo
  {author} {\bibfnamefont {M.}~\bibnamefont {van~der Grinten}}, \bibinfo
  {author} {\bibfnamefont {Z.~D.}\ \bibnamefont {Grujic}}, \bibinfo {author}
  {\bibfnamefont {P.~G.}\ \bibnamefont {Harris}}, \bibinfo {author}
  {\bibfnamefont {N.}~\bibnamefont {Hild}}, \bibinfo {author} {\bibfnamefont
  {P.}~\bibnamefont {Iaydjiev}}, \bibinfo {author} {\bibfnamefont {S.~N.}\
  \bibnamefont {Ivanov}}, \bibinfo {author} {\bibfnamefont {M.}~\bibnamefont
  {Kasprzak}}, \bibinfo {author} {\bibfnamefont {Y.}~\bibnamefont {Kermaidic}},
  \bibinfo {author} {\bibfnamefont {K.}~\bibnamefont {Kirch}}, \bibinfo
  {author} {\bibfnamefont {H.-C.}\ \bibnamefont {Koch}}, \bibinfo {author}
  {\bibfnamefont {S.}~\bibnamefont {Komposch}}, \bibinfo {author}
  {\bibfnamefont {P.~A.}\ \bibnamefont {Koss}}, \bibinfo {author}
  {\bibfnamefont {A.}~\bibnamefont {Kozela}}, \bibinfo {author} {\bibfnamefont
  {J.}~\bibnamefont {Krempel}}, \bibinfo {author} {\bibfnamefont
  {B.}~\bibnamefont {Lauss}}, \bibinfo {author} {\bibfnamefont
  {T.}~\bibnamefont {Lefort}}, \bibinfo {author} {\bibfnamefont
  {Y.}~\bibnamefont {Lemi\`ere}}, \bibinfo {author} {\bibfnamefont {D.~J.~E.}\
  \bibnamefont {Marsh}}, \bibinfo {author} {\bibfnamefont {P.}~\bibnamefont
  {Mohanmurthy}}, \bibinfo {author} {\bibfnamefont {A.}~\bibnamefont
  {Mtchedlishvili}}, \bibinfo {author} {\bibfnamefont {M.}~\bibnamefont
  {Musgrave}}, \bibinfo {author} {\bibfnamefont {F.~M.}\ \bibnamefont
  {Piegsa}}, \bibinfo {author} {\bibfnamefont {G.}~\bibnamefont {Pignol}},
  \bibinfo {author} {\bibfnamefont {M.}~\bibnamefont {Rawlik}}, \bibinfo
  {author} {\bibfnamefont {D.}~\bibnamefont {Rebreyend}}, \bibinfo {author}
  {\bibfnamefont {D.}~\bibnamefont {Ries}}, \bibinfo {author} {\bibfnamefont
  {S.}~\bibnamefont {Roccia}}, \bibinfo {author} {\bibfnamefont
  {D.}~\bibnamefont {Rozpedzik}}, \bibinfo {author} {\bibfnamefont
  {P.}~\bibnamefont {Schmidt-Wellenburg}}, \bibinfo {author} {\bibfnamefont
  {N.}~\bibnamefont {Severijns}}, \bibinfo {author} {\bibfnamefont
  {D.}~\bibnamefont {Shiers}}, \bibinfo {author} {\bibfnamefont {Y.~V.}\
  \bibnamefont {Stadnik}}, \bibinfo {author} {\bibfnamefont {A.}~\bibnamefont
  {Weis}}, \bibinfo {author} {\bibfnamefont {E.}~\bibnamefont {Wursten}},
  \bibinfo {author} {\bibfnamefont {J.}~\bibnamefont {Zejma}}, \ and\ \bibinfo
  {author} {\bibfnamefont {G.}~\bibnamefont {Zsigmond}},\ }\href {\doibase
  10.1103/PhysRevX.7.041034} {\bibfield  {journal} {\bibinfo  {journal} {Phys.
  Rev. X}\ }\textbf {\bibinfo {volume} {7}},\ \bibinfo {pages} {041034}
  (\bibinfo {year} {2017})}\BibitemShut {NoStop}%
\bibitem [{\citenamefont {Gooth}\ \emph {et~al.}(2019)\citenamefont {Gooth},
  \citenamefont {Bradlyn}, \citenamefont {Honnali}, \citenamefont {Schindler},
  \citenamefont {Kumar}, \citenamefont {Noky}, \citenamefont {Qi},
  \citenamefont {Shekhar}, \citenamefont {Sun}, \citenamefont {Wang},
  \citenamefont {Bernevig},\ and\ \citenamefont {Felser}}]{Gooth:2019np}%
  \BibitemOpen
  \bibfield  {author} {\bibinfo {author} {\bibfnamefont {J.}~\bibnamefont
  {Gooth}}, \bibinfo {author} {\bibfnamefont {B.}~\bibnamefont {Bradlyn}},
  \bibinfo {author} {\bibfnamefont {S.}~\bibnamefont {Honnali}}, \bibinfo
  {author} {\bibfnamefont {C.}~\bibnamefont {Schindler}}, \bibinfo {author}
  {\bibfnamefont {N.}~\bibnamefont {Kumar}}, \bibinfo {author} {\bibfnamefont
  {J.}~\bibnamefont {Noky}}, \bibinfo {author} {\bibfnamefont {Y.}~\bibnamefont
  {Qi}}, \bibinfo {author} {\bibfnamefont {C.}~\bibnamefont {Shekhar}},
  \bibinfo {author} {\bibfnamefont {Y.}~\bibnamefont {Sun}}, \bibinfo {author}
  {\bibfnamefont {Z.}~\bibnamefont {Wang}}, \bibinfo {author} {\bibfnamefont
  {B.~A.}\ \bibnamefont {Bernevig}}, \ and\ \bibinfo {author} {\bibfnamefont
  {C.}~\bibnamefont {Felser}},\ }\href {\doibase 10.1038/s41586-019-1630-4}
  {\bibfield  {journal} {\bibinfo  {journal} {Nature}\ }\textbf {\bibinfo
  {volume} {575}},\ \bibinfo {pages} {315} (\bibinfo {year}
  {2019})}\BibitemShut {NoStop}%
\bibitem [{\citenamefont {Cao}\ and\ \citenamefont
  {Zhitnitsky}(2017)}]{Cao2017}%
  \BibitemOpen
  \bibfield  {author} {\bibinfo {author} {\bibfnamefont {C.}~\bibnamefont
  {Cao}}\ and\ \bibinfo {author} {\bibfnamefont {A.}~\bibnamefont
  {Zhitnitsky}},\ }\href {\doibase 10.1103/PhysRevD.96.015013} {\bibfield
  {journal} {\bibinfo  {journal} {Phys. Rev. D}\ }\textbf {\bibinfo {volume}
  {96}},\ \bibinfo {pages} {015013} (\bibinfo {year} {2017})}\BibitemShut
  {NoStop}%
\bibitem [{\citenamefont {Anton}\ \emph {et~al.}(2013)\citenamefont {Anton},
  \citenamefont {Birenbaum}, \citenamefont {O'Kelley}, \citenamefont
  {Bolkhovsky}, \citenamefont {Braje}, \citenamefont {Fitch}, \citenamefont
  {Neeley}, \citenamefont {Hilton}, \citenamefont {Cho}, \citenamefont {Irwin},
  \citenamefont {Wellstood}, \citenamefont {Oliver}, \citenamefont {Shnirman},\
  and\ \citenamefont {Clarke}}]{FlickSquid}%
  \BibitemOpen
  \bibfield  {author} {\bibinfo {author} {\bibfnamefont {S.~M.}\ \bibnamefont
  {Anton}}, \bibinfo {author} {\bibfnamefont {J.~S.}\ \bibnamefont
  {Birenbaum}}, \bibinfo {author} {\bibfnamefont {S.~R.}\ \bibnamefont
  {O'Kelley}}, \bibinfo {author} {\bibfnamefont {V.}~\bibnamefont
  {Bolkhovsky}}, \bibinfo {author} {\bibfnamefont {D.~A.}\ \bibnamefont
  {Braje}}, \bibinfo {author} {\bibfnamefont {G.}~\bibnamefont {Fitch}},
  \bibinfo {author} {\bibfnamefont {M.}~\bibnamefont {Neeley}}, \bibinfo
  {author} {\bibfnamefont {G.~C.}\ \bibnamefont {Hilton}}, \bibinfo {author}
  {\bibfnamefont {H.-M.}\ \bibnamefont {Cho}}, \bibinfo {author} {\bibfnamefont
  {K.~D.}\ \bibnamefont {Irwin}}, \bibinfo {author} {\bibfnamefont {F.~C.}\
  \bibnamefont {Wellstood}}, \bibinfo {author} {\bibfnamefont {W.~D.}\
  \bibnamefont {Oliver}}, \bibinfo {author} {\bibfnamefont {A.}~\bibnamefont
  {Shnirman}}, \ and\ \bibinfo {author} {\bibfnamefont {J.}~\bibnamefont
  {Clarke}},\ }\href {\doibase 10.1103/PhysRevLett.110.147002} {\bibfield
  {journal} {\bibinfo  {journal} {Phys. Rev. Lett.}\ }\textbf {\bibinfo
  {volume} {110}},\ \bibinfo {pages} {147002} (\bibinfo {year}
  {2013})}\BibitemShut {NoStop}%
\bibitem [{\citenamefont {Goryachev}\ \emph {et~al.}(2014)\citenamefont
  {Goryachev}, \citenamefont {Ivanov}, \citenamefont {van Kann}, \citenamefont
  {Galliou},\ and\ \citenamefont {Tobar}}]{SQUIDQuartz}%
  \BibitemOpen
  \bibfield  {author} {\bibinfo {author} {\bibfnamefont {M.}~\bibnamefont
  {Goryachev}}, \bibinfo {author} {\bibfnamefont {E.~N.}\ \bibnamefont
  {Ivanov}}, \bibinfo {author} {\bibfnamefont {F.}~\bibnamefont {van Kann}},
  \bibinfo {author} {\bibfnamefont {S.}~\bibnamefont {Galliou}}, \ and\
  \bibinfo {author} {\bibfnamefont {M.~E.}\ \bibnamefont {Tobar}},\ }\href
  {\doibase 10.1063/1.4898813} {\bibfield  {journal} {\bibinfo  {journal}
  {Applied Physics Letters}\ }\textbf {\bibinfo {volume} {105}},\ \bibinfo
  {pages} {153505} (\bibinfo {year} {2014})},\ \Eprint
  {http://arxiv.org/abs/https://doi.org/10.1063/1.4898813}
  {https://doi.org/10.1063/1.4898813} \BibitemShut {NoStop}%
\bibitem [{\citenamefont {Ouellet}\ \emph
  {et~al.}(2019{\natexlab{b}})\citenamefont {Ouellet}, \citenamefont {Salemi},
  \citenamefont {Foster}, \citenamefont {Henning}, \citenamefont {Bogorad},
  \citenamefont {Conrad}, \citenamefont {Formaggio}, \citenamefont {Kahn},
  \citenamefont {Minervini}, \citenamefont {Radovinsky}, \citenamefont {Rodd},
  \citenamefont {Safdi}, \citenamefont {Thaler}, \citenamefont {Winklehner},\
  and\ \citenamefont {Winslow}}]{Oue19}%
  \BibitemOpen
  \bibfield  {author} {\bibinfo {author} {\bibfnamefont {J.~L.}\ \bibnamefont
  {Ouellet}}, \bibinfo {author} {\bibfnamefont {C.~P.}\ \bibnamefont {Salemi}},
  \bibinfo {author} {\bibfnamefont {J.~W.}\ \bibnamefont {Foster}}, \bibinfo
  {author} {\bibfnamefont {R.}~\bibnamefont {Henning}}, \bibinfo {author}
  {\bibfnamefont {Z.}~\bibnamefont {Bogorad}}, \bibinfo {author} {\bibfnamefont
  {J.~M.}\ \bibnamefont {Conrad}}, \bibinfo {author} {\bibfnamefont {J.~A.}\
  \bibnamefont {Formaggio}}, \bibinfo {author} {\bibfnamefont {Y.}~\bibnamefont
  {Kahn}}, \bibinfo {author} {\bibfnamefont {J.}~\bibnamefont {Minervini}},
  \bibinfo {author} {\bibfnamefont {A.}~\bibnamefont {Radovinsky}}, \bibinfo
  {author} {\bibfnamefont {N.~L.}\ \bibnamefont {Rodd}}, \bibinfo {author}
  {\bibfnamefont {B.~R.}\ \bibnamefont {Safdi}}, \bibinfo {author}
  {\bibfnamefont {J.}~\bibnamefont {Thaler}}, \bibinfo {author} {\bibfnamefont
  {D.}~\bibnamefont {Winklehner}}, \ and\ \bibinfo {author} {\bibfnamefont
  {L.}~\bibnamefont {Winslow}},\ }\href {\doibase 10.1103/PhysRevD.99.052012}
  {\bibfield  {journal} {\bibinfo  {journal} {Phys. Rev. D}\ }\textbf {\bibinfo
  {volume} {99}},\ \bibinfo {pages} {052012} (\bibinfo {year}
  {2019}{\natexlab{b}})}\BibitemShut {NoStop}%
\bibitem [{\citenamefont {McAllister}\ \emph {et~al.}(2017)\citenamefont
  {McAllister}, \citenamefont {Flower}, \citenamefont {Ivanov}, \citenamefont
  {Goryachev}, \citenamefont {Bourhill},\ and\ \citenamefont {Tobar}}]{ORGAN}%
  \BibitemOpen
  \bibfield  {author} {\bibinfo {author} {\bibfnamefont {B.~T.}\ \bibnamefont
  {McAllister}}, \bibinfo {author} {\bibfnamefont {G.}~\bibnamefont {Flower}},
  \bibinfo {author} {\bibfnamefont {E.~N.}\ \bibnamefont {Ivanov}}, \bibinfo
  {author} {\bibfnamefont {M.}~\bibnamefont {Goryachev}}, \bibinfo {author}
  {\bibfnamefont {J.}~\bibnamefont {Bourhill}}, \ and\ \bibinfo {author}
  {\bibfnamefont {M.~E.}\ \bibnamefont {Tobar}},\ }\href {\doibase
  https://doi.org/10.1016/j.dark.2017.09.010} {\bibfield  {journal} {\bibinfo
  {journal} {Physics of the Dark Universe}\ }\textbf {\bibinfo {volume} {18}},\
  \bibinfo {pages} {67 } (\bibinfo {year} {2017})}\BibitemShut {NoStop}%
\bibitem [{HIA(2016)}]{HIA}%
  \BibitemOpen
  \href@noop {} {\emph {\bibinfo {title} {HFC 50 D / E Dual Cryogenic Ultra Low
  Noise RF-Amplifier}}},\ \bibinfo {organization} {Stahl Electronics} (\bibinfo
  {year} {2016}),\ \bibinfo {note} {version 2.38}\BibitemShut {NoStop}%
\bibitem [{\citenamefont {Damour}\ \emph {et~al.}(1990)\citenamefont {Damour},
  \citenamefont {Gibbons},\ and\ \citenamefont {Gundlach}}]{Damour1990}%
  \BibitemOpen
  \bibfield  {author} {\bibinfo {author} {\bibfnamefont {T.}~\bibnamefont
  {Damour}}, \bibinfo {author} {\bibfnamefont {G.~W.}\ \bibnamefont {Gibbons}},
  \ and\ \bibinfo {author} {\bibfnamefont {C.}~\bibnamefont {Gundlach}},\
  }\href {\doibase 10.1103/physrevlett.64.123} {\bibfield  {journal} {\bibinfo
  {journal} {Physical Review Letters}\ }\textbf {\bibinfo {volume} {64}},\
  \bibinfo {pages} {123} (\bibinfo {year} {1990})}\BibitemShut {NoStop}%
\bibitem [{\citenamefont {Damour}\ and\ \citenamefont
  {Polyakov}(1994)}]{Damour1994}%
  \BibitemOpen
  \bibfield  {author} {\bibinfo {author} {\bibfnamefont {T.}~\bibnamefont
  {Damour}}\ and\ \bibinfo {author} {\bibfnamefont {A.}~\bibnamefont
  {Polyakov}},\ }\href {\doibase 10.1016/0550-3213(94)90143-0} {\bibfield
  {journal} {\bibinfo  {journal} {Nuclear Physics B}\ }\textbf {\bibinfo
  {volume} {423}},\ \bibinfo {pages} {532} (\bibinfo {year}
  {1994})}\BibitemShut {NoStop}%
\bibitem [{\citenamefont {Flambaum}\ \emph {et~al.}(2004)\citenamefont
  {Flambaum}, \citenamefont {Leinweber}, \citenamefont {Thomas},\ and\
  \citenamefont {Young}}]{Flambaum2004}%
  \BibitemOpen
  \bibfield  {author} {\bibinfo {author} {\bibfnamefont {V.~V.}\ \bibnamefont
  {Flambaum}}, \bibinfo {author} {\bibfnamefont {D.~B.}\ \bibnamefont
  {Leinweber}}, \bibinfo {author} {\bibfnamefont {A.~W.}\ \bibnamefont
  {Thomas}}, \ and\ \bibinfo {author} {\bibfnamefont {R.~D.}\ \bibnamefont
  {Young}},\ }\href {\doibase 10.1103/physrevd.69.115006} {\bibfield  {journal}
  {\bibinfo  {journal} {Physical Review D}\ }\textbf {\bibinfo {volume} {69}}
  (\bibinfo {year} {2004}),\ 10.1103/physrevd.69.115006}\BibitemShut {NoStop}%
\bibitem [{\citenamefont {Stadnik}\ and\ \citenamefont
  {Flambaum}(2015)}]{Stadnik2015}%
  \BibitemOpen
  \bibfield  {author} {\bibinfo {author} {\bibfnamefont {Y.}~\bibnamefont
  {Stadnik}}\ and\ \bibinfo {author} {\bibfnamefont {V.}~\bibnamefont
  {Flambaum}},\ }\href {\doibase 10.1103/physrevlett.115.201301} {\bibfield
  {journal} {\bibinfo  {journal} {Physical Review Letters}\ }\textbf {\bibinfo
  {volume} {115}} (\bibinfo {year} {2015}),\
  10.1103/physrevlett.115.201301}\BibitemShut {NoStop}%
\bibitem [{\citenamefont {Turneaure}\ \emph {et~al.}(1983)\citenamefont
  {Turneaure}, \citenamefont {Will}, \citenamefont {Farrell}, \citenamefont
  {Mattison},\ and\ \citenamefont {Vessot}}]{Turneaure1983}%
  \BibitemOpen
  \bibfield  {author} {\bibinfo {author} {\bibfnamefont {J.~P.}\ \bibnamefont
  {Turneaure}}, \bibinfo {author} {\bibfnamefont {C.~M.}\ \bibnamefont {Will}},
  \bibinfo {author} {\bibfnamefont {B.~F.}\ \bibnamefont {Farrell}}, \bibinfo
  {author} {\bibfnamefont {E.~M.}\ \bibnamefont {Mattison}}, \ and\ \bibinfo
  {author} {\bibfnamefont {R.~F.~C.}\ \bibnamefont {Vessot}},\ }\href {\doibase
  10.1103/physrevd.27.1705} {\bibfield  {journal} {\bibinfo  {journal}
  {Physical Review D}\ }\textbf {\bibinfo {volume} {27}},\ \bibinfo {pages}
  {1705} (\bibinfo {year} {1983})}\BibitemShut {NoStop}%
\bibitem [{\citenamefont {Tobar}\ \emph {et~al.}(2010)\citenamefont {Tobar},
  \citenamefont {Wolf}, \citenamefont {Bize}, \citenamefont {Santarelli},\ and\
  \citenamefont {Flambaum}}]{Tobar2009}%
  \BibitemOpen
  \bibfield  {author} {\bibinfo {author} {\bibfnamefont {M.~E.}\ \bibnamefont
  {Tobar}}, \bibinfo {author} {\bibfnamefont {P.}~\bibnamefont {Wolf}},
  \bibinfo {author} {\bibfnamefont {S.}~\bibnamefont {Bize}}, \bibinfo {author}
  {\bibfnamefont {G.}~\bibnamefont {Santarelli}}, \ and\ \bibinfo {author}
  {\bibfnamefont {V.}~\bibnamefont {Flambaum}},\ }\href {\doibase
  10.1103/PhysRevD.81.022003} {\bibfield  {journal} {\bibinfo  {journal} {Phys.
  Rev. D}\ }\textbf {\bibinfo {volume} {81}},\ \bibinfo {pages} {022003}
  (\bibinfo {year} {2010})}\BibitemShut {NoStop}%
\bibitem [{\citenamefont {Tilburg}\ \emph {et~al.}(2015)\citenamefont
  {Tilburg}, \citenamefont {Leefer}, \citenamefont {Bougas},\ and\
  \citenamefont {Budker}}]{Tilburg2015}%
  \BibitemOpen
  \bibfield  {author} {\bibinfo {author} {\bibfnamefont {K.~V.}\ \bibnamefont
  {Tilburg}}, \bibinfo {author} {\bibfnamefont {N.}~\bibnamefont {Leefer}},
  \bibinfo {author} {\bibfnamefont {L.}~\bibnamefont {Bougas}}, \ and\ \bibinfo
  {author} {\bibfnamefont {D.}~\bibnamefont {Budker}},\ }\href {\doibase
  10.1103/physrevlett.115.011802} {\bibfield  {journal} {\bibinfo  {journal}
  {Physical Review Letters}\ }\textbf {\bibinfo {volume} {115}} (\bibinfo
  {year} {2015}),\ 10.1103/physrevlett.115.011802}\BibitemShut {NoStop}%
\bibitem [{\citenamefont {Hees}\ \emph {et~al.}(2018)\citenamefont {Hees},
  \citenamefont {Minazzoli}, \citenamefont {Savalle}, \citenamefont {Stadnik},\
  and\ \citenamefont {Wolf}}]{Hees2018}%
  \BibitemOpen
  \bibfield  {author} {\bibinfo {author} {\bibfnamefont {A.}~\bibnamefont
  {Hees}}, \bibinfo {author} {\bibfnamefont {O.}~\bibnamefont {Minazzoli}},
  \bibinfo {author} {\bibfnamefont {E.}~\bibnamefont {Savalle}}, \bibinfo
  {author} {\bibfnamefont {Y.~V.}\ \bibnamefont {Stadnik}}, \ and\ \bibinfo
  {author} {\bibfnamefont {P.}~\bibnamefont {Wolf}},\ }\href {\doibase
  10.1103/physrevd.98.064051} {\bibfield  {journal} {\bibinfo  {journal}
  {Physical Review D}\ }\textbf {\bibinfo {volume} {98}} (\bibinfo {year}
  {2018}),\ 10.1103/physrevd.98.064051}\BibitemShut {NoStop}%
\bibitem [{\citenamefont {Arvanitaki}\ \emph {et~al.}(2015)\citenamefont
  {Arvanitaki}, \citenamefont {Huang},\ and\ \citenamefont
  {Tilburg}}]{Arvanitaki2015}%
  \BibitemOpen
  \bibfield  {author} {\bibinfo {author} {\bibfnamefont {A.}~\bibnamefont
  {Arvanitaki}}, \bibinfo {author} {\bibfnamefont {J.}~\bibnamefont {Huang}}, \
  and\ \bibinfo {author} {\bibfnamefont {K.~V.}\ \bibnamefont {Tilburg}},\
  }\href {\doibase 10.1103/physrevd.91.015015} {\bibfield  {journal} {\bibinfo
  {journal} {Physical Review D}\ }\textbf {\bibinfo {volume} {91}} (\bibinfo
  {year} {2015}),\ 10.1103/physrevd.91.015015}\BibitemShut {NoStop}%
\bibitem [{\citenamefont {Schlamminger}\ \emph {et~al.}(2008)\citenamefont
  {Schlamminger}, \citenamefont {Choi}, \citenamefont {Wagner}, \citenamefont
  {Gundlach},\ and\ \citenamefont {Adelberger}}]{Schlamminger2008}%
  \BibitemOpen
  \bibfield  {author} {\bibinfo {author} {\bibfnamefont {S.}~\bibnamefont
  {Schlamminger}}, \bibinfo {author} {\bibfnamefont {K.-Y.}\ \bibnamefont
  {Choi}}, \bibinfo {author} {\bibfnamefont {T.~A.}\ \bibnamefont {Wagner}},
  \bibinfo {author} {\bibfnamefont {J.~H.}\ \bibnamefont {Gundlach}}, \ and\
  \bibinfo {author} {\bibfnamefont {E.~G.}\ \bibnamefont {Adelberger}},\ }\href
  {\doibase 10.1103/physrevlett.100.041101} {\bibfield  {journal} {\bibinfo
  {journal} {Physical Review Letters}\ }\textbf {\bibinfo {volume} {100}}
  (\bibinfo {year} {2008}),\ 10.1103/physrevlett.100.041101}\BibitemShut
  {NoStop}%
\bibitem [{\citenamefont {Smith}\ \emph {et~al.}(1999)\citenamefont {Smith},
  \citenamefont {Hoyle}, \citenamefont {Gundlach}, \citenamefont {Adelberger},
  \citenamefont {Heckel},\ and\ \citenamefont {Swanson}}]{Smith1999}%
  \BibitemOpen
  \bibfield  {author} {\bibinfo {author} {\bibfnamefont {G.~L.}\ \bibnamefont
  {Smith}}, \bibinfo {author} {\bibfnamefont {C.~D.}\ \bibnamefont {Hoyle}},
  \bibinfo {author} {\bibfnamefont {J.~H.}\ \bibnamefont {Gundlach}}, \bibinfo
  {author} {\bibfnamefont {E.~G.}\ \bibnamefont {Adelberger}}, \bibinfo
  {author} {\bibfnamefont {B.~R.}\ \bibnamefont {Heckel}}, \ and\ \bibinfo
  {author} {\bibfnamefont {H.~E.}\ \bibnamefont {Swanson}},\ }\href {\doibase
  10.1103/physrevd.61.022001} {\bibfield  {journal} {\bibinfo  {journal}
  {Physical Review D}\ }\textbf {\bibinfo {volume} {61}} (\bibinfo {year}
  {1999}),\ 10.1103/physrevd.61.022001}\BibitemShut {NoStop}%
\bibitem [{\citenamefont {Berg{\'{e}}}\ \emph {et~al.}(2018)\citenamefont
  {Berg{\'{e}}}, \citenamefont {Brax}, \citenamefont {M{\'{e}}tris},
  \citenamefont {Pernot-Borr{\`{a}}s}, \citenamefont {Touboul},\ and\
  \citenamefont {Uzan}}]{Berge2018}%
  \BibitemOpen
  \bibfield  {author} {\bibinfo {author} {\bibfnamefont {J.}~\bibnamefont
  {Berg{\'{e}}}}, \bibinfo {author} {\bibfnamefont {P.}~\bibnamefont {Brax}},
  \bibinfo {author} {\bibfnamefont {G.}~\bibnamefont {M{\'{e}}tris}}, \bibinfo
  {author} {\bibfnamefont {M.}~\bibnamefont {Pernot-Borr{\`{a}}s}}, \bibinfo
  {author} {\bibfnamefont {P.}~\bibnamefont {Touboul}}, \ and\ \bibinfo
  {author} {\bibfnamefont {J.-P.}\ \bibnamefont {Uzan}},\ }\href {\doibase
  10.1103/physrevlett.120.141101} {\bibfield  {journal} {\bibinfo  {journal}
  {Physical Review Letters}\ }\textbf {\bibinfo {volume} {120}} (\bibinfo
  {year} {2018}),\ 10.1103/physrevlett.120.141101}\BibitemShut {NoStop}%
\bibitem [{\citenamefont {Hlo\ifmmode~\check{z}\else \v{z}\fi{}ek}\ \emph
  {et~al.}(2017)\citenamefont {Hlo\ifmmode~\check{z}\else \v{z}\fi{}ek},
  \citenamefont {Marsh}, \citenamefont {Grin}, \citenamefont {Allison},
  \citenamefont {Dunkley},\ and\ \citenamefont {Calabrese}}]{ULACMB2017}%
  \BibitemOpen
  \bibfield  {author} {\bibinfo {author} {\bibfnamefont {R.}~\bibnamefont
  {Hlo\ifmmode~\check{z}\else \v{z}\fi{}ek}}, \bibinfo {author} {\bibfnamefont
  {D.~J.~E.}\ \bibnamefont {Marsh}}, \bibinfo {author} {\bibfnamefont
  {D.}~\bibnamefont {Grin}}, \bibinfo {author} {\bibfnamefont {R.}~\bibnamefont
  {Allison}}, \bibinfo {author} {\bibfnamefont {J.}~\bibnamefont {Dunkley}}, \
  and\ \bibinfo {author} {\bibfnamefont {E.}~\bibnamefont {Calabrese}},\ }\href
  {\doibase 10.1103/PhysRevD.95.123511} {\bibfield  {journal} {\bibinfo
  {journal} {Phys. Rev. D}\ }\textbf {\bibinfo {volume} {95}},\ \bibinfo
  {pages} {123511} (\bibinfo {year} {2017})}\BibitemShut {NoStop}%
\bibitem [{\citenamefont {Diez-Tejedor}\ and\ \citenamefont
  {Marsh}(2017)}]{Marsh17}%
  \BibitemOpen
  \bibfield  {author} {\bibinfo {author} {\bibfnamefont {A.}~\bibnamefont
  {Diez-Tejedor}}\ and\ \bibinfo {author} {\bibfnamefont {D.~J.~E.}\
  \bibnamefont {Marsh}},\ }\href@noop {} {\bibfield  {journal} {\bibinfo
  {journal} {arXiv:1702.02116 [hep-ph]}\ } (\bibinfo {year}
  {2017})}\BibitemShut {NoStop}%
\bibitem [{\citenamefont {Zhang}\ \emph {et~al.}(2018)\citenamefont {Zhang},
  \citenamefont {Tsai}, \citenamefont {Kuo}, \citenamefont {Cheung},\ and\
  \citenamefont {Chu}}]{Zhang_2018}%
  \BibitemOpen
  \bibfield  {author} {\bibinfo {author} {\bibfnamefont {J.}~\bibnamefont
  {Zhang}}, \bibinfo {author} {\bibfnamefont {Y.-L.~S.}\ \bibnamefont {Tsai}},
  \bibinfo {author} {\bibfnamefont {J.-L.}\ \bibnamefont {Kuo}}, \bibinfo
  {author} {\bibfnamefont {K.}~\bibnamefont {Cheung}}, \ and\ \bibinfo {author}
  {\bibfnamefont {M.-C.}\ \bibnamefont {Chu}},\ }\href {\doibase
  10.3847/1538-4357/aaa485} {\bibfield  {journal} {\bibinfo  {journal} {The
  Astrophysical Journal}\ }\textbf {\bibinfo {volume} {853}},\ \bibinfo {pages}
  {51} (\bibinfo {year} {2018})}\BibitemShut {NoStop}%
\bibitem [{\citenamefont {Fedderke}\ \emph {et~al.}(2019)\citenamefont
  {Fedderke}, \citenamefont {Graham},\ and\ \citenamefont
  {Rajendran}}]{Fedderke19}%
  \BibitemOpen
  \bibfield  {author} {\bibinfo {author} {\bibfnamefont {M.~A.}\ \bibnamefont
  {Fedderke}}, \bibinfo {author} {\bibfnamefont {P.~W.}\ \bibnamefont
  {Graham}}, \ and\ \bibinfo {author} {\bibfnamefont {S.}~\bibnamefont
  {Rajendran}},\ }\href {\doibase 10.1103/PhysRevD.100.015040} {\bibfield
  {journal} {\bibinfo  {journal} {Phys. Rev. D}\ }\textbf {\bibinfo {volume}
  {100}},\ \bibinfo {pages} {015040} (\bibinfo {year} {2019})}\BibitemShut
  {NoStop}%
\bibitem [{\citenamefont {Rosa}(1908)}]{Rosa1908}%
  \BibitemOpen
  \bibfield  {author} {\bibinfo {author} {\bibfnamefont {E.~B.}\ \bibnamefont
  {Rosa}},\ }\href@noop {} {\bibfield  {journal} {\bibinfo  {journal} {Bulletin
  of the Bureau of Standards}\ }\textbf {\bibinfo {volume} {4}},\ \bibinfo
  {pages} {301} (\bibinfo {year} {1908})}\BibitemShut {NoStop}%
\bibitem [{\citenamefont {Grover}(1946)}]{Grover46}%
  \BibitemOpen
  \bibfield  {author} {\bibinfo {author} {\bibfnamefont {F.~W.}\ \bibnamefont
  {Grover}},\ }\href@noop {} {\emph {\bibinfo {title} {Inductance Calculations:
  Working formulas and tables}}}\ (\bibinfo  {publisher} {New York: Dover
  Publications, Inc.},\ \bibinfo {year} {1946})\BibitemShut {NoStop}%
\bibitem [{\citenamefont {Fickett}(1982)}]{NBS1053}%
  \BibitemOpen
  \bibfield  {author} {\bibinfo {author} {\bibfnamefont {F.~R.}\ \bibnamefont
  {Fickett}},\ }\href@noop {} {\bibfield  {journal} {\bibinfo  {journal} {NBS
  Technical Note}\ }\textbf {\bibinfo {volume} {1053}} (\bibinfo {year}
  {1982})}\BibitemShut {NoStop}%
\bibitem [{\citenamefont {McAllister}\ \emph {et~al.}(2018)\citenamefont
  {McAllister}, \citenamefont {Goryachev}, \citenamefont {Bourhill},
  \citenamefont {Ivanov},\ and\ \citenamefont {Tobar}}]{BEAST}%
  \BibitemOpen
  \bibfield  {author} {\bibinfo {author} {\bibfnamefont {B.~T.}\ \bibnamefont
  {McAllister}}, \bibinfo {author} {\bibfnamefont {M.}~\bibnamefont
  {Goryachev}}, \bibinfo {author} {\bibfnamefont {J.}~\bibnamefont {Bourhill}},
  \bibinfo {author} {\bibfnamefont {E.~N.}\ \bibnamefont {Ivanov}}, \ and\
  \bibinfo {author} {\bibfnamefont {M.~E.}\ \bibnamefont {Tobar}},\ }\href@noop
  {} {\bibfield  {journal} {\bibinfo  {journal} {arXiv:1803.07755
  [physics.ins-det]}\ } (\bibinfo {year} {2018})}\BibitemShut {NoStop}%
\end{thebibliography}
\end{document}